%% file: odca.tex
\makeatletter
\def\input@path{ {./} {TeX/} }
\makeatother
\documentclass[a4paper,UKenglish,cleveref, autoref, thm-restate]{lipics-v2021}
\title{Weighted one-deterministic-counter automata}
\author{Prince Mathew}{Indian Institute of Technology, Goa \and \url{prince-iitgoa.github.io} }{prince@iitgoa.ac.in}{https://orcid.org/0000-0001-6410-1474}{}%TODO mandatory, please use full name; only 1 author per \author macro; first two parameters are mandatory, other parameters can be empty. Please provide at least the name of the affiliation and the country. The full address is optional. Use additional curly braces to indicate the correct name splitting when the last name consists of multiple name parts.
\author{Vincent Penelle}{Univ. Bordeaux, CNRS,  Bordeaux INP, LaBRI, UMR 5800, F-33400, Talence, France \and \url{https://www.labri.fr/perso/vpenelle/}}{vincent.penelle@u-bordeaux.fr}{}{}
\author{Prakash Saivasan}{The Institute of Mathematical Sciences, HBNI \and \url{https://www.imsc.res.in/prakash_saivasan}}{prakashs@imsc.res.in}{}{}
\author{A.V. Sreejith}{Indian Institute of Technology, Goa \and \url{https://www.iitgoa.ac.in/~sreejithav/}}{sreejithav@iitgoa.ac.in}{}{}%I would like to thank DST Matrics grant for the project ``Probabilistic Pushdown Automata". }

\authorrunning{P. Mathew, V. Penelle, P. Saivasan and A.V. Sreejith} %TODO mandatory. First: Use abbreviated first/middle names. Second (only in severe cases): Use first author plus 'et al.'

\Copyright{Prince Mathew, Vincent Penelle, Prakash Saivasan and A.V Sreejith} %TODO mandatory, please use full first names. LIPIcs license is "CC-BY";  http://creativecommons.org/licenses/by/3.0/

 %TODO mandatory: Please choose ACM 2012 classifications from https://dl.acm.org/ccs/ccs_flat.cfm 

 %TODO mandatory; please add comma-separated list of keywords

\category{} %optional, e.g. invited paper

\relatedversion{} %optional, e.g. full version hosted on arXiv, HAL, or other respository/website
\acknowledgements{A.V. Sreejith would like to thank the DST Matrics grant for the project ``Probabilistic Pushdown Automata". The authors would like to thank Dr.~Rahul C S, SMCS, IIT Goa, for his valuable and intuitive suggestions that helped in proving \Cref{specialword}.}%optional

\nolinenumbers %uncomment to disable line numbering

\ArticleNo{1}

\usepackage{hyperref}

\usepackage{graphicx} % Allows including images
\usepackage{booktabs} % Allows the use of \toprule, \midrule and \bottomrule in tables
\usepackage{mathtools}
\usepackage[margin=.15cm]{caption}
\usepackage{bm}
\usepackage{todonotes}
\usepackage{xspace}
\usepackage{soul}
\usepackage{amsthm}
\usepackage{amsfonts}
\usepackage{setspace}
\usepackage{extpfeil}
\usepackage{xcolor}
\usepackage{macros}
\usepackage{lineno}
\usepackage{wrapfig}

\usepackage{ipa}

\usepackage{ifthen}
\newcounter{num}
\usetikzlibrary{arrows,automata,positioning}
\begin{document}
\maketitle 

%\tableofcontents

\input{abstract.tex}
\input{introduction.tex}
\input{prelims.tex}

\input{laprelims.tex}
\input{automatamodel.tex}
\input{reachability.tex}

%\input{equiv1.tex}
%\input{equiv2.tex}
\label{apxequiv}

\input{equiv.tex}

%\input{covering.tex}
\input{regularity.tex}

\input{covering.tex}
\input{nodca.tex}

\section{Conclusion}
\label{sec:conclusion}
We introduced a new model called one-deterministic-counter automata. The model ``separates'' the machine into two components, (1) counter structure - that can modify the counter, and (2) finite state machine - that can access the counter. This separation of the ``writing'' and ``reading'' part gives some natural advantages to the model. We show that the equivalence problem for weighted \odca is in $\CF{P}$ if the weights are from a field while that of non-deterministic \odca is in $\CF{PSPACE}$. Note that the equivalence problems on weighted automata (where weights are from a field) and non-deterministic finite automata are in $\CF{P}$ and $\CF{PSPACE}$ respectively. On the other hand, the equivalence problem for non-deterministic \textsc{oca} is undecidable and that of weighted \textsc{oca} (weights from a field) is not-known. It will be interesting to look at other models where we can separate the ``writing'' and the ``reading'' parts. For example, a natural extension is to consider \emph{stack-deterministic} pushdown automata - where a deterministic machine updates the stack. We also leave open the question of learning of weighted \odcas.

\onehalfspacing
\addcontentsline{toc}{chapter}{Bibliography}
{\small
\bibliographystyle{plainurl}
\bibliography{paper}
}

%\sav{Why is the font different in Appendix? It is ok to leave it as it is, if there is no quick fix.}

\end{document}

%% file: TeX/abstract.tex
% !TEX root = ../odca.tex
\begin{abstract}
We introduce weighted one-deterministic-counter automata (\odca). These are weighted one-counter automata (\textsc{oca}) with the property of counter-determinacy, meaning that all paths labelled by a given word starting from the initial configuration have the same counter-effect. Weighted \odcas are a strict extension of weighted visibly \textsc{oca}s, which are weighted \textsc{oca}s where the input alphabet determines the actions on the counter. 

%The paper focuses on weighted \odcas, where the weights are from a field. 
We present a novel problem called the co-VS (complement to a vector space) reachability problem for weighted \odcas over fields, which seeks to determine if there exists a run from a given configuration of a weighted \odca to another configuration whose weight vector lies outside a given vector space. We establish two significant properties of witnesses for co-VS reachability: they satisfy a pseudo-pumping lemma, and the lexicographically minimal witness has a special form. It follows that the co-VS reachability problem is in $\CF{P}$.

These reachability problems help us to show that the equivalence problem of weighted \odcas over fields is in $\CF{P}$ by adapting the equivalence proof of deterministic real-time \textsc{oca}s~\cite{droca} by B\"ohm et al.  This is a step towards resolving the open question of the equivalence problem of weighted \textsc{oca}s. Furthermore, we demonstrate that the regularity problem, the problem of checking whether an input weighted \odca over a field is equivalent to some weighted automaton, is in $\CF{P}$. Finally, we show that the covering and coverable equivalence problems for uninitialised weighted \odcas are decidable in polynomial time.
We also consider boolean \odcas and show that the equivalence problem for (non-deterministic) boolean \odcas is in $\CF{PSPACE}$, whereas it is undecidable for (non-deterministic) boolean \textsc{oca}s.

%The equivalence problem for weighted OCAs is not known.
%
%Counter determinacy is 
%a natural way to introduce non-determinism/weights to \textsc{oca}s while maintaining the decidability of crucial problems, that are undecidable on general \textsc{oca}s. For example, the equivalence problem is decidable for deterministic \textsc{oca}s whereas it is undecidable for non-deterministic \textsc{oca}s.
%
%
%We consider both non-deterministic and weighted \odcas. 
%
%This work shows that the equivalence problem is decidable in polynomial time for weighted \odcas over a field and polynomial space for non-deterministic \odcas. As a corollary, we get that the regularity problem, i.e., the problem of checking whether an input weighted \odca is equivalent to some weighted automaton, is also in polynomial time.

%
%
%
%We also introduce a few reachability problems that are of independent interest and show that they are in P. % the unary versions of these problems are in polynomial time, and the binary versions are in non-deterministic polynomial time.
%These reachability problems later help in solving the equivalence problem.
\end{abstract}

%% file: TeX/introduction.tex
% !TEX root = ../odca.tex

\newcommand{\avs}[1]{\emph{\color{blue}#1} - SAV}

\section{Introduction}
%\prm{Remove repeated lemma statements.}
\label{appendIntro}
%\sav{We should decide on one of the two things. \\
%1. Mention weights are from a semiring in intro and prelims. In equiv and reachability section mention fields. The reader of intro is not confused at the beginning on where the weights are coming from. \\
%2. Mention weights are from a field from the beginning. In the non-deterministic odca case talk about the boolean semiring and say that some of the previous results (like equivalence between oca with counter determinacy and odca) holds for semirings too.
%}

%\vincent{Sample comment}
%\prakash{Sample comment}
%\sav{Sample comment}
%\prm{Sample comment}
This paper investigates a restriction on weighted one-counter automata (\textsc{oca}). Like weighted finite automata, weighted \textsc{oca}s recognise functions - every word over a finite alphabet is mapped to a weight. 
%We assume that the weights come from some fixed field. 
We say that a weighted \textsc{oca} has \emph{counter-determinacy} (see \Cref{def:odcasem}) if ``all paths labelled by a given word, starting from the initial configuration, have the same counter-effect''.
%\prm{Upon reading a word from the initial configuration, the counter can have only a single value.} 
%We now give a syntactic model (see \Cref{def:odcasyn}) equivalent to weighted \textsc{oca} with counter-determinacy. 
Weighted one-deterministic-counter automata (\odca) is a syntactic model equivalent to weighted \textsc{oca} with counter-determinacy (see \Cref{def:odcasyn}). It consists of,
% three parts (see \Cref{ocaimage}):
\begin{enumerate}
\item Counter: A counter that stays non-negative and allows zero tests.
\item Counter structure: A finite state deterministic machine where the transitions depend only on its current state, the input letter, and whether the counter is zero. The counter structure can increment/decrement the counter by one.%In 
\item Finite state machine: A finite state weighted automaton whose transitions depend on its current state, the input letter, and whether the counter value is zero. This machine cannot modify the counter.
\end{enumerate}
\begin{figure}[!h]
\centering
\scalebox{.95}{
\begin{tikzpicture}[shorten >=1pt,node distance=3cm,on grid,auto]
\tikzset{every path/.style={line width=.4mm}}
\filldraw [fill=gray!30, draw=black] (2,-2.3) rectangle node[xshift=2.1cm,yshift=.28cm]{\small One way read-only input tape} (-7,-1.8);
\filldraw [fill=gray!30, draw=black] (2,-4.8) rectangle node[xshift=.82cm,yshift=.4cm]{\scriptsize Finite state} (0,-3.3);
\filldraw [fill=gray!30, draw=black] (-5,-4.8) rectangle node[xshift=.6cm,yshift=.4cm]{\scriptsize Counter} (-7,-3.3);
\filldraw [fill=gray!30, draw=black] (-1.8,-4.5) rectangle node[xshift=.6cm,yshift=.28cm]{\scriptsize Counter} (-3.2,-3.5);
\draw[stealth-](-5,-3.8)--(-3.2,-3.8)node[anchor=south, xshift=-.9cm, yshift=0cm]{\scriptsize zero test};
\draw[-stealth](-5,-4.1)--(-3.2,-4.1)node[anchor=north, xshift=-.9cm, yshift=0cm]{\scriptsize $\{+1,-1,0\}$};

\draw[-stealth](-1.8,-3.8)--(-0,-3.8)node[anchor=south, xshift=-.9cm, yshift=0cm]{\scriptsize zero test};
\draw[-](-2.5,-2.3)--(-2.5,-2.85);
\draw[-](1,-2.8)--(-6,-2.8);
\draw[-stealth](1,-2.78)--(1,-3.3);
\draw[-stealth](-5.98,-2.78)--(-5.98,-3.3);
\draw (-6,-4.2) node {\scriptsize structure};
\draw (1,-4.2) node {\scriptsize machine};
\end{tikzpicture}}
\caption{One-deterministic-counter automata}
\label{ocaimage}
\end{figure}
The counter structure and the finite state machine run synchronously on any word. The finite state machine computes the weight associated with the word. 
%\sav{Go through the example.} 
Our first observation is:%\sav{Where do the weights come from? semiring, ring, field?}\prm{change in abstract.}
\begin{theoremrep}
\label{thm:semanticvsyntax}
There is a polynomial time translation from a weighted \textsc{oca} with counter-determinacy to a weighted \odca and vice versa.
\end{theoremrep}
The proof is given in \Cref{introappen}.
%There is a polynomial time algorithm that takes as input a non-deterministic (resp. weighted) \textsc{oca} with counter-determinacy and outputs an equivalent non-deterministic (resp. weighted) \odca. Similarly, \odcas are equivalent to \textsc{oca}s with counter-determinacy.
%\end{theoremrep}
\begin{proof}
\label{introappen}
First, we show that we can construct an equivalent weighted \odca from a given weighted \textsc{oca} with counter-determinacy in polynomial time.
Let $\Autom=(Q, \vc \lambda, \delta_0, \delta_1, \vc\eta)$ be a weighted \textsc{oca} with counter-determinacy.
For this purpose, we define a function $color: [1,\mq] \to [1,\mq]$ as follows.
$$color(i)= min\{j \mid\forall w\in\Sigma^*, n\in \N, \text{ counter-effect of $w$ from } (q_i,n) \text{ and } (q_j,n) \text{ are equal} \}$$
%\begin{observation}
%\label{counterdet}

\final{Given a weighted \odca with an initial configuration $(\vc \lambda,0)$, we can find this coloring function in polynomial time. First, we look at the smallest $i\in[1,\mq]$ such that $\vc \lambda[i]\neq 0$. For all $j\in[0,\mq]$, where $\vc \lambda[j]\neq 0$, we say $color(j)=i$. Now, we look at the configuration reachable from $(\vc \lambda,0)$ in one step. Let $(\vc x, c)$ for some $c\in\N$ be this configuration. If there exists a $j\in[1,\mq]$ with $\vc x[j]\neq 0$ and $color(j)=i$ for some $i\in[1,\mq]$, then for all $k\in[0,\mq]$, where $\vc x[k]\neq 0$, we say $color(k)=i$. If for all $j\in[1,\mq]$ with $\vc x[j]\neq 0$, $color(j)$ is not defined, then we look at the smallest $i\in[1,\mq]$ such that $\vc x[i]\neq 0$. For all $j\in[0,\mq]$, where $\vc x[j]\neq 0$, we say $color(j)=i$. This process is repeated until $color(i)$ is defined for all $i\in[1,\mq]$. This terminates after polynomial steps as all reachable states from the initial configuration can be reached by reading a polynomial length word.}
Let $(\vc x, n)$ be a configuration reachable from the initial configuration of a weighted \textsc{oca} $\Autom$ with counter-determinacy.

\begin{clam}
 If $\vc x[i]\neq 0$ and $\vc x[j] \neq 0$, then $color(i)=color(j)$.
 \end{clam}
 \begin{clamproof}
%\end{observation}
This follows from the notion of counter-determinacy. {If $\vc x[i]\neq 0$ and $\vc x[j] \neq 0$ and $color(i)\neq color(j)$, there there exists a $w\in\Sigma^*$ such that $(q_i,n)\hookrightarrow^{w|s_1} (p_1,n_1)$ and $(q_j,n)\hookrightarrow^{w|s_2} (p_2,n_2)$ for some $p_1,p_2\in Q, s_1,s_2\in\Sring$ and $n_1,n_2\in\N$ such that $n_1\neq n_2$. This contradicts the fact that the machine is counter-deterministic.}
\end{clamproof}
Therefore, for all $w\in \Sigma^*$, if $(q_i,n)\hookrightarrow^{w|s_1} (p_1,n_1)$ and $(q_j,n)\hookrightarrow^{w|s_2} (p_2,n_1)$ for some $q,p_1,p_2 \in Q, n,n_1\in\N$ and $s_1,s_2\in\Sring$, then $color(p_1)=color(p_2)$.
The colors are analogous to the counter-states in the syntactic definition. The transition from one color to another depends solely on the current input symbol and whether the current counter value is zero or non-zero. Hence, in a weighted \odca the counter transitions are determined by a deterministic one-counter automata where the states represent these colors and transitions represent the transition from one color to another.
Now, we formally define the equivalent weighted \odca $((C, \delta^\prime_0, \delta^\prime_1, p_0), (Q, \vc \lambda, \Delta, \vc\eta))$ as follows:
\begin{itemize}
\item $C=\{j \mid color(i)=j, i\in[1,\mq]\}$ is the set of counter states.
\item $\delta^\prime_0: C \times \Sigma \times \{0\} \rightarrow C \times \{0, +1\}$ and $\delta^\prime_1: C \times \Sigma \times \{1\} \rightarrow C \times \{-1,0, +1\}$ are the deterministic counter transitions. For all $q\in \mq, a\in\Sigma$ we define $\delta^\prime_1(q,a)$ and $\delta^\prime_0(q,a)$ as:
 $\delta^\prime_1(q,a)= (p,d)\text{ if }(q,1)\xhookrightarrow{a|s} (p,1+d) \text{ and }\delta^\prime_0(q,a)= (p,d)\text{ if }(q,0)\xhookrightarrow{a|s} (p,d)$ 
for some $p \in Q, s\in\Sring$, and $d\in\{-1,0,+1\}$.
\item Let $i\in[1,\mq]$ such that $\lambda[i] \neq 0$. $p_0= j$, is the start state for counter transition, where $j=color(i)$.
\item $\Delta: \Sigma \times \{0,1\} \to \Sring^{\mq \times \mq}$ gives the transition matrix for all $a \in \Sigma$ and $d \in \{0,1\}$. For $i,j\in[1,\mq]$, $\Delta(a,d)[i][j] = s$, if $\delta_d(q_i,a,q_j,e)=s$ for $q_i,q_j\in Q, s\in\Sring$ and $e\in\{-1,0,+1\}$.
\end{itemize}
Hence, we can construct an equivalent weighted \odca from a given weighted \textsc{oca} with counter-determinacy in polynomial time. Proving the converse is straightforward.
\end{proof}
In the following example, the functions $\texttt{prefixAwareDecimal}$ and $\texttt{equalPrefixPower}$ are recognised by weighted \textsc{oca} with counter-determinacy.
\begin{example} 
\label{examplesodca}
The functions are defined over the alphabet $\Sigma=\{a,b\}$. The transition weights of the \odcas are from the field of rational numbers. 
\begin{enumerate}[(a)]
\item The function $\texttt{prefixAwareDecimal}:\Sigma^*\to\N$ is defined as follows: \\
$\texttt{prefixAwareDecimal}(w)= \texttt{decimal}(w_2)$ 
 if $w=w_1w_2$, $w_1\in \{a^nba^n\mid n >0\}$, and the number of $a$'s $\geq$ number of $b$'s for any prefix of $w_2 \text{, and }
 0 \text{ otherwise.}$
 Here, $\texttt{decimal}(w_2)$ represents the decimal equivalent of $w_2$ when interpreted as a binary number, where `a' is treated as a one and `b' as a zero.
\item The function $\texttt{equalPrefixPower}:\Sigma^* \to \N$ is defined as follows: for all $w \in \Sigma^*$, $\texttt{equalPrefixPower}(w) = 2^k$ where $k$ is the number of proper prefixes of $w$ with equal number of $a$'s and  $b$'s.
\end{enumerate}
\label{eg:oca-cd}
\end{example}
The weighted \odcas recognising these functions are given in \Cref{figureweightedodca}.
 In the figure, if a transition from $p_i$ to $p_j$ of the \emph{counter structure} is labelled $(A,R,D)$ and $(a,r,d)\in A \times R \times D$, then there is a transition from $p_i$ to $p_j$ on reading the symbol $a$ with counter action $d$. %The current counter value should be $0$ if $r=0$ and greater than $0$ if $r=1$.
     If a transition from $q_i$ to $q_j$  of the \emph{finite state machine} is labelled $(A,R,s)$ and $(a,r,s)\in A \times R \times \{s\}$, then there is a transition from $q_i$ to $q_j$ on reading the symbol $a$ with weight $s$. In both cases, the current counter value should be $0$ if $r=0$ and greater than $0$ if $r=1$. For the finite state machine, the initial (resp. output) weight is marked using an inward (resp. outward) arrow. The weight of a path is the product of transition weights along that path. 
The accepting weight of a word is the sum of weights of all the paths from an initial state to an output state labelled by that word. 
%\prm{For instance, the weighted \textsc{oca} with counter-determinacy recognising the function $g$ in \Cref{examplesodca} checks whether the number of a's is equal to the number of b's for any prefix of the input word and multiplies the current weight by $2$ on reading the next symbol.}
%Consider the weighted \odca recognizing the functions $f$ and $g$ in Example \ref{eg:oca-cd} given in \Cref{figureweightedodca}. 
 \begin{figure}
   \fbox{
   \begin{subfigure}[b]{0.45\textwidth}   
  \centering\scalebox{.5}{
\begin{tikzpicture}[shorten >=1pt,node distance=3cm,on grid,auto,
every state/.style={draw,thick},
    every edge/.style={draw,thick}]
\node[state,initial,initial text=] at (0,3) (q_0) {$q_0$};
\node[state] at (5.5,3) (q_1) {$q_1$};
\node[state] at (0,0) (q_2) {$q_2$};
\node[state] at (5.5,0) (q_3) {$q_3$};
\path[->]
(q_0) edge [loop above] node {$(\{a\},\{0,1\},1)$} ()
edge [above] node {$(\{b\},\{1\},1)$} (q_1)
(q_1) edge [loop above] node {$(\{a\},\{1\},1)$} ()
edge [above] node[xshift=-.2cm, yshift=.2cm] {$(\{a\},\{0\},1)$} (q_2)
(q_2) edge [loop above] node {$(\{a\},\{0,1\},1)$} ()
edge [loop below] node {$(\{b\},\{1\},1)$} ()
%edge [loop below] node [xshift=-.4cm]{$(\{b,c\},\{1\},-1)$} ()
edge [above] node {$(\{a\},\{0,1\},1)$} (q_3)
(q_3) edge [loop above] node {$(\{a\},\{0,1\},2)$} ()
edge [loop below] node {$(\{b\},\{1\},2)$} ();
\draw (-2.2,1.15) node[rotate=90] {{Finite state machine}};

\draw (-2.2,-5) node[rotate=90] {{Counter structure}};
\draw[<-] (6.4,0)-- (5.9,0) node[anchor=south,xshift=.2cm]{1};
\node[state,initial,initial text=] at (0,-4) (p_0) {$p_0$};
\node[state] at (5.5,-4) (p_1) {$p_1$};
\node[state] at (5.5,-6) (p_2) {$p_2$};
\path[->]
(p_0) edge [loop above] node {$(\{a\},\{0,1\},+1)$} ()
edge [above] node {$(\{b\},\{1\},-1)$} (p_1)
(p_1) edge [loop above] node {$(\{a\},\{1\},-1)$} ()
edge [left] node {$(\{a\},\{0\},0)$} (p_2)
(p_2) edge [loop left] node{$(\{a\},\{0,1\},+1)$} ()
edge [loop below] node {$(\{b\},\{1\},-1)$} ();
%(p_2) edge[loop below] node {$(\{a,b\},\{0\},0)$}();

%\draw (-.6,-2) node {{Counter structure}};
\draw (-.6,3) node[anchor=south,xshift=-.2cm] {1};
%\draw[->] (5.95,0)-- (6.45,0) node[anchor=west]{1};
%%\draw[<-] (.6,-4)-- (1.1,-4) node[anchor=east,xshift=-.5cm]{1};
%%
%\node[state,initial,initial text=] at (0,-4) (p_0) {$p_0$};
%\path[->]
%(p_0) edge [loop above] node {$(\{a\},\{0,1\},+1)$} ()
%edge [loop below] node{$(\{b\},\{1\},-1)$} ();
\end{tikzpicture}}
\caption{\texttt{prefixAwareDecimal}$(w_1w_2)=$ decimal value of $w_2$'s binary interpretation, if $w_1 \in \{a^nba^n\mid n >0\}$ and \#$a$'s $\geq$ \#$b$'s for any prefix of $w_2$; $0$ otherwise.}
\label{figureex4} 
 \end{subfigure}  
 }
  \hfill
   \fbox{
   \begin{subfigure}[b]{0.45\textwidth}   
  \centering\scalebox{.5}{
\begin{tikzpicture}[shorten >=1pt,node distance=3cm,on grid,auto,
every state/.style={draw,thick},
    every edge/.style={draw,thick}]
\node[state,initial,initial text=] at (0,1.8) (q_0) {$q_0$};
\node[state] at (5.5,3.6) (q_1) {$q_1$};
\node[state] at (5.5,0) (q_2) {$q_2$};
\path[->]
(q_0) edge [above] node[yshift=.2cm] {$(\{a\},\{0\},1)$} (q_1)
edge [below] node[yshift=-.2cm] {$(\{b\},\{0\},1)$} (q_2)
(q_1) edge [loop above] node {$(\{a\},\{0\},2)$} ()
edge [loop right] node {$(\{a,b\},\{1\},1)$} ()
edge [bend right] node[xshift=-2cm, yshift=.2cm] {$(\{b\},\{0\},2)$} (q_2)
(q_2) edge [loop below] node {$(\{b\},\{0\},2)$} ()
edge [loop right] node {$(\{a,b\},\{1\},1)$} ()
edge [bend right] node[xshift=2cm, yshift=.2cm] {$(\{a\},\{0\},2)$} (q_1);
\draw (-2.2,1.75) node[rotate=90] {{Finite state machine}};

\draw (-2.2,-4.4) node[rotate=90] {{Counter structure}};
\draw[<-] (6.1,-.65)-- (5.75,-.3) node[anchor=north,xshift=.5cm, yshift=-.15cm]{1};
\draw[<-] (6.1,4.25)-- (5.75,3.9) node[anchor=south,xshift=.5cm, yshift=.15cm]{1};
\draw[gray!5] [-] (0,-7.56)-- (0,-7.56);
\node[state,initial,initial text=] at (0,-4.3) (p_0) {$p_0$};
\node[state] at (5.5,-4.3) (p_1) {$p_1$};
\path[->]
(p_0) edge [loop above] node {$(\{a\},\{0,1\},+1)$} ()
 edge [loop below] node {$(\{b\},\{1\},-1)$} ()
edge [bend left] node {$(\{b\},\{0\},+1)$} (p_1)
(p_1) edge [loop above] node {$(\{b\},\{0,1\},+1)$} ()
edge [loop below] node {$(\{a\},\{1\},-1)$} ()
edge [bend left] node {$(\{a\},\{0\},+1)$} (p_0);
%(p_2) edge[loop below] node {$(\{a,b\},\{0\},0)$}();

%\draw (-.6,-2) node {{Counter structure}};
\draw (-.6,1.9) node[anchor=south,xshift=-.2cm] {1};
%\draw[->] (5.95,0)-- (6.45,0) node[anchor=west]{1};
%%\draw[<-] (.6,-4)-- (1.1,-4) node[anchor=east,xshift=-.5cm]{1};
%%
%\node[state,initial,initial text=] at (0,-4) (p_0) {$p_0$};
%\path[->]
%(p_0) edge [loop above] node {$(\{a\},\{0,1\},+1)$} ()
%edge [loop below] node{$(\{b\},\{1\},-1)$} ();
\end{tikzpicture}}
\caption{\texttt{equalPrefixPower}$(w) = 2^k$ where $k$ is the number of proper prefixes of $w$ with equal number of $a$'s and $b$'s.}
\label{figureex5} 
 \end{subfigure}  
 }
 
    \caption{The figure shows weighted \odcas recognising the functions given in \Cref{examplesodca}. %Let $A\subseteq \Sigma, R\subseteq\{0,1\}$ and $D \in\{-1,0,+1\}$ be non-empty sets. 
  %For \Cref{figureex4}, the state $q_0$ is having an input weight $1$ and the state $q_3$ has an output weight $1$.
     }
   %Let $S$ be a finite subset of a field $\Sring$. For \Cref{figureex4}, the transitions of the finite state machine behave the same except for an additional weight associated with the transitions. If a transition of the finite state machine is labelled $(A,R,S)$ and $(a,r,s)\in A \times R \times S$, then that particular transition has a weight $s$. The state $q_0$ is having an input weight $1$ and the state $q_3$ has an output weight $1$.}
    \label{figureweightedodca}
\end{figure}
%\prm{Move figure to after definition? Mention which field $f$ is defined on. Give figure for only the weighted case.}

\subsection{Comparisons with other models}
%\vspace{.2cm}\\
%\noindent\textit{Comparisons with other models}
%\vspace{.1cm}\\
{Visibly pushdown automata (\textsc{vpda}) were introduced by Alur and Madhusudan in 2004~\cite{AlurM04}. They have received much attention as they are a strict subclass of pushdown automata suitable for program analysis. \textsc{vpda}s enjoy tractable decidable properties, which are undecidable in the general case. The visibly restriction, in essence, is that the stack operations are \emph{input-driven}, i.e., only depends on the letter read. Weighted \textsc{vpda} is a natural extension to the weighted setting. 
Counter-determinacy can be seen as a relaxation in the visibly constraint on \textsc{oca}s, as the counter actions are no longer input-driven but are deterministic. The fact that weighted \odcas are strictly more expressive than
weighted visibly \textsc{oca} can be noted from the fact that the functions in \Cref{examplesodca} are not recognised by a weighted visibly \textsc{oca}.

Nowotka et al.~\cite{Nowotka} introduced height-deterministic pushdown automata, where the input string determines the stack height. 
Weighted \odcas can be seen as weighted height-deterministic pushdown automata over a single stack alphabet and a bottom-of-stack symbol. 

The reader might feel that a weighted \odca is equivalent to a cartesian product of a deterministic \textsc{oca} and a weighted finite automata. However, one can note that the functions $\texttt{prefixAwareDecimal}$ and $\texttt{equalPrefixPower}$ in \Cref{examplesodca} are not definable by the cartesian product of deterministic \textsc{oca} and a weighted automaton. The reason is that the weighted automaton cannot ``see'' the counter values, so its power is restricted.

\subsection{Motivation}
%\vspace{.2cm}
%\noindent\textit{Motivation}
%\vspace{.1cm}\\
Probabilistic pushdown automata (\ppda) have been studied for the analysis of stochastic programs with recursion~\cite{Javier, Olmedo}. They are equivalent to recursive Markov chains~\cite{temporal,Antonin}. \ppdas are also a generalisation of stochastic context-free grammars~\cite{Abney} used in natural language processing and many variants of one-dimensional random walks~\cite{analprob}.

%Probabilistic pushdown automata (\ppda) is equivalent to probabilistic recursive state machines or recursive Markov chains \cite{temporal} \cite{Antonin}. These models have been studied extensively for the analysis and model checking of procedural programs \cite{qbdp}. \ppda can model probabilistic sequential programs with recursive procedure calls.
%They are also a generalisation of stochastic context-free grammars \cite{Abney} used in natural language processing, molecular biology, and many variants of one-dimensional random walks \cite{analprob}. 

The decidability of equivalence of probabilistic pushdown automata is a long-standing open problem~\cite{openpda}. The problem is inter-reducible to multiplicity equivalence of context-free grammars. In fact, the decidability is only known for some special subclasses of \ppda. It is known that the equivalence problem for \ppda is in $\CF{PSPACE}$ if the alphabet contains only one letter and is at least as hard as polynomial identity testing~\cite{openpda}. There is a randomised polynomial time algorithm that determines the non-equivalence of two visibly \ppda over the alphabet triple $(\Sigma_{call}, \Sigma_{ret}, \Sigma_{int})$ where both machines perform push, pop, and no-action on the stack over the symbols in $\Sigma_{call}$, $\Sigma_{ret}$, and $\Sigma_{int}$ respectively~\cite{Kiefer}. There is a polynomial-time reduction from polynomial identity testing to this problem. Hence it is highly unlikely that the problem is in $\CF{P}$.

Since the equivalence problem for \ppda is unknown, the natural question to ask is the equivalence problem for probabilistic one-counter automata. However, this problem is also unresolved. In this paper, we identify a subclass of probabilistic \textsc{oca}s (probabilistic \odcas are also a superclass of visibly probabilistic \textsc{oca}s) for which the equivalence problem is decidable. In particular, we show that the problem is in $\CF{P}$. Note that our results are slightly more general since we consider weighted \odcas where weights are from a field. 

%In this paper, all sections except \Cref{sec:bool} we consider weighted \odcas where weights are from a field.

\subsection{Our contributions on weighted ODCA (weights from a field)}
%\vspace{.2cm}
%\noindent\textit{Our contributions on weighted ODCA (weights from a field)}
%\vspace{.1cm}\\
The paper's primary focus is on the equivalence problem for weighted \odcas where the weights are from a field. 

{
We first introduce a novel reachability problem on weighted \odca, called the complement to vector space (co-VS) reachability problem. %, to solve the equivalence problem for weighted \odca. %These problems are also of independent interest \vincent{why?}. 
The co-VS reachability problem (see \Cref{sec:reachability}) takes a weighted \odca, an initial configuration, a vector space, a final counter state, and a final counter value as input. It asks, starting from the initial configuration, whether it is possible to reach a configuration with the final counter state, final counter value, and weight distribution over the states that is not in the vector space. %\vincent{Instead of explaining that (which intuitively is obvious), you should rather explain (intuitively) why we need that special reachability.} %given an input \odca and a vector space whether it is possible to reach a configuration in the complement to the vector space.

%We develop novel ideas to show that the unary (resp. binary) co-VS reachability problem is in $\CF{P}$ (resp. $\CF{NP}$). 
Let us call a word a \emph{witness} if the run of the word `reaches' a configuration desired by the reachability problem. We identify two interesting properties of witnesses.
\begin{enumerate}
\item pseudo-pumping lemma (\Cref{lem:uturn}): If the run of a witness encounters a `large' counter value, then it can be pumped-down (resp. pumped-up) to get a run where the maximum counter value encountered is smaller (resp. larger). However, the lemma is distinct from a traditional pumping lemma, where the same subword can be pumped-down (or pumped-up) multiple times while maintaining reachability. In the case of a weighted \odca, we only claim that a subword can be pumped, but the same subword may not be repeatedly pumped. It follows from the pseudo-pumping lemma that the co-VS reachability problem is in $\CF{P}$ (\Cref{reachabilityandcoverability}).
\item special-word lemma (\Cref{specialwordnonfloating}): The lexicographically smallest witness 
is of the form $uy_1^{r_1} vy_2^{r_2}w$ where $u, v, w,y_1$ and $y_2$ are `small' words and $r_1, r_2 \in\N$. The length of the word $uy_1vy_2w$ is bounded by a polynomial in the number of states of the \odca, whereas $r_1$ and $r_2$ also depend on the counter values of the initial and final configurations.
\end{enumerate}

Comparing the above properties with that of deterministic one-counter automata will be interesting. In a deterministic \textsc{oca}, the reachability problem is equivalent to asking whether there is a path to a final state (rather than a weight distribution over states) and a counter value from an initial state and counter value. Let $z$ be an arbitrary `long' witness. Consider the run on $z$ of the deterministic \textsc{oca}. By the Pigeon-hole principle (see Valiant and Paterson~\cite{doca}), there will be words $u, y_1, v, y_2$, and $w$ such that $z = uy_1vy_2w$, and $y_1$ (and similarly $y_2$) starts and ends in the same state and the effect of $y_1$ on the counter is minus of the effect of $y_2$ on the counter. 
In short, $y_1$ and $y_2$ form loops with inverse counter-effects and can be pumped simultaneously.
Therefore, for all $r \in \N$, the word $uy_1^{r} vy_2^{r}w$ is a witness. One can view this as a pumping lemma for deterministic \textsc{oca}. Such a property does not hold in the case of weighted \odca. The presence of weights at each state makes the problem inherently complex, necessitating a more sophisticated approach. 

The proofs of \Cref{lem:uturn} and \Cref{specialwordnonfloating} use linear algebra and combinatorics on words and are distinct from those employed for deterministic \textsc{oca}.
We also introduce a similar problem called co-VS coverability (see \Cref{sec:reachability}). The two properties of the witness and co-VS coverability are crucial along with the ideas developed by B\"ohm et al.~\cite{droca,bisim,bisimjour} and Valiant and Paterson~\cite{doca} in solving the equivalence problem. 
 \begin{theorem}%[main] 
\label{main}
There is a polynomial time algorithm that decides if two weighted \odcas (weights from a field) are equivalent and outputs a word that distinguishes them otherwise.
\end{theorem}
Consider two non-equivalent weighted \odcas. Let $z$ be a minimal word that distinguishes them. We show that the counter values in the run of $z$ are bounded by a polynomial in the number of states. A polynomial-time machine can then simulate the run of both machines up to this counter value to check for distinguishing words. It is sufficient to show that the counter values can be bounded in the run of the minimal distinguishing word. Let $\Autom$ be the disjoint union of the two machines. It is easy to show the existence of a vector space $\lsv$ such that the \odcas are non-equivalent if there is a reachable configuration (in $\Autom$) that is not in $\lsv$. But $\Autom$ contains two counters; hence, we cannot directly apply the properties of reachability witnesses. However, there are subruns in the run of $z$ similar to runs of an \odca. The pseudo-pumping lemma holds in these subruns, and therefore the counter values are bounded during these subruns of $z$ (if the counter values are not bounded, then one can pump-down and generate a shorter witness violating the minimality of $z$). We use the special-word lemma to show the existence of such subruns. 
\Cref{sec:ptime} provides a full proof. 

Next, we consider the regularity problem - the problem of deciding whether a weighted \odca is equivalent to some weighted automaton. %The pseudo-pumping lemma (particularly pumping-up) plays a crucial role in proving the following theorem.
The proof technique is adapted from the ideas developed by B\"ohm et al.~\cite{bisimjour} in the context of real-time \textsc{oca}. The crucial idea in proving regularity is to check for the existence of infinitely many equivalence classes.
The pseudo-pumping lemma (particularly pumping-up) is used in proving this. A detailed proof can be found in \Cref{sec:regularity}.
%We use the pseudo-pumping lemma to show the following theorem.

\begin{theorem} \label{regularity}
{The regularity problem of weighted \odca (weights from a field) is in $\CF{P}$.}
\end{theorem}
%\prm{we can mention a line or two about this, since it is not mentioned anywhere else in the main paper.}

%Let $\Autom$ be a weighted \odca that is not equivalent to any weighted automata. We show that there exists a vector space $\lsv$ and a run from the initial configuration to a configuration not in $\lsv$ where the maximum counter value encountered is `large'. From the quasi-pumping lemma, it follows that we can generate a run where the maximum counter value is larger (while maintaining reachability to complement of $\lsv$). 
%This process can be repeated as many number of times as required. We can now show that the \odca has infinite number of ``equivalence classes'' and therefore the \odca is not equivalent to any weighted automata.
%
%We use the quasi-pumping lemma to prove the above theorem. In particular, we use the fact that a long witness can be pumped-up to get a longer witness. 
%
%This is done by showing the existence of infinitely many equivalence classes by ``pumping up" some parts of a run.
%
{Finally, we look at uninitialised \odcas\ - an \odca without initial finite state distribution and initial counter state. We show that the ``equivalence'' problem for unitialised \odcas are in polynomial time.}
Given two uninitialised \odcas $\Autom_1$ and $\Autom_2$, we say $\Autom_2$ covers ${\Autom_1}$ if for all initial configurations of $\Autom_1$ there exists an initial configuration of $\Autom_2$ such that they are equivalent. The coverable equivalence problem asks whether $\Autom_1$ covers ${\Autom_2}$ and $\Autom_2$ covers ${\Autom_1}$.}%From \Cref{main}, it follows that the covering and coverable equivalence of uninitialised weighted \odcas is polynomial-time solvable. 
\begin{theorem}
Covering and coverable equivalence problems of uninitialised weighted \odcas are in $\CF{P}$.
\end{theorem}
The proof relies on the algorithm to check the equivalence of two weighted \odcas. A detailed proof can be found in \Cref{sec:covering}.

\subsection{Related work}
Extensive studies have been conducted on weighted automata with weights from semirings. Tzeng~\cite{prob} (also see Sch\"{u}tzenberger~\cite{berger}) gave a polynomial time algorithm to decide the equivalence of two probabilistic automata. The result has been extended to weighted automata with weights over a field. On the other hand, the problem is undecidable if the weights are over the semiring $(\N,\min,+)$~\cite{waundec}. Unlike the extensive literature on weighted automata, the study on weighted versions of pushdown or one-counter machines is limited~\cite{probbisim, Juraj, Javier}. %One of the major bottlenecks is the undecidability of many interesting problems. 
The undecidability of several interesting problems creates a major bottleneck.

Moving on to the non-weighted models,
the equivalence problem for non-deterministic pushdown automata is known to be undecidable. On the other hand, from the seminal result by S\'enizergues~\cite{dpdaequiv}, we know that the equivalence problem for deterministic pushdown automata is decidable. The lower bound, though, is primitive recursive~\cite{dpdalowerbound}. 
The language equivalence of synchronised real-time height-deterministic pushdown automata is in $\CF{EXPTIME}$~\cite{Nowotka}.
The equivalence problem for deterministic one-counter automata {(with and without $\epsilon$ transitions)}, similar to that of deterministic finite automata, is $\CF {NL}$-complete~\cite{docaequiv}.

%% file: TeX/prelims.tex
\section{Preliminaries}
\label{sec:prelims}
%In the first two subsections, we will look at some basic notations, definitions, and lemmas from linear algebra. We conclude the section by giving a formal definition of an \odca and a few quick observations regarding \odcas.
\subsection{Basic notations}

%We use the symbol $\N$ to denote the set of natural numbers and $\Z$ to denote the set of integers.

%A monoid is a triple $(S, \circ, e)$ where $S$ is a non-empty set, $\circ$ is an associative binary operation and $e \in S$ is an identity element. The set $S$ is closed under the operation $\circ$ and for all $s\in S$, $e\circ s=s\circ e=s$. A monoid $(S, \circ, e)$ is commutative if $s\circ t = t\circ s$ for all $s,t\in S$.

%An alphabet is a non-empty finite set of letters. 
In this paper, we fix an alphabet $\Sigma$.
We use $\Sigma^*$ to denote the set of finite length words over $\Sigma$.
Given a word $w\in\Sigma^*$, we use $|w|$ to denote the length of the word $w$.
For any set $S$, we use $|S|$ to denote the number of elements in $S$.
We use the notation $[i,j]$ to denote the interval $\{i, i+1, \ldots , j\}$.
We say that a word $u = a_{1}\cdots a_{k}$ is a subword of a word $w$, if $w=u_0 a_{1} u_1 a_2 \cdots a_k u_k$, where $a_i \in \Sigma$, $u_j\in \Sigma^*$ for all $i\in[1,k]$ and $j\in[0,k]$. We call $u$ a proper subword of $w$ if $u \neq w$. %and denote it by $u \prec w$.
We say that a word $u$ is a prefix of a word $w$ if there exists $v\in\Sigma^*$ such that $w=uv$.
Given a word $w = a_0\cdots a_n$, we write $w{[i\cdots j]}$ to denote the factor $a_i\cdots a_j$.
%Given a tuple $e=(e_1,e_2)$, we use $\fst(e)$ to denote $e_1$ and $\snd(e)$ to denote $e_2$. 
{For a $d\in\N$, the sign of $d$ (denoted by $\sgn(d)$) is defined as $\sgn(d)=0$ if $d=0$ and is $1$ otherwise.}
%\begin{toappendix}
 For all $l \in \N$, we use $\Sigma^{\leq l}$ (resp. $\Sigma^l$) to denote the set of words over $\Sigma$ having length less than or equal to $l$ (resp. exactly equal to $l$). %We use the letters $u,v,w,x,y,z$ to denote words over the alphabet $\Sigma$.
% \end{toappendix}
%\begin{center}
%\begin{tabular}{ |l|l| }
% \hline
% Symbols & Meaning \\
% \hline
% $p, q$ & states \\
% $m, n$ & counter values \\
% $i, j, l ,d,k, \ell, h$ & integers\\
% $\con, \conD, \conE, \conF, \conG, \conH$ & configurations\\
%$ \Autom, \Butom, \Cutom$ & Machines\\
%$u,v, w,x,y,z$ & words\\
%$\vc x, \vc y, \vc z$ & vectors\\
%$s, t, r$ & scalars\\
%$a,b$ & alphabets\\
%$\alpha, \beta$ & fractions\\
%$\K$ & number of states\\
%$\mata, \matb, \matm$ & matrices\\
%$\Sring$ & field\\
%$\lsv,\lsu$ & vector space\\
%$X, S, Z$ & sets\\
%$\ucon$ & configurations of underlying machine\\
%
% \hline
%\end{tabular}
%\end{center}

%% file: TeX/laprelims.tex
% !TEX root = ../odca.tex
%\begin{toappendix}
\subsection{Linear algebra}
%A field $\Sring = (S,+,\cdot,0,1)$ is a set $S$ with operations $+$ and $\cdot$ and distinguished elements $0$ and $1$ such that $(S, + , 0)$ and $(S, \cdot, 1)$ are groups.
%\item $\times$ distributes over $+$ (i.e., for all $ s,t,r\in\Sring$, $s \times (t+r) = s\times t + s \times r$)
%\item For all $s \in S\backslash\{0\}$ there exists $s^{-1}, \bar s \in S$ such that $s \times s^{-1} = 1$ and $s + \bar s = 0$
%\item $s\times 0=0\times s=0$ for all $s \in S$
%\end{itemize}
%In the rest of the paper, the distributivity property is never used, so we could relax that condition on the field and still get the same result, though we stick to the usual field definition.
In this paper, we use $\vc x, \vc y, \vc z$ to denote vectors over a field $\Sring$, $s,t,r$ to denote elements in a field $\Sring$ and $\mata,\matb,\matm$ to denote matrices over a field $\Sring$. We use $\lsu,\lsv$ to denote vector spaces.
We recall the following facts.

%\sav{Maybe we can recall the following known facts about vector spaces. \\
%1. Any vector space in $F^K$ will have at most $K$ basis. \\
%2. (gaussian elimination) Given $K$ basis and a vector $v$, there is a polynomial time algorithm to check if $v$ is in the vector space.
%3. (matrix powering) $M^r$ can be computed in time polynomial in $log (r)$.
%}

\begin{lemma}%[\cite{strang}]
\label{lalemmas}
The following are true for a field $\Sring$.
\begin{enumerate}
\item {For any set $X$ of $n$ vectors in $\Sring^m$ with $n>m$, there exists a vector $\vc x\in X$ that is a linear combination of the other vectors in $X$.}
\item Given a set $B$ of $n$ vectors in $\Sring^m$ and a vector $\vc x\in \Sring^m$, we can check if $\vc x$ is a linear combination of vectors in $B$ in time polynomial in $m$ and $n$.
%\item Let $k,r\in \N$ and $\matm \in \Sring^{k \times k}$. The matrix $\matm^r$ can be computed in time polynomial in $k$ and $\log(r)$.
\end{enumerate}

\end{lemma}

The following properties of vector spaces are important.
%We use the following lemmas on linear algebra in this paper. % for the sake of completeness.

%\prm{change $0$ to $\overline \lsv$}
\begin{lemma}
\label{lem:vec}
{Let $\lsv$ be a vector space, $k\in\N$ and for all $r\in[0,k]$ $\vc z_r \in \Sring^{k}$ and $\matm_r \in \Sring^{k\times k}$.}
{%Let $k\in\N, r\in[0,k]$, $\vc z_r \in \Sring^{k}, \matm_r \in \Sring^{k\times k}$  be matrices over $\Sring$ and $\lsv$ be a vector space. 
Then, there exists an $i\in[1, k]$ such that the following conditions are true:}
\begin{enumerate}
\item $\vc z_i$ is a linear combination of $\vc z_0, \dots \vc z_{i-1}$, and
%\item for all $\vc x \in\Sring^k$ such that ${\vc x} \trans{\vc z}_i \neq 0$, there exists a $j < i$ such that $ {\vc x} \trans{\vc z}_j \neq 0$, and
\item if $\vc z_i\matm_i \notin \lsv$, then there exists $j < i$ such that $\vc z_j\matm_i \notin \lsv$.
\end{enumerate}
\end{lemma}
\begin{proof}
Let $k\in\N, r\in[0,k]$, $\vc z_r \in \Sring^{k}, \matm_r \in \Sring^{k\times k}$  be matrices and $\lsv$ be a vector space. 
\begin{point}
Consider the set $\{\vc z_0, \vc z_1, \dots, \vc z_{k}\}$ of $k+1$ vectors of dimension $k$. It follows from \Cref{lalemmas} that
 there are at most $k$ independent vectors of dimension $k$, and hence not all elements of the set can be independent.
 \end{point}
% \begin{point}
% Let $i\in[1,k]$ be such that $\vc z_i$ is dependent on $\vc z_0, \dots \vc z_{i-1}$ and let $\vc x \in\Sring^k$ such that ${\vc x} \trans{\vc z}_i \neq 0$.
% Since $\vc z_i$ is dependent on $\vc z_0, \dots \vc z_{i-1}$, there exists $s_0, \dots s_{i-1} \in \Sring$ such that
%\[
%\vc z_i = s_0\cdot \vc z_0 + s_1\cdot \vc z_1 + \dots + s_{i-1}\cdot \vc z_{i-1}
%\]
%Let us assume for contradiction that ${\vc x} \trans{\vc z}_j = 0$ for all $j \in [0,i-1]$ and therefore ${\vc x} \trans{\vc z}_i= \sum_{j=0}^{i-1} s_j \cdot {\vc x}\trans{\vc z}_j \ =0$. This is a contradiction.
% \end{point}
 \begin{point} 
 Let $i\in[1,k]$ be such that $\vc z_i$ is a linear combination of $\vc z_0, \dots \vc z_{i-1}$ and ${\vc z_i} \matm_i\notin \lsv$.
%Since $\vc z_i$ is a linear combination of $\vc z_0, \dots \vc z_{i-1}$, we prove that there exists $j<i$ such that ${\vc z_j} \matm_i \notin \lsv$.
 %Let $\vc z_i \matm_i\not\in \lsv$. 
 Let us assume for contradiction that $\vc z_j \matm_i\in \lsv$ for all $j \in [0,i-1]$. Since $\vc z_i$ is a linear combination on $\vc z_0, \dots \vc z_{i-1}$, there exists $s_0, \dots s_{i-1} \in \Sring$ such that
\[
\vc z_i = s_0\cdot \vc z_0 + s_1\cdot \vc z_1 + \dots + s_{i-1}\cdot \vc z_{i-1}
\]

 Since $\vc z_i\matm_i=\sum_{j=0}^{i-1} s_j \cdot\vc z_j \matm_i $ and $\lsv$ is closed under linear combinations, we get that $\vc z_i\matm_i \in \lsv$ contradicting our initial assumption. %$= \sum_{j=0}^{i-1} s_j \cdot\vc z_j \in \lsv$, which contradicts our initial assumption that $\vc z_i \not\in \lsv$.
 \end{point}
\end{proof}

\begin{lemma}
\label{lem:mat}
{Let $\lsv$ be a vector space, $k\in\N$ and for all $r\in[0,k^2]$ $\mata_r,\matm_r, \matb_r
\in \Sring^{k\times k}$.}
 %Let $k\in\N, r\in[0,k^2], \mata_r,\matm_r, \matb_r
%\in \Sring^{k\times k}$ be matrices over $\Sring$, and $\lsv$ be a vector space. 
Then, there exists an $i \in[1, k^2]$ such that for all $\vc x\in \Sring^k$ the following conditions are true: 
\begin{enumerate}
\item $\matm_i$ is a linear combination of $\matm_0, \dots, \matm_{i-1}$, and
\item if ${\vc x} \mata_i \matm_i \matb_i \notin \lsv$, then there exists a $j < i$ such that ${\vc x} \mata_i \matm_j \matb_i \notin \lsv$.
%\item if ${\vc x} \mata_i \notin \lsv$, then there exists a $j < i$ such that ${\vc x} \mata_j \notin \lsv$.
\end{enumerate}
\end{lemma}
\begin{proof}
Let $\mata_r,\matm_r, \matb_r
\in \Sring^{k\times k}$ for $r\in[0,k^2]$, be matrices over $\Sring$ and $\lsv$ a vector space.
%\sav{The numbering is all wrong!}
\begin{point}
Consider the set $\{\matm_0, \matm_1, \dots, \matm_{k^2}\}$ of $k^2+1$ matrices of dimension $k^2$. It follows from \Cref{lalemmas} that
there are at most $k^2$ independent vectors of dimension $k^2$, and hence not all elements of this set can be independent.
\end{point}
\begin{lastpoint} Let $i\in[1,k^2]$ be such that $\matm_i$ is a linear combination of $\matm_0, \dots, \matm_{i-1}$ and ${\vc x} \mata_i\matm_i\matb_i \notin \lsv$.
Since $\vc \matm_i$ is dependent on $\matm_0, \dots, \matm_{i-1}$, we prove that there exists $j<i$ such that ${\vc x} \mata_i \matm_j \matb_i \notin \lsv$.
Let us assume for contradiction that this is not the case.
Since $\matm_i$ is a linear combination on $\matm_0, \dots \matm_{i-1}$, there exists $s_0, \dots s_{i-1} \in \Sring$ such that
\[
\vc \matm_i = s_0\cdot \matm_0 + s_1\cdot \matm_1 + \dots + s_{i-1}\cdot \matm_{i-1}
\]
Since ${\vc x} \mata_i \matm_j \matb_i \in \lsv$ for all $j \in [0,i-1]$ we get that ${\vc x} \mata_i \matm_i \matb_i= \sum_{j=0}^{i-1} s_j \cdot {\vc x} \mata_i \matm_j \matb_i \in \lsv$, which is a contradiction.
\end{lastpoint}
% \begin{point}
% Let ${\vc x} \mata_i \notin \lsv$. Let us assume for contradiction that ${\vc x} \mata_j \in \lsv$ for all $j \in [0,i-1]$. Since $\mata_i=\sum_{j=0}^{i-1} s_j \cdot\mata_j $ and $\lsv$ is closed under linear combinations, we get that ${\vc x} \mata_i= \sum_{j=0}^{i-1} s_j \cdot {\vc x} \mata_j \in \lsv$, which contradicts our initial assumption that $\mata_i \not\in \lsv$.
% \end{point}
\end{proof}

\begin{lemma}
\label{lem:comp}
%\prm{It suffices to show only the two properties, additive closure, and multiplicative closure}
Let $k\in\N, \mata \in \Sring^{k\times k}$ and $\lsv \subseteq \Sring^k$ be a vector space. Then the following set is a vector space,
\[
\lsu=\{ \vc y \in \Sring^{k} \mid {\vc y} \mata \in \lsv \}.
\]
%$+$ is the vector addition operator and $\cdot$ is the scalar multiplication operator.
\end{lemma}
% \begin{proofsketch}
%\prince{proof sketch if required.}
%\end{proofsketch}
%\sav{What is this lemma? $\vc x^{T} \vc y$ is a scalar quantity. How is it in $\lsv$? It makes no sense.}
\begin{proof}
To prove that $\lsu$ is a vector space, it suffices to show that it is closed under vector addition and scalar multiplication. First, we prove that $\lsu$ is closed under vector addition.
Let $\vc z_1, \vc z_2 \in \lsu$ be two vectors, since ${\vc z}_1 \mata, {\vc z}_2 \mata \in \lsv$, ${({\vc z_1}+{\vc z_2})} \mata= {\vc z}_1\mata + {\vc z}_2 \mata \in \lsv$. Therefore, $\vc z_1+ \vc z_2\in \lsu$.
Now we prove that $\lsu$ is closed under scalar multiplication. For any vector $\vc z_1 \in \lsu$, we know that ${\vc z}_1 \mata \in\lsv$. Since $\lsv$ is a vector space, for any scalar $r\in\Sring$, $ { (r\cdot {\vc z_1}) }\mata\in \lsv$, and therefore $r\cdot \vc z_1 \in \lsu$. This concludes the proof.
\end{proof}
In particular, the above lemma holds for the vector space $\{\vc 0 \in \Sring^k\}$. %The proofs are given in the Appendix. 
%\end{toappendix}

%% file: TeX/automatamodel.tex
% !TEX root = ../odca.tex

%\prm{Give semantic definition instead}
%\prm{Recheck: Equivalence odcwa $\to$ equivalence odcna}
\subsection{Weighted one-deterministic-counter automata}
%\prm{mention weights from a field}
In this section, we define weighted \odca, where the weights are from a semiring. However, our results require that the weights come from some field $\Sring$ except for \Cref{sec:nondet}, where the weights are from the boolean semiring.
%We provide a syntactic and semantic interpretation of weighted \odca and demonstrate their equivalence. 
First, we define a weighted one-counter automata.
\begin{definition}
A weighted one-counter automata $\Autom=(Q, \vc \lambda, \delta_0, \delta_1, \vc\eta)$, is defined over an alphabet $\Sigma$ where, %\prm{Added concept of final state in oca part.}
%\[
%\Autom=(Q, \vc \lambda, \delta, \vc\eta)
%\]
%\item $F\subseteq C$ is a set of non-empty final counter states.
$Q$ is a non-empty finite set of states, %{We assume $|C|=|Q|$ and use $\K$ to denote $|Q|$.}
$\vc \lambda \in \Sring^{\mq}$ is the initial distribution where the $i^{th}$ component of $\vc \lambda$ indicates the initial weight on state $q_i\in Q$,
%\prm{Should we remove C from the following definition?}
%\item \rem{$\Delta: \Sigma \times \{0,1\} \to \Sring^{\mq \times \mq \times \{-1, 0, +1\}}$ gives the transition matrix for all $a \in \Sigma$ and $d \in \{0,1\}$. 
%Let $a\in\Sigma, d \in \{0,1\}, e\in \{-1, 0, +1\}, p,q\in Q, n \in \N$, $\mata \in \Sring^{\mq \times \mq \times \{-1, 0, +1\}}$, $\sgn(n)=d$, and $\Delta(a,d)= \mata$. The component of $\mata$ indexed by $(p,q,e)$ denotes the weight on the transition from state $p\in Q$ to state $q\in Q$ with counter effect $e$ on reading symbol $a$ from counter value $n$.} %As the machine is counter deterministic, for any given row of $\matm$, all elements in that row will have the same counter-effect.
{$\delta_0: Q \times \Sigma \times Q \times \{0,+1\} \to \Sring$ and $\delta_1: Q \times \Sigma \times Q \times \{-1,0,+1\} \to \Sring$ are the transition functions, and}
%{ $\delta: Q \times \Sigma \times \{0,1\}\to \Sring^{Q \times \{-1,0,+1\}}$ is the transition function where the rows of the matrix $\Sring^{Q \times \{-1,0,+1\}}$ is indexed by the states in $Q$ and the columns are indexed by counter actions in $ \{-1,0,+1\}$, and }
$\vc\eta\in \Sring^{\mq}$ is the final distribution, where the $i^{th}$ component of $\vc \eta$ indicates the output weight on state $q_i\in Q$.
%\prm{state what is counter determinacy.}
\end{definition}
Note that the counter values do not go below zero. 
Let $p,q\in Q, a\in \Sigma, n\in\N, e \in \{-1,0,+1\}$, and $s \in \Sring$. We say $(q,n) \hookrightarrow^{a|s} (p,n+e)$ if $\delta_{\sgn(n)}(q,a,p,e) = s$. 
 %If $\delta(q,a,d)[p][e]=s$, then there is a transition on symbol $a$ from state $q$ with counter value $n$ to state $p$ with counter value $n+e$ with weight $s$. We denote this by $(q,n) \xhookrightarrow{a|s} (p,n+e)$. 
 Let $w=a_1a_2\cdots a_t\in \Sigma^*$ for some $t\in\N$. For a $q_0 \in Q$ and $n_0 \in \N$, we say $(q_0,n_0) \hookrightarrow^{w|s} (q_t,n_t)$ if for all $i \in [1,t]$, there are $q_i \in Q, n_i \in \N$, $s_i \in \Sring$ such that $(q_{i-1},n_{i-1})\hookrightarrow^{a_i|s_i} (q_{i},n_{i})$ and 
 $s= \prod_{i=1}^t s_i$.
 %, $q_i\in Q, n_i\in\N, s_i \in \Sring$ for $i\in[0,t]$. We extend the notion of transitions to that over words and write $(q_0,n_0) \xhookrightarrow{w|s} (q_t,n_t)$ if there are transitions $(q_i,n_i)\xhookrightarrow{a_i|s_i} (q_{i+1},n_{i+1})$ for all $i\in[0,t]$ and $s= \prod_{i=0}^t s_i$. %We call $n_t-n_0$ the counter-effect on reading $w$ from $(q_0,n_0)$.
% Now, we provide a definition of weighted \textsc{oca} with counter-determinacy.
\begin{definition}
\label{def:odcasem}
A weighted \textsc{oca} with counter-determinacy is a weighted one-counter automata $\Autom=(Q, \vc \lambda, \delta_0, \delta_1, \vc\eta)$ with the following restriction: if $\vc \lambda[i]$ and $\vc \lambda[j]$ are non-zero for some $i,j\in[1,\mq]$, then for all $w\in \Sigma^*$, if $(q_i,0)\hookrightarrow^{w|s_1} (p_1,n_1)$ and $(q_j,0)\hookrightarrow^{w|s_2} (p_2,n_2)$ for some $p_1,p_2 \in Q,n_1,n_2\in\N$ and $s_1,s_2\in\Sring$, then $n_1=n_2$.  
\end{definition}
%\begin{toappendix}
The configuration of a weighted \textsc{oca} with counter-determinacy is therefore of the form $(\vc x, n)$ where $\vc x \in \Sring^{\mq}$ such that for $i\in[0,\mq], \vc x[i]$ denotes the weight with which the machine is in state $q_i$ and $n\in\N$ denotes the current counter value.
%\prm{Definition, this is a more general model than the previous, (runs, transitions, etc.)}
%\end{toappendix}
We present a definition for weighted \odca, which is an equivalent syntactic model.
%A weighted \odca consists of two parts: a finite state machine, which is a weighted automaton over a field, and a counter structure, which is a deterministic \textsc{oca}. %The transitions of the control structure depend on the current configuration of the data structure. 
%A weighted \odca is defined as follows:

%\prm{Give a semantic definition instead.}
%\prm{A weighted \odca, $\Autom=((C, \delta_0,\delta_1, p_0);(\ Q, \vc \lambda, \Delta, \vc\eta))$, mention: $\delta_0,\delta_1$ are deterministic transitions after definition.}
\begin{definition}
\label{def:odcasyn}
A weighted \odca, $\Autom=((C, \delta_0,\delta_1, p_0),(Q, \vc \lambda, \Delta, \vc\eta))$ is defined over an alphabet $\Sigma$ where,
\begin{itemize}
%\item Counter structure:
%\begin{itemize}
\item $(C, \delta_0,\delta_1, p_0)$ represents the counter structure and $(Q, \vc \lambda, \Delta, \vc\eta)$ represents the finite state machine.
\item $C$ is a non-empty finite set of counter states.
\item $\delta_0: C \times \Sigma \rightarrow C \times \{0, +1\}$, $\delta_1: C \times \Sigma \rightarrow C \times \{-1, 0, +1\}$ are counter transitions.
\item $p_0 \in C$ is the start state for the counter structure.
%\end{itemize}
%\item Finite state machine:
%\begin{itemize}
%\item $F\subseteq C$ is a set of non-empty final counter states.
\item $Q$ is a non-empty finite set of states of the finite state machine. %{We assume $|C|=|Q|$ and use $\K$ to denote $|Q|$.}
\item $\vc \lambda \in \Sring^{\mq}$ is the initial distribution where the $i^{th}$ component of $\vc \lambda$ indicates the initial weight on state $q_i\in Q$.
%\prm{Should we remove C from the following definition?}
\item $\Delta: \Sigma \times \{0,1\} \to \Sring^{\mq \times \mq}$ gives the transition matrix. For all $a \in \Sigma$ and $d \in \{0,1\}$, the component in the $i^{th}$ row and $j^{th}$ column of $\Delta(a,d)$ denotes the weight on the transition from state $q_i\in Q$ to state $q_j\in Q$ on reading symbol $a$ from counter value $n$ with $\sgn(n)=d$.
\item $\vc\eta\in \Sring^{\mq}$ is the final distribution, where the $i^{th}$ component of $\vc \eta$ indicates the output weight on state $q_i\in Q$.
%\end{itemize}
\end{itemize}
\end{definition}

Note that $\delta_0$ and $\delta_1$ are deterministic transition functions. The counter structure and the finite state machine run synchronously on any given word. 
%\prm{We consider vectors as column vectors.}
%Note that if $\Sring$ is the boolean field $\{0,1\}$, then we can view the machine as a non-deterministic model.
A configuration $\con$ of an \odca is of the form $\configodc{x_\con}{p_\con}{n_\con} \in \Sring^{\mq} \times C \times \N$. We use the notation $\wgtvec{\con}$ to denote $\vc x_\con$, $\cntstate{\con}$ to denote $p_\con$, and $\cntval{\con}$ to denote $n_\con$. The initial configuration is
$\configodc{\lambda}{p_0}{0}$.
%\sav{Can we make this tuple smaller (transition). }%Just use $(c_i, d, a, c_o)$ for transition where $c_i, c_o$ are configurations and $a$ is letter and $d$ is counter zero or not.}
%\prm{Including configurations in the definition of transition will stop us from saying why a sequence of transition is valid after performing a cut. This was the reason why we did a similar modification to dwroca.}
A \emph{transition} is a tuple $\tau = \transodcx$ where %$\pmb{\iota_\tau},{\theta_\tau}\in \Sring^\K$,
$\iota, \theta \in C$ are counter states, $d \in \{0,1\}$ denotes whether the counter value is zero or not, $a \in \Sigma$, $\ce \in \{-1,0,1\}$ is the \emph{counter-effect}, $\mata \in \Sring^{\mq\times\mq}$ such that $\Delta(a,{d}) = \mata$, and $\delta_{d}(\iota, a)=(
\theta, \ce)$. % and $\pmb{\theta_\tau}=\pmb{{\iota_\tau}} \mata_\tau$.
Given a transition $\tau = \transodcx$ and a configuration $\con=(\vc x,n,p)$, we denote the application of $\tau$ to $\con$ as $\tau(\con) = \configodc{\vc x \mata}{\theta}{n+\ce}$ if $p=\iota$ and $d = \sgn(n)$;  $\tau(\con)$ is undefined otherwise. Note that the counter values are always non-negative.

Consider a sequence of transitions $T = \tau_0\cdots \tau_{\ell}$ where $\tau_i = \transodc{i}$ for all $i \in [0,\ell]$. We denote by $\word(T) = a_{0}\cdots a_{{\ell}}$ the word labelling it, $\we(T) = \mata_{0} \cdots \mata_{{\ell}}$ its weight-effect matrix, and $\ce(T) = \ce_{0} + \cdots + \ce_{{\ell}}$ its counter-effect. For all $0 \leq i < j \leq \ell$, we use $T_{i\cdots j}$ to denote the sequence of transitions $\tau_i\cdots \tau_j$ and $|T|$ to denote its length $\ell+1$.
We call $T$ \emph{floating} if for all $i\in[0,\ell-1]$, $d_{i} = 1$ and \emph{non-floating} otherwise.
%\begin{toappendix}
We denote $\min_{\ce}(T) = \min_i(\ce(\tau_0\cdots \tau_i))$ the minimal effect of its prefixes and $\max_{\ce}(T) = \max_i(\ce(\tau_0\cdots \tau_i))$ is the maximal effect of its prefixes.
%\end{toappendix}
%We say that $T$ is \emph{valid} if for every $i\in[0,\ell-1]$, $\theta_{i}=\iota_{{i+1}}$. We will only consider valid sequences of transitions.

A \emph{run} $\pi$ is an alternating sequence of configurations and transitions denoted as $\pi = \con_0 \tau_0 \con_1 \cdots \tau_{\ell-1} \con_\ell$ such that for every $i$, $\con_{i+1} = \tau_i(\con_i)$. The word labelling, length, weight-effect, and counter-effect of the run are those of its underlying sequence of transitions.
Given a sequence of transitions $T= \tau_0\cdots \tau_{\ell-1}$ and a configuration $\con$, we denote by $T(\con)$ the run (if it is defined) $\con_0 \tau_0 \con_1 \cdots \tau_{\ell-1} \con_\ell$ where $\con_0 = \con$.

%\begin{toappendix}
%\subsection{Weighted one-deterministic-counter automata}
Let $\con$ be a configuration with $\cntval{\con}=n$ for some $n\in\N$.
{Observe that, for a valid floating sequence of transitions, $T(\con)$ is defined if and only if $n > - \min_\ce(T)$, and for a valid non-floating sequence of transitions, $T(\con)$ is defined if and only if $n = - \min_\ce(T)$ and for every $i$, $d_{\tau_i} = 0$ if and only if $\ce(\tau_0\cdots \tau_{i-1}) = \min_\ce(T)$.
In particular, observe that if a valid floating sequence of transition $T$ is applicable to a configuration $\con=\configodc{x}{p}{n}$, then for every $n^\prime \geq n$ and vector $\vc x^\prime \in \Sring^{\mq}$, it is applicable to $\configodc{x^\prime}{p}{n^\prime}$.}
%\end{toappendix}

For any word $w$, there is at most one run labelled by $w$ starting from a given configuration $\con_0$. We denote this run $\pi(w,\con_0)$.
%A run $\pi$ is called an \emph{execution} if $\con_0=\configodc{\lambda}{p_0}{0}$ is the initial configuration of $\Autom$. We use $\pi(w)$ to denote the execution labelled by $w$.
%A \emph{run} $\pi_1$ is called a \emph{sub-execution} of an execution $\pi$ if $\pi=\con_0 \cdots \tau_i\pi_1\tau_j \cdots \con_\ell$ for some $i,j, 0 \leq i <j\leq \ell$.
A run $\pi(w,\con_0)=\con_0 \tau_0 \con_1 \cdots \tau_{\ell-1} \con_\ell$ is also represented as %$\con_0 \xrightarrow{\we(\pi(w,\con_0))|w}\con_\ell$ and
$\con_0 \xrightarrow{w}\con_\ell$. We say %$\Phi(w,\con_0)$ to denote the configuration $\con_\ell$,
$\con_0 \rightarrow^* \con_\ell$ if there is some word $w$ such that $\con_0 \xrightarrow{w}\con_\ell$. %and $\con_0 \rightarrow^+ \con_\ell$ to denote the existence of a floating run from $\con_0$ to $\con_\ell$.
%The counter effect of a word $w$ on a floating run $\con_0 \xrightarrow{w}\con_\ell$ is counter value of ${\con_\ell}$ minus counter value of ${\con_0}$.
For a weighted \odca $\Autom$, the accepting weight of $w$ is denoted by $f_\Autom(w,\con)= \vc{\lambda} \we(\pi(w,\con)) \trans{\vc\eta}$, where $\con$ is the initial configuration of $\Autom$. 
Two weighted \odcas $\Autom$ and $\Butom$ are equivalent if for all $w\in\Sigma^*$, $f_\Autom(w,\con)= f_\Butom(w,\conD)$ where $\con$ and $\conD$ are the initial configurations of $\Autom$ and $\Butom$ respectively.
Let $\con$ and $\conD$ be configurations of \odcas $\Autom$ and $\Butom$ respectively.  We say that
$\con\equiv_l \conD$ if and only if
for all $w\in\Sigma^{\leq l}$, $f_\Autom(w, \con)= f_\Butom(w, \conD)$ otherwise $\con\not\equiv_l \conD$. 
%An uninitialised weighted \odca $\Autom$ is a weighted \odca without an initial counter state and initial distribution.
%Weighted automata (\WA) is a restricted form of weighted \odca where the counter value is fixed at zero. The above notions of transitions, runs, acceptance, etc.\ are used for $\WA$ also.
%\begin{toappendix}
We use the notation $f_\Autom(w)$ to denote $f_\Autom(w,{\configodc{\lambda}{p_0}{0}})$.
{

We say that two configurations $\con$ and $\conD$ are equivalent if and only if $\con \equiv_l \conD$ for all $l\in \N$ and we denote this by $\con\equiv \conD$.
{If the weighted \odca is defined over %a semiring which is also a field, then we call the model a weighted \odca, and if it is 
the boolean semiring, then we call it a non-deterministic/ deterministic \odca.}
The class of weighted \odcas includes deterministic \textsc{oca}, visibly weighted \textsc{oca}, and deterministic weighted \textsc{oca}.}
%\sav{We have to bring in known examples from literature. Have others considered weighted oca? Does our model capture it?
%Also, note that the $\delta$ need not be a function and $A$ is always a function. In that case, the control states are non-deterministic. Like in the previous section, we can determinise it (with an exponential blow-up). The question is, Is equivalence checking in this model (as well as the weighted case) in $\CF{PTIME}$? }
%\begin{toappendix}

{An uninitialised weighted \odca $\Autom$ is a weighted \odca without an initial counter state and initial distribution. Formally, $\Autom=((C, \delta_0,\delta_1),(Q, \Delta, \vc\eta))$. Given an uninitialised weighted \odca $\Autom$ and an initial configuration $\con_0=(\vc x, p,0)$, we define the weighted \odca $\un{\Autom}{\con_0}= ((C, \delta_0,\delta_1, p), (Q, \vc x, \Delta, \vc\eta))$.}
%\end{toappendix}

In \Cref{thm:semanticvsyntax} (proof in \Cref{introappen}), we show that weighted \textsc{oca} with counter-determinacy is equivalent to weighted \odca. %\final{All omitted proofs are given in the Appendix.}
%\prm{ look at example, here initial config. two example runs ( a ,aba), weight associated, weight effect, counter effect, path}

%{Consider the weighted \odca $\Cutom$ recognising the function \texttt{prefixAwareDecimal} given in \Cref{figureex4}. Here $\vc \lambda = [1,0,0,0]$ and $\vc \eta= [0,0,0,1]$.
%The configuration $\con_0= ([1,0,0,0], p_0, 0)$ is the initial configuration of this machine. Let $w=abaaab$. %The run of this machine on the word $w$ can be written as:
%%\begin{align*}\pi(w,\con_0)= ([1,0,0,0], p_0, 0)\xrightarrow{a} &([1,0,0,0],p_0,1) \xrightarrow{b} ([0,1,0,0],p_1,0) \xrightarrow{a} ([0,0,1,0],p_2,0) \xrightarrow{a}\\
%% & ([0,0,1,1],p_2,1) \xrightarrow{a} ([0,0,1,2],p_2,2) \xrightarrow{b} ([0,0,1,6],p_2,1).\end{align*}
%The counter-effect of the run on $w$ from $\con_0$ is $\ce(\pi(w,\con_0))=1$. % the weight-effect matrix is given by \[\we(\pi(w,\con_0))=\Delta(a,{0})\ \Delta(b,{1})\ \Delta(a,{0})\ \Delta(a,{1})\ \Delta(a,{1})\ \Delta(b,{1})= \begin{bmatrix} 0&0&1&6\\0&0&1&14\\0&0&1&46\\0&0&0&64\end{bmatrix}.\] 
%The accepting weight of $w$ is $f_{\Cutom}(w, \con_0)= \vc{\lambda} \we(\pi(w,\con_0)) \trans{\vc\eta}= 6$.
%}
%\begin{toappendix}

{ Consider the weighted \odca $\Cutom$ recognising the function \texttt{prefixAwareDecimal} given in \Cref{figureex4}. Here, $\vc \lambda = [1,0,0,0]$ and $\vc \eta= [0,0,0,1]$.
The configuration $\con_0= ([1,0,0,0], p_0, 0)$ is the initial configuration of this machine. Let $w=abaaab$. The run of this machine on the word $w$ can be written as:
\begin{align*}\pi(w,\con_0)= ([1,0,0,0], p_0, 0)\xrightarrow{a} &([1,0,0,0],p_0,1) \xrightarrow{b} ([0,1,0,0],p_1,0) \xrightarrow{a} ([0,0,1,0],p_2,0) \xrightarrow{a}\\
 & ([0,0,1,1],p_2,1) \xrightarrow{a} ([0,0,1,2],p_2,2) \xrightarrow{b} ([0,0,1,6],p_2,1).\end{align*}
The counter-effect of this run is $\ce(\pi(w,\con_0))=1$ and the weight-effect matrix is given by \[\we(\pi(w,\con_0))=\Delta(a,{0})\ \Delta(b,{1})\ \Delta(a,{0})\ \Delta(a,{1})\ \Delta(a,{1})\ \Delta(b,{1})= \begin{bmatrix} 0&0&1&6\\0&0&1&14\\0&0&1&46\\0&0&0&64\end{bmatrix}.\] The accepting weight of the word $w$ is $f_{\Cutom}(w, \con_0)= \vc{\lambda} \we(\pi(w,\con_0)) \trans{\vc\eta}= 6$.
}
%\end{toappendix}
%\final{We get that the semantic definition of \textsc{oca} with counter-determinacy (\Cref{def:odcasem}) is equivalent to the syntactic definition of weighted \odca (\Cref{def:odcasyn}) from \Cref{thm:semanticvsyntax} (proof in Appendix).
%For all $i\in[0,\mq]$, we define the vector $\vc z_i\in \Sring^{\mq}$ as follows:
%\[
%\vc z_i[j]= \begin{cases} 1, \text{ if }i=j\\
% 0, \text{ otherwise} \end{cases}
%\]
%\prm{define counter loss}
%A weighted automata (\WA) is a restricted form of an \odca where the counter value is fixed at zero. The above notions of transitions, runs, acceptance, etc.\ are used for weighted automata also.
%}
%\begin{toappendix}

Weighted automata (\WA) is a restricted form of an \odca where the counter value is fixed at zero. The above notions of transitions, runs, acceptance, etc.\ are used for $\WA$ also.
We also use the classical notion and represent weighted automata as $\Autom= (Q, \lambda, \Delta, \vc\eta)$, without counter states. 
%\end{toappendix}
%Note that the equivalence problem of \odca defined over an arbitrary semiring is undecidable because of the undecidability of equivalence of weighted automata over semirings.
%\begin{toappendix}
Given a weighted \odca $\Autom$ over the alphabet $\Sigma$ and a field $\Sring$, we define its \emph{$M$-unfolding} weighted automaton $\Autom^M$ as {a finite state weighted automaton that recognises the same function as $\Autom$ for all runs where the counter value does not exceed $M$. A formal definition is given in \Cref{unfolding}.}
\begin{definition}[$M$-unfolding weighted automata]
\label{unfolding}
Let $\Autom=((C,\delta_0,\delta_1, p_0), (Q, \vc \lambda, \Delta, \vc\eta))$ %$(Q, \Sigma, q_0, s_0,\delta_0, \delta_1, \eta_F)$
be a weighted \odca over the alphabet $\Sigma$ and a field $\Sring$. % let $\K= |Q|= |C|$. 
For a given $M\in \N$, we define an $M$-unfolding weighted automata $\Autom^M$ of $\Autom$ as follows,
$\Autom^M = (C^\prime, \delta^\prime, p^\prime_0;\ Q^\prime, \vc \lambda^\prime, \Delta^\prime, \vc\eta^\prime_F)$
%$\Autom^M=(Q^\prime, \Sigma,\vc\lambda^\prime,\delta,\vc\xi_{F})$
where,
\begin{itemize}
\item $C^\prime = C \times [0,M]$ is the finite set of counter states.
\item $\delta^\prime: C^\prime \times \Sigma \to C^\prime$ is the deterministic counter transition. Let $p,q\in C, m\in\N$, $a\in\Sigma$ and $d\in\{-1,0,+1\}$. $\delta^\prime((p,m),a)= (q,m+d) $, if $\delta_{\sgn(m)}(p,a)=(q,d)$.
\item $p^\prime_0 = (p_0,0)$ is the initial counter state.
%\item $F^\prime= \{(p,i)\mid p\in F \text{ and } i\in[0,M]\}$ is the set of final counter states.
\item $Q^\prime = Q \times [0,M]$ is the finite set of states.
\item $\lambda^\prime\in\Sring^{|Q^\prime|}$ is the initial distribution.% where the $i^{th}$ component of $\vc \lambda^\prime$ indicates the initial weight on state $q^\prime_i\in Q^\prime$
$$\lambda^\prime[i]=
\begin{cases} \lambda[i], \text{ if }i<\mq\\ 0, \text{ otherwise} \end{cases}$$
% \item $\vc \lambda^\prime : \Sring^{|Q^\prime|}$ is the initial distribution, where
. %For any $q\in Q, p\in C$ and $n\in[0,M]$, $\lambda^\prime(q,p,n)=\begin{cases}
%\item $\Delta^\prime: C^\prime \times \Sigma \times \{0\} \to \Sring^{\K\times\K}$ gives the transition matrix. %The component in the $i^{th}$ row and $j^{th}$ column of
%For $p\in C, m\in\N$ and $a\in\Sigma$,
%$$\Delta^\prime((p,m),a,0)= \begin{cases} \Delta(p,a,0)\text{ if } m=0\\
% \Delta(p,a,1) \text{ if } 0<m<M\end{cases}$$

\item \final{$\Delta^\prime: \Sigma \to \Sring^{|Q|^\prime\times|Q^\prime|}$ gives the transition matrix. %The component in the $i^{th}$ row and $j^{th}$ column of
For $i,j\in|Q^\prime|$ and $a\in\Sigma$,
%$$\Delta^\prime((p,m),a,0)= \begin{cases} \Delta(p,a,0)\text{ if } m=0\\
% \Delta(p,a,1) \text{ if } 0<m<M\end{cases}$$

$$\hspace{-.15cm}\Delta^\prime(a)[i][j]=\begin{cases} \Delta(a,0)[i][j], \text{ if } i,j<\mq\\
\Delta(a,1)[i\bmod\mq][j\bmod\mq], \text{ if }\frac{i}{\mq}= \frac{j}{\mq}\\
0, \text{ otherwise}\end{cases}$$}
%denotes the weight on the transition from state $q_i\in Q$ to state $q_j\in Q$ on reading a symbol $a$ from counter state $p$.%For any $p\in C,\ a\in\Sigma, q\in Q$ and $m\in \[0,M]$,$$\Delta^\prime(p,a)
\item $\vc\eta^\prime_F\in \Sring^{|Q^\prime|}$ is the final distribution.% where the $i^{th}$ component of $\vc \eta^\prime_F$ indicates the output weight on state $q^\prime_i\in Q^\prime$.
$$\vc\eta^\prime_F[i]=\vc \eta[i\bmod\mq]$$
\end{itemize}
\end{definition}

%\end{toappendix}
%A \dwlts with finitely many states is called a deterministic weighted automata (\dwa).
%\prm{Not required}
%The following result shows that if two weighted automata (\WA) over a field are not equivalent, then there is a linear-sized word to distinguish them.
%\begin{lemma} \cite[Lemma 3.4]{prob}
%\label{WAequivalence} \label{ptimewaequiv}
%Let $\Autom$ and $\Butom$ be two weighted automata with weights from a field. Then the following holds:
%\begin{enumerate}
%\item $\Autom$ and $\Butom$ are not equivalent if and only if there exists $w\in\Sigma^{< |\Autom|+|\Butom|}$ with $f_\Autom(w)\neq f_\Butom(w)$.
%\item There is a polynomial time algorithm that decides if $\Autom$ and $\Butom$ are equivalent. If the machines are not equivalent, the algorithm outputs a word that distinguishes the machines.
%\end{enumerate}
%\end{lemma}
%Tzeng \cite{prob} shows the above lemma for two probabilistic automata. The proof can easily be extended to weighted automata over fields. %For an alternate proof see \cite[Theorem 5.30]{handbook} \label{ptimewaequiv}.

%% file: TeX/reachability.tex
% !TEX root = ../odca.tex
%\section{Tools and Techniques}
%\clearpage
\section{Reachability problems of weighted ODCA}
%\prm{Give a clear and high level picture.}
%\prm{reachability witness to witness}
\label{sec:reachability}
%\prm{Based on LICS feedback\\
%}
%\prm{why this particular reachability problem.}
In this section, we introduce the co-VS reachability and co-VS coverability problems for weighted \odcas over a field $\Sring$.
%\vincent{These two sentences are redundant -- you should rather explain why they are useful for proving equivalence.}
%\final{In this section, we examine two reachability problems of weighted \odcas. These problems lie at the very heart of solving the equivalence of \odcas.
%The identification of these reachability problems as the fundamental component in proving equivalence of \odcas is a notable finding.}
%
%In the subsequent section, we develop techniques that play a key role in proving equivalence of weighted \odca. }
 %{In the first subsection, we prove that there is a polynomial time algorithm to solve these reachability problems when the input counter values are specified in unary notation. In the second subsection, we prove that these reachability problems are in $\CF{NP}$ when the counter values are specified in binary notation.}
%\sav{We should start with a short paragraph describing in simple English what this section is all about. The technical details can be skipped in this paragraph.}
%\sav{Is this reachability problem over some transition system (is this the correct terminology)? Since we are not really using the final state of an odca.}
%\prm{We can call the minimal word, a minimal witness, if required.}
%\sav{You should have a brief writeup on what this section will do.}
%\sav{I am quickly noting down informally the lemma statements/proof idea to be used. It has to be formalised.}
We fix a weighted \odca $\Autom=((C, \delta_0,\delta_1, p_0),(Q, \vc \lambda, \Delta, \vc\eta))$. % for this section. %Without loss of generality, assume $|C|=|Q|$ and denote $|Q|$ by $\K$. 
We use $\lsv\subseteq\Sring^\mq$ to denote a vector space and
%\sav{Use some macro for complement of $\lsv$.}
{$\clsv$ its complement}.
%\rem{{$\clsv= \Sring^\mq \setminus \lsv$ to denote the set complement of $\lsv$}.} %That is, $ \clsv = \{\vc x \in \Sring^\K \mid \vc x \not\in \lsv \}$. 
Let $S\subseteq C$ be a subset of the set of counter states, $X\subseteq \N$ a set of counter values, and $w\in\Sigma^*$. The notation $\con\xrightarrow{w} \clsv \times S \times X$ denotes the run $\con\xrightarrow{w}\conD$ where $\conD \in \clsv \times S \times X$ if it exists. We use $\con\xrightarrow{*} \clsv \times S \times X$ to denote that there exists a word $u\in\Sigma^*$ such that $\con\xrightarrow{u} \clsv \times S \times X$. 
%\prm{Do we have to mention that the run need not exist.}
 %\prince{Can be removed: Note that a minimal basis for $\lsv$ can contain at most $\K$ vectors. A polynomial time algorithm (Gaussian elimination) checks whether a given vector belongs to a vector space $\lsv$.} In this section, we look at two problems.

% These will aid us in proving the equivalence of \odcas. %\final{Reducing  the equivalence problem into reachability problems is a crucial step towards its resolution. -- redundant with intro, remove -- Vincent}

  \vspace{.2cm}
 \begin{tabular}{|p{0.9\textwidth}|}
        \hline
       \textsc {co-VS reachability problem}\\
           \textsc{{Input:}} a weighted \odca $\Autom$, an initial configuration $\con$, a vector space $\lsv$, a set of counter states $S$, and a counter value $m$.\\       
      \textsc{{Output:}} \textit{Yes}, if there exists a run $ \con \xrightarrow{*} \clsv\times S\times \{m\}$ in $\Autom$. \textit{No}, otherwise.\\
        \hline
        \end{tabular}
        \vspace{.1cm}
        
\vspace{.1cm}
 \begin{tabular}{|p{.9\textwidth}|}
        \hline
        {\textsc{co-VS coverability problem}} \\
\textsc{{Input:}} a weighted \odca $\Autom$, an initial configuration $\con$, a vector space $\lsv$, and a set of counter states $S$. \\
\textsc{{Output:}} \textit{Yes}, if there exists a run $ \con \xrightarrow{*} \clsv\times S \times \N$ in $\Autom$. \textit{No}, otherwise.\\
\hline
\end{tabular}
\vspace{.2cm}

%\begin{example}
%\rem{
%Let $\Autom$ be the weighted \odca given in \Cref{figureex4}. 
%Let $\Butom$ be a weighted automaton recognising the function $f: (a+b)^* \to \N$ such that for all $w\in(a+b)^*$, $f(w)=$ decimal value of $w$ when interpreted as a binary number. For any $k\in\N$ and a given configuration $\con$ of $\Autom$, we say that $\Butom$ has a $k$-equivalent distribution, if there is an initial distribution of $\Butom$ which accepts all words of length at most $k$ with the same weight as $\Autom$ does from $\con$.
%
%Let $k>0$ and $\lsv=\{\vc x \mid$ the configuration $(\vc x,p_2,0)$ has a $k$-equivalent distribution in $\Butom\}$ be a vector space. Given the initial configuration $(\vc \lambda, p_0, 0)$ of $\Autom$, where $\lambda$ denotes the initial weight vector $[1,0,0,0]$, the vector space $\lsv$, set of counter state $\{p_2\}$, and counter value $0$ as input for the co-VS reachability problem. The minimal witness of reachability is the word $aba^{k+1}$. This is because of the fact that the weighted automaton $\Butom$ and weighted \odca $\Autom$ behaves the same way for all words of length $k$ after reading the string $aba^{k+1}$, as the counter value cannot become zero.}
%\end{example}

Unlike the co-VS reachability problem, the final configuration's counter value is not considered part of the input for co-VS coverability problem.  
We assume that the vector space $\lsv\subseteq\Sring^\mq$ is provided by giving a suitable basis. 
%We assume that the counter value of the initial configuration and the final counter value, if part of the input, are given in unary notation.
We call $z\in\Sigma^*$ a \emph{witness} of $(\con,\clsv,S,X)$ if $\con\xrightarrow{z} \clsv \times S \times X$. 
Furthermore, $z$ is called a \emph{minimal}  witness for $(\con,\clsv,S,X)$ if for all $u\in\Sigma^*$ with $\con\xrightarrow{u} \clsv \times S \times X$, $|u|\geq|z|$.
\label{appendReach}
%\prm{To be removed}
First, we look at the particular case of co-VS reachability problem for weighted automata.
Note that for weighted automata, the counter value is always zero. Given a weighted automata $\Butom$, with $k$ states, an initial configuration $\ucon$, a vector space $\lsu\subseteq \Sring^{k}$ and a set of counter states $S$, the co-VS reachability problem asks whether there exists a run $\ucon\xrightarrow{*} \clsu \times S \times \{0\}$.
\begin{theorem}\label{wacovs}
There is a polynomial time algorithm that decides the co-VS reachability problem for weighted automata and outputs a minimal reachability witness if it exists.
\end{theorem}
%\prm{Cite sch\"{u}tzenberger} 
\begin{proof}
The idea of equivalence checking of weighted automata goes back to the seminal paper by Sch\"{u}tzenberger~\cite{berger}.
 %The algorithm used there for solving the equivalence of two probabilistic automata and to find a minimal distinguishing word can be used to find a minimal reachability witness for the co-VS reachability problem in polynomial time for any vector space $\lsu$ when the transition weights are taken from a field.
 Tzeng~\cite{prob} provided a polynomial time algorithm for the equivalence of two probabilistic automata by reducing the problem to the co-VS reachability problem where $\lsv = \{\vc 0\}$. The same algorithm can be modified to solve the general co-VS reachability problem.
% 
%The equivalence of two probabilistic automata is solved by Tzeng in \cite{prob} by reducing it to a co-VS reachability problem, where $\lsv= \{\vc 0\}$. Given a weighted automaton $\Butom$, we can use their algorithm to find a minimal reachability witness for any co-VS reachability problem of  $\Butom$ in polynomial time.
%\sav{Rewrite this. Not written properly. One line contains around 50 words.}
%\prm{reachability problem for weighted automata. standard proof idea.}
\end{proof}
%\end{toappendix}

%binary version of both problems are in $\CF{NP}$.
%\prm{The next subsection gives a pumping lemma for floating runs and use them to bound the length of the witness whose run is a floating run.}

In the upcoming subsection, we give some interesting properties of minimal witnesses.  In \Cref{sec:reachcover}, we provide a pseudo-pumping lemma which helps us show that co-VS reachability and co-VS coverability are in $\CF{P}$ if the counter values are given in unary notation. Finally, in \Cref{sec:lex_min_witness}, we demonstrate that the lexicographically minimal witness has a canonical form. In the following subsections, $\lsv$ denote a vector space, $\con$ a configuration, $S$ a subset of counter states, and $X \subseteq \N$. We also denote by $K = \mq\cdot\mc$, where $C$ is the set of counter states, and $Q$ is the set of states of the finite state machine.

\subsection{Minimal witness and its properties}
The following observation helps in breaking down the reachability problem into sub-problems. %This approach facilitates the development of an effective solution to the reachability problem.}
%
%\prm{Add a subsection ``witness and its properties". }
If $z \in \Sigma^*$ is a minimal witness for $(\con,\clsv,S,X)$, then for every $z_1,z_2$ such that $z= z_1 z_2$, there is a computable vector space $\lsu$ such that $z_1$ is a minimal witness for $(\con, \clsu, \{p\},\{n\})$ where $p$ is the counter state and $n$ is the counter value reached after reading $z_1$ from $\con$. %, and $\clsu$ is defined such that one can always reach $\clsv \times S \times X$ on reading the word $z_2$ from $\clsu \times \{p\} \times \{n\}$ (see \Cref{figure}). 

%The following lemma helps us in breaking down both the reachability problems into smaller sub-problems. %The proof is given in the Appendix.
%\prm{Explain why it is needed. Mention that the path has changed but $z_2$ remains the same.}
\begin{observation}\label{newvs}
Consider arbitrary $z,z_1, z_2\in\Sigma^*$ such that $z= z_1z_2$. Let $\conD= \configodc{\vc x_\conD}{p_\conD}{n_\conD}$ and  $\conE=\configodc{\vc x_\conE}{p_\conE}{n_\conE}$ be configurations such that $\con\xrightarrow{z_1}\conD\xrightarrow{z_2}\conE$ and $\mata\in\Sring^{\mq\times\mq}$ be such that $\vc x_\conD\mata= \vc x_\conE$. If  $z$ is a minimal witness for $(\con,\clsv,S,X)$, then $z_1$ is a minimal witness for $(\con, \clsu, \{p_\conD\},\{n_\conD\})$, where $\lsu=\{ \vc y \in \Sring^{\mq} \mid {\vc y} \mata \in \lsv \}$.
\end{observation}
\begin{proof}
Let $z\in \Sigma^*$ be a minimal reachability witness for $(\con,\clsv,S,X)$, $\conD= \configodc{\vc x_\conD}{p_\conD}{n_\conD}$ and  $\conE=\configodc{\vc x_\conE}{p_\conE}{n_\conE}$ be configurations such that $\con\xrightarrow{z_1}\conD\xrightarrow{z_2}\conE$ where $z_1,z_2\in \Sigma^*$ with $z=z_1z_2$ and $\mata\in\Sring^{\mq\times\mq}$ be such that $\vc x_\conD\mata= \vc x_\conE$.
Let $\lsu=\{ \vc y \in \Sring^{\mq} \mid {\vc y} \mata \in \lsv \}$.
Assume for contradiction that there exists $z_1^\prime\in\Sigma^*$ smaller than $z_1$ and $\con\xrightarrow{z_1^\prime} \conF$ for some configuration $\conF\in \clsu\times \{p_\conD\} \times \{n_\conD\}$.
Note that for all $\vc y\in\clsu$, the vector $\vc y \mata \in \clsv$. Since the configurations $\conF$ and $\conD$ have the same counter state and counter value, $\con\xrightarrow{z^\prime_1}\conF\xrightarrow{z_2}\clsv\times \{p_\conE\} \times \{n_\conE\}$ is a run and the word $z^\prime_1z_2$ contradicts the minimality of $z$.
\end{proof}
%\sav{combining witnesses} 
%Even though the path and final configuration have changed, it should be noted that the word $z_2$ remains the same in the above lemma. 
%\rem{Now, we show that the length of a minimal witness can be bounded by a polynomial in the number of distinct counter values encountered during its run, as stated in the following lemma.}
We aim to show that the length of a minimal witness for $(\con,\clsv,S,X)$ is polynomially bounded. The following lemma shows that if the counter values are polynomially bounded during the run of a minimal witness, then its length is also polynomially bounded.
%\begin{lemma}[cut]
%\label{lem:cut}
%Let $w\in\Sigma^*$ and $\con, d$ be configurations of $\Autom$ where $\vc x_{d} \in \clsv$, and $\con \xrightarrow{w} d$ is a floating run. If $|n_{d}- n_{\con}|> \K^2$ then there exists a word $u$ which is a subword of $w$ and a configuration $e$ such that $\con \xrightarrow{u} e$ is a floating run with $\vc x_{e} \in \clsv$ and $p_{d}= p_{e}$.
%\end{lemma}
%

\begin{lemma}\label{lengthboundcounter}
Let $z\in \Sigma^*$ be a minimal witness for $(\con,\clsv,S,X)$.
If the number of distinct counter values encountered during the run $\con\xrightarrow{z} \clsv\times S \times X$ is $t$, then $|z| \leq t \cdot K$.
\end{lemma}
\begin{proof}
Let $\con=\con_1$ and $T(\con_1)= \con_1 \tau_1 \con_2 \cdots \tau_{h-1} \con_h$ be the run on word $z$ from $\con_1$ and $T$ the corresponding sequence of transitions. Let $t$ be the number of distinct counter values encountered during this run. Now assume for contradiction that $h> \mq\cdot\mc\cdot t$, then by Pigeon-hole principle, there are $\mq+1$ many configurations $\con_{i_0}, \con_{i_1}, \ldots , \con_{i_\mq}$ with the same counter state and counter value during this run.
%Let $\con_{i_0}, \con_{i_1}, \ldots , \con_{i_\K}$ be these $\K+1$ configurations.
Given a configuration $\con$, let $\vc x_\con$ denote $\wgtvec{\con}$.
Let $\mata_{j}$ denote the matrix such that ${\vc x}_{\con_{i_j}} \mata_j = \vc x_{\con_{h}}$ for all $j \in [0,\mq]$. From \Cref{lem:mat} we get that there exists $r\leq \mq$, %such that $\vc x_{\con_{i_r}}$ is dependent on $\vc x_{\con_{i_0}}, \vc x_{\con_{i_1}}, \ldots , \vc x_{\con_{i_{r-1}}}$.
%Also, since ${\vc x}_{\con_{i_r}} \mata_r \in \clsv$, we get that there exists
and $t \in [0,r-1]$ such that ${\vc x}_{\con_{i_{t}}} \mata_r \in \clsv$.
Consider the sequence of transitions $T^\prime=\tau_{1\cdots {i_{t}}}\tau_{r \cdots {\ell-1}}$ and $v=\word(T^\prime)$. The run $\pi(v,\con_1)=T^\prime(\con_1)$ is a  run since configurations ${\con_t}$ and ${\con_r}$ have the same counter state and counter value. This is a shorter run than $\pi(z,\con_1)$ and $\con_1\xrightarrow{v}\clsv \times S \times X$. This is a contradiction since $z$ was assumed to be minimal.% contradicts the minimality of $z$.
\end{proof}
It now suffices to show that the counter values encountered during the run of a minimal witness are polynomially bounded.

%\subsection{co-VS reachability}
\subsection{Pseudo-pumping lemma}
\label{sec:reachcover}
%\sav{We can pump up and pump down a sufficiently long word.}

{The pseudo-pumping lemma is a valuable tool in our analysis, allowing us to pump up or down a sufficiently long word while maintaining the reachability conditions.}
%
%We first show that if the run of a minimal reachability witness of $(\con, \clsv, S,\{m\})$ is a floating run, %of a minimal word $z$ which reaches a configuration in $ \clsv \times \{q\} \times \{m\}$ for some $q\in C, m\in\N$ from a given configuration $\con$,
%then the maximum and minimum counter values encountered during this run are bounded by a polynomial in $\mq,\mc$ and the initial and final counter values. %This will in turn help us in bounding the length of the word $z$.
%%\prm{combined u-turn and the other lemma. Need to do another clean up. Add figure.}

%\prm{downward hill can be mentioned inside the proof where it is used.}%  Refer previous lemma statement. i.e., include ideas of \Cref{mincounter} also.}
%\prm{Should we replace even these polynomials by $\p{i}{\mq}{\mc}$.}
%\vincent{You should not give polynomials at all, simply say it is polynomial}

\begin{lemma}[pseudo-pumping lemma]
\label{lem:uturn}
\label{lengthboundfloating}
Let $m,R \in \N$, be such that $\cntval{\con}=m$ and $z \in \Sigma^*$ be such that $\con \xrightarrow{z} \clsv \times S \times \{m\}$ is a floating run, and the maximum counter value encountered during this run is $m+R$. If $R > K^2$, then there exists $z_{sub}, z_{sup} \in \Sigma^*$ such that the following hold:
\begin{enumerate}
\item \label{pumpdown} there exist $x,y,u,v,w\in \Sigma^*$ such that $z=xyuvw, z_{sub}=xuw$, $\con \xrightarrow{z_{sub}} \clsv \times S \times \{m\}$ is a floating run, and the counter values encountered during this run are less than $m+R$, and 
\item there exist $x,y,u,v,w \in \Sigma^*$ such that $z=xyuvw, z_{sup}=xy^2uv^2w$, $\con \xrightarrow{z_{sup}} \clsv \times S \times \{m\}$ is a floating run, and the maximum counter value encountered in this run exceeds $m+R$. \label{pumpup}
\end{enumerate}
\end{lemma}
%\prm{Note length of $x,y,u,v$ and $r$ need not be polynomial in the size of the machine. The pumping is not true for any $i\in\N$. Holds only for 2.}
%\rem{
%\begin{lemma}%[uturn]
%\label{lem:uturn}
%\label{lengthboundfloating}
%Let $z\in\Sigma^*$ be a reachability witness for $(\con, \clsv,S, \{m\})$. If $\con \xrightarrow{z} \clsv \times S \times \{m\}$ is a floating run and
%%If $|u| >\K^2\cdot(t+ 2\K^4)$ then there exists $ v \prec u$ and a configuration $\con_t$ such that
%%$\con_1 \xrightarrow{v} \con_t$ % \xrightarrow{u_3} \con_6$
%%with $ p_{\con_t}=q, n_{\con_t}=m$ and $\vc x_{\con_t} \in \clsv$ and 
% the maximum counter value encountered during this run is greater than $max(n_{\con},m)+(\mq\cdot\mc)^2$ then,%, and
%%\item \label{mincounter}The minimum counter value during this run is greater than $min(n_{\con},m)-\K^4$.
%%\item $|z| \leq \K^2\cdot(|m-n_{\con}|+ 2\K^4)$.
%\begin{enumerate}
%\item $z$ is not the minimal reachability witness for $(\con, \clsv,S, \{m\})$, and
%\item There exists $z^\prime \in \Sigma^*$ such that $\con \xrightarrow{z^\prime} \clsv \times S \times \{m\}$ is a floating run and the maximum counter value encountered during this run is greater than the maximum counter value encountered during the run $\con \xrightarrow{z} \clsv \times S \times \{m\}$.
%\end{enumerate}
%\end{lemma}}
\begin{proof}
Let $z\in\Sigma^*$ be a  witness for $(\con,\clsv, S,\{m\})$ and $\conE\in\clsv \times S \times \{m\}$ be such that $\con \xrightarrow{z}\conE$ is a floating run, and the maximum counter value encountered in this run be $m+R$ where $R> K^2$. Let $\cntval{\con}=m$. 
%Let us assume that $n\geq m$. The case where $m<n$ can be proven analogously. 
%\begin{point}\label{maxcounter}
%Let the maximum counter value encountered during this run be greater than $m+\mq^2\cdot\mc^2$.
There exist $z_1,z_2\in\Sigma^*$ and  configuration $\conF$ such that $z=z_1z_2$ and %the run on  $z$ from $\con$ can be written as follows:
$
\con \xrightarrow{z_1}\conF\xrightarrow{z_2} \conE% \xrightarrow{z_3} \conF
$, % $n_\conE=m$ and
where $\cntval{\conF}=m+R$ % is a configuration with the maximum counter value in this run (see \Cref{fig:ucut}). %Let $\matm \in \Sring^{\mq\times\mq}$ such that $\vc x_{\conF}={\vc x}_{\conE}\matm$.
%From \Cref{lem:comp}, we know that the set $\lsu=\{ \vc y \in \Sring^{\mq} \mid {\vc y} \matm \in \lsv \}$
%is a vector space and hence the vector ${\vc x}_{\conE}\in\clsu$.
(see \Cref{fig:ucut}).

%Let $R\in\N$, such that the counter value of $\conD$ is $m+R$.
Let $\con_1=\con$ and $\pi= \con_1 \tau_1 \con_2 \cdots \tau_{\ell-1} \con_\ell$ denote the run on word $z$ from the configuration $\con_1$ and $T = \tau_1\tau_2\cdots \tau_{\ell-1}$ the sequence of transitions of $\pi$.
%Let $M = max_\ce(\pi(z,\con))$. Note that $M= n_\conD-n_\con$.
For any $i \in [0,R]$, we denote by $l_i$ and $d_i$ the indices such that a configuration with counter value $m+i$ is encountered for the last (resp. first) time before (resp. after) reaching counter value $m+R$ in $\pi$. 
That is, $\cntval{\con_{l_i}} = \cntval{\con_{d_i}} = m+i$, and for any $j$ where $l_i < j < d_i$, $\cntval{\con_{j}} > m+i$.
%$\ce({T_{1 \cdots l_i-1}}) = \ce({T_{1 \cdots d_i-1}}) = i$, and for any $j \in [l_i,d_i-2]$, $\ce({T_{1 \cdots j}}) > i$.
To simplify the notation, we denote by $\conG_i = \con_{l_i}$ and $\conG^\prime_i = \con_{d_i}$.

Consider the pairs of configurations $(\conG_1, \conG^\prime_1), (\conG_2, \conG^\prime_2), \dots, (\conG_R, \conG^\prime_R)$. 
 Since $R>(\mq\cdot\mc)^2$, by the Pigeonhole principle, there exist two counter states $p, q$, and a set of indices $X \subseteq [0,R]$ where $|X| = \mq^2+1$ such that for all $h \in X$, $\cntstate{\conG_h} = p$ and $\cntstate{\conG^\prime_h} = q$. 
%
%Let $r=\mc^2+1$. Since $R>(\mq\cdot\mc)^2$, by Pigeonhole principle, there exists set of indices $X=\{i_1,i_2, \cdots , i_{r}\}\subseteq [0,R]$ such that for any $1\leq k<l \leq r$, we have $i_k<i_l$ and for all $h,j\in X$, ${\conG_{h}},{\conG^\prime_{h}},{\conG_{j}}$ and ${\conG^\prime_{j}}$ have the same counter state.
%Consider the configuration pairs $(\con_{i_r}, \con^\prime_{i_r})$, $r\in[1, m]$.
%Let $m=\K^2$, if $k>\K^4$ then we can find at least $m+1$ pairs of configurations (on both sides of the configuration $\con_2$) of the following form. \prm{rewrite, clarity}
%\[
%[\con_{l}, \con_{i^\prime}],[\con_{i+1}, \con_{i^\prime+1}], \ldots, [\con_{i+m}, \con_{i^\prime+m}]
%%[(\vc q_1, b, l_1), (\vc q_1',b, l_1)], [(\vc q_2, b, l_2), (\vc q_2',b, l_2)], \dots, [(\vc q_m, b, l_m), (\vc q_m',b, l_m)]
%\]
%such that $p_{\con_j}= p_{\con_{j^\prime}}$ and $n_{\con_j}= n_{\con_{j^\prime}}$ for all $j\in[i,i+m]$.
%/The run of $u$ can be written in the following form
%
%\[
%\con_1 \xrightarrow{*} \con_{i_1} \xrightarrow{*} \con_{i_2} \xrightarrow{*} \con_{i_3} \cdots \xrightarrow{*} \con_{i_m} \xrightarrow{*} \con_2 \xrightarrow{*} \con^\prime_{i_m}\xrightarrow{*} \con^\prime_{i_{m-1} } \cdots \xrightarrow{*} \con^\prime_{i_1} \xrightarrow{*} \con^\prime_1
%%(\vc p_1, s_1, l) \rightarrow^* (\vc q_1, b, l_1) \rightarrow^* (\vc q_2, b, l_2) \dots (\vc q_m, b, l_m) \rightarrow^* (\vc p_k, s_k, l+k) \rightarrow^* (\vc q_m', b, l_m) \rightarrow^* (\vc q_1', b, l_1) \rightarrow^* (\vc p_t, s_t, l)
%\]
%Let $X=\{i_1,i_2, \cdots , i_{r}\}$ be a set containing these indices.
For all $j \in X$, let $u_j, v_j, w_j \in \Sigma^*$ be such that 
$\con_1 \xrightarrow{u_j} \conG_{j} \xrightarrow{v_j} \conG^\prime_{j} \xrightarrow{w_j} \conE
%(\vc p_1, s_1, l) \rightarrow^{x_i} (\vc q_i, b, l_i) \rightarrow^{y_i} (\vc q_i', b, l_i) \rightarrow^{z_i} (\vc p_t, b, l)
$.  We use the following shorthand for any configuration $\conG$: $\vc x_{\conG} = \wgtvec{\conG}$.
%For any configuration $\conG$, let $\vc x_\conG$ denote its weight vector.
%Note that if $i<j$, we have that $u_i$ is a prefix of $u_j$ and $w_i$ is a suffix of $w_j$.
For all $j \in X$, let matrix $\mata_j$ and $\matb_j$ be such that $\vc x_{\conG_j^\prime} = {\vc x}_{\conG_j} \mata_j$ and $\vc x_{\conE} = {\vc x}_{\conG^\prime_j} \matb_j$. Since ${\vc x}_{\conE} \in \clsv$, for all $j\in X$, ${\vc x}_{\conG_j}\mata_j\matb_j \in \clsv$. %Now we list the matrices in the following sequence $
%\mata_{i_r}, \mata_{i_{r-1}}, \dots, \mata_{i_1}
%$.
%From \Cref{lem:mat}, it follows that, there exists $h,j\in X$ with $h< j $ such that ${\vc x}_{\conG_h}\mata_j\matb_h \in \clsu$.
%
Let $r = \mq^2+1$, and $i_1 < i_2 < \dots < i_r$ be the indices in $X$.
We prove the two cases separately.
\begin{point}
%Assume for contradiction that $z$ is a minimal reachability witness for $(\con,\clsv, S,\{m\})$.
%We prove that the maximum counter value encountered during this run is bounded.
 %From \Cref{newvs}, we know that $z_1z_2$ is a minimal reachability witness for $(\con,\clsu,\{p_\conE\},\{n_\conE\})$. We contradict the minimality of $z_1z_2$.
%we can replace the run as follows
%\[
%\con_1 \xrightarrow{u_t} \con_{l}\xrightarrow{v_d} \con^{\prime\prime}_{l}\xrightarrow{w_t} \con^{\prime\prime}_{1}
%%(\vc p_1, s_1, l) \rightarrow^{x_i} (\vc q_i, b, l_i) \rightarrow^{y_j} (\vc q_i'', b,l) \rightarrow^{z_i} (\vc p_r, s_r, l)
%\]
%where $\con^{\prime\prime}_{l}= (\trans{\vc x_{\con_{l}}}\mata_d, n_{\con_t}, p_{\con_t})$
%and $\vc x_{\con^{\prime\prime}_1} \in \clsv$.
Consider the sequence of matrices: $
\mata_{i_r}, \mata_{i_{r-1}}, \dots, \mata_{i_1}
$.
Since there can be at most $\mq^2$ independent matrices, there exists $k\in[1,r]$ such that $\mata_{i_k}$ is a linear combination of $\mata_{i_r}, \dots, \mata_{i_{k+1}}$.
%Since $\vc \matm_i$ is dependent on $\matm_0, \dots, \matm_{i-1}$, we prove that there exists $j<i$ such that ${\vc x} \mata_i \matm_j \matb_i \notin \lsv$.
Hence,  there exists $h \in \{i_r, \dots, i_{k+1} \}$ such that ${\vc x}_{\conG_{i_k}}\mata_{h} \matb_{i_k} \in \clsv$. Let $z_{sub} = u_{i_k}v_hw_{i_k}$. It is easy to observe that $z_{sub}$ is a subword of $z$ as mentioned in the lemma. To conclude the proof, it now suffices to show that $z_{sub}$ is a witness for $(\con,\clsv, S,\{m\})$ and the counter values encountered during the run $\con \xrightarrow{z_{sub}} \conH$ are less than $m+R$. Consider the floating run $\conG_{h} \xrightarrow{v_h} \conG^\prime_{h}$. From the choice of $\conG_h$ and $\conG^{\prime}_h$ we know that $\cntval{\conG_h} = \cntval{\conG^{\prime}_h} = m+h$ and for all $j$ where $l_h < j < d_h$, $\cntval{\con_j} > m+h$. Since $\cntstate{\conG_h} = \cntstate{\conG_{i_k}}$, $\pi(v_h,\conG_{i_k})$ is also a floating run $\conG_{i_k} \xrightarrow{v_h} \conD$ such that $\cntstate{\conG^\prime_{h}} = \cntstate{\conD}$, $\cntval{\conG_{i_k}} = \cntval{\conD} = m+i_k < m+h$, and the minimum and maximum counter values encountered in the run is $m+i_k$ and $m+R - (h-i_k)$ respectively (see \Cref{fig:ucut}). Furthermore, $\vc{x}_{d} = \vc{x}_{\conG_{i_k}} \mata_{h}$. Since $\cntstate{\conG^\prime_{i_k}} = \cntstate{\conG^\prime_h}$, we get that $\cntstate{\conG^\prime_{i_k}} = \cntstate{\conD}$. Moreover, since $\cntval{\conG^\prime_{i_k}} = \cntval{\conG_{i_k}}$, we have $\cntval{\conG^\prime_{i_k}} = \cntval{\conD}$. Therefore, $\pi(w_{i_k},\conD)$ is the run $\conD \xrightarrow{w_{i_k}} \conH$ where $\vc{x}_{\conH} = \vc{x}_{\conD} \matb_{i_k}$ and hence $\vc{x}_{\conH} = \vc{x}_{\conG_{i_k}} \mata_{h} \matb_{i_k}\in \clsv$. This concludes that $z_{sub}$ is a witness for $(\con,\clsv, S,\{m\})$ and satisfies the properties mentioned in the lemma.
\end{point}
\begin{point}
\label{proofofpumpup}
Consider the sequence of matrices: $\mata_{i_1}, \mata_{i_2}, \ldots , \mata_{i_r}$. Since there can be at most $\mq^2$ independent matrices, there exists $k\in[1,r]$ such that $\mata_{i_k}$ is a linear combination of $\mata_{i_1}, \dots, \mata_{i_{k-1}}$.
%Since $\vc \matm_i$ is dependent on $\matm_0, \dots, \matm_{i-1}$, we prove that there exists $j<i$ such that ${\vc x} \mata_i \matm_j \matb_i \notin \lsv$.
Hence,  there exists an $h \in \{i_1, \dots, i_{k-1} \}$ such that ${\vc x}_{\conG_{i_k}}\mata_{h} \matb_{i_k} \in \clsv$. Let $z_{sup} = u_{i_k}v_hw_{i_k}$. It is easy to observe that $z_{sup}$ is a superword of $z$ as mentioned in the lemma. To conclude the proof, it now suffices to show that $z_{sup}$ is a witness for $(\con,\clsv, S,\{m\})$ and the counter values encountered during the run $\con \xrightarrow{z_{sup}} \conH$ is greater than $m+R$. Consider the floating run $\conG_{h} \xrightarrow{v_h} \conG^\prime_{h}$. From the choice of $\conG_h$ and $\conG^{\prime}_h$ we know that $\cntval{\conG_h} = \cntval{\conG^{\prime}_h} = m+h$ and for all $j$ where $l_h < j < d_h$, $\cntval{\con_j} > m+h$. Since $\cntstate{\conG_h} = \cntstate{\conG_{i_k}}$, $\pi(\conG_{i_k},v_h)$ is also a floating run $\conG_{i_k} \xrightarrow{v_h} \conD$ such that $\cntstate{\conG^\prime_{h}} = \cntstate{\conD}$, $\cntval{\conG_{i_k}} = \cntval{\conD} = m+i_k > m+h$, and the minimum and maximum counter values encountered in the run is $m+i_k$ and $m+R + (i_k-h)$ respectively. Furthermore, $\vc{x}_{d} = \vc{x}_{\conG_{i_k}} \mata_{h}$. Since $\cntstate{\conG^\prime_{i_k}} = \cntstate{\conG^\prime_h}$, $\cntstate{\conG^\prime_{i_k}} = \cntstate{\conD}$. Moreover, since $\cntval{\conG^\prime_{i_k}} = \cntval{\conG_{i_k}}$, we have $\cntval{\conG^\prime_{i_k}} = \cntval{\conD}$. Therefore $\pi(\conD,w_{i_k})$ is the run $\conD \xrightarrow{w_{i_k}} \conH$ where $\vc{x}_{\conH} = \vc{x}_{\conD} \matb_{i_k}$ and hence $\vc{x}_{\conH} = \vc{x}_{\conG_{i_k}} \mata_{h} \matb_{i_k}\in \clsv$. This concludes that $z_{sup}$ is a witness for $(\con,\clsv, S,\{m\})$ and satisfies the properties mentioned in the lemma.
\end{point}
\end{proof}

\begin{figure}
%\centering
%\begin{minipage}{.53\textwidth}
\centering
\resizebox{.55\columnwidth}{!}{%
\begin{tikzpicture}
\tikzset{every path/.style={line width=.3mm}}
\draw plot [smooth, tension=.6] coordinates { (0,0) (.3,1) (.7,2) (1.35,1) (1.64,1.5) };
\draw [dashed] plot [smooth, tension=.6] coordinates{(1.64,1.5) (2,2.5)(2.25,3.3)};
\draw plot [smooth, tension=.6] coordinates{(2.25,3.3) (3,5) (3.8,3.3)};
\draw [dashed] plot [smooth, tension=.6] coordinates{(3.8,3.3) (4.2,2.8) (4.7,3.5) (4.95,3.3)(5.2,2.5) (5.53,1.5)};
\draw (3,4.3) node {$v_{h}$};
\draw plot [smooth, tension=.6] coordinates{ (5.53,1.5)(5.7,1) (6,0) }; %(6.3,-.4)(6.7,1) (6.9,2)
%\filldraw (6.9,2) circle[radius=1.5pt] node[anchor=south] {$\conF$} node[xshift=.15cm, yshift=-1.1cm]{$z_3$};
\filldraw (3,5) circle[radius=1.5pt] node[anchor=north,yshift=0.6cm] {$\conF$};
\filldraw (0,0) circle[radius=1.5pt] node[anchor=south,yshift=-0.6cm] {$\con$};
\filldraw (6,0) circle[radius=1.5pt] node[anchor=south,yshift=-0.6cm] {$\conE$};
\filldraw (1.64,1.5) circle[radius=1.5pt] node[anchor=east] {$\conG_{i_k}$} node[xshift=-1.65cm, yshift=-.5cm]{$u_{i_k}$};%rotate=70,
\filldraw (5.53,1.5) circle[radius=1.5pt] node[anchor=west] {$\conG^\prime_{i_k}$} node[xshift=.45cm, yshift=-.5cm]{$w_{i_k}$};%yshift=-.3cm
\filldraw (2.25,3.3) circle[radius=1.5pt] node[anchor=east] {$\conG_{h}$};
\filldraw (3.8,3.3) circle[radius=1.5pt] node[anchor=west] {$\conG^\prime_{h}$};
\draw [line width=0.38mm] [-stealth](-1,-1) -- (7,-1);%node[anchor=south, xshift=-4.5cm, yshift=-1cm] {word length};
%\draw [gray] [|<->|](-.2,0) -- (-.2,5)node[anchor=south,rotate=90, xshift=-2.3cm] {$>\K^4$};
\draw [dotted] (-1,5)node[anchor=east]{$m+R$} -- (7,5) ;%dashed
\draw[dotted](-1,0) node[anchor=east]{$m$}-- (7,0);
\draw[dotted](-1,1.5)node[anchor=east]{$m+i_k$} -- (7,1.5);
\draw[dotted](-1,3.3)node[anchor=east]{$m+h$} -- (7,3.3);
\draw [line width=0.38mm] [-stealth](-1,-1)node[anchor=south,rotate=90, xshift=3.5cm, yshift=1.5cm] {counter value} -- (-1,5.8);\end{tikzpicture}
}
\caption{The figure shows the floating run from a configuration $\con$ with $\cntval{\con}=m$ to a configuration $\conE= (\vc x, p,m)$ such that $\vc x\in\clsu$. Configurations $\conG_{i_k}$ and $\conG_h$ (resp. $\conG^\prime_{i_k}$ and $\conG^\prime_h$) are where the counter values $m+i_k$ and $m+h$ are encountered for the last (resp. first) time before (resp. after) reaching $m+R$. Also, $\cntstate{\conG_{i_k}}= \cntstate{\conG_{h}}$ and $\cntstate{\conG^\prime_{h}}=\cntstate{\conG^\prime_{i_k}}$. The dashed line denotes the part of the run that can be removed to get a shorter witness for $(\con, \clsu, \{p\}, \{n\})$.}
\label{fig:ucut}
%\end{minipage}
\end{figure}
%\final{Note that the lengths of the words $x$, $y$, $u$, $v$, and $r$ are not necessarily polynomially bounded by the size of the machine.}
%\vincent{What is this remark for?}
It is important to note that we do not end up in the same configuration while pumping up/down, but we ensure that we reach a configuration with the same counter state, counter value, and whose weight vector is in the complement of the given vector space.
%In deterministic \textsc{oca}, the pumping maintains the same final configuration and the same word pairs can be pumped up any number of times, maintaining the reachability to a given configuration.

Now, we prove that for any run (it need not necessarily be a floating run)
of a minimal reachability witness $z$ for $(\con,\clsv,S,\{m\})$, the maximum counter value encountered during the run $\con\xrightarrow{z} \clsv \times S \times \{m\}$ is bounded by a polynomial in the number of states of the machine, and the initial and final counter values. This can be achieved by iteratively applying \Cref{lengthboundfloating} on the run of the minimal witness (refer  \Cref{fig:nonfloatingrun}) and using \Cref{newvs} and \Cref{lengthboundcounter}.%A detailed proof is given in the Appendix.
%We prove this in \Cref{lengthbound}.
%\begin{figure}[!h]
%
%\end{figure}
%\prm{Next 2 lemmas, are corollary of the pumping lemmas. Combine them?}
%The next lemma can be regarded as a corollary of the pseudo-pumping lemma.
\begin{corollary}
\label{lengthbound} \label{smp}
%The counter values (and therefore length) of a run to a complement to a vector space can be bounded.
If $z\in \Sigma^*$ is a minimal witness for $(\con, \clsv, S, \{m\})$, then
%where $\cntval{\con}=n$, then
\begin{enumerate}
\item the maximum counter value encountered during the run
$\con \xrightarrow{z} \clsv\times S\times \{m\}$ is less than $max(\cntval{\con},m)+ K^2$, and
\item $|z| \leq K^3 + max(\cntval{\con},m) \cdot K$.
\end{enumerate}
%\item $|w| \leq \K^2\cdot(max(n_{\con},m)+ \K^4)$.
%\end{enumerate} %\sav{counter bound, say what is $ \clsv$ in all the cuts.}
\end{corollary}
\begin{proof}
%[proof of \Cref{smpreachability} Point \ref{covssmp}]
Let $z\in \Sigma^*$ be a minimal reachability witness for
$(\con,\clsv, S, \{m\})$, where $\con$ is a configuration with counter value $n$.
%Let $\con\xrightarrow{w}\conD$ be the run of the minimal word $w$ from $\con$ .
\begin{point}
Consider the run of word $z$ from $\con$. Let $\conD\in\clsv\times S\times \{m\}$ such that $\con \xrightarrow{z} \conD$.
Assume for contradiction that the maximum counter value encountered during the run $\con \xrightarrow{z} \conD$ is greater than $max(n,m)+ (\mq\cdot\mc)^2$. % then the run on word $w$ from $\con$ can be written as follows:
Let $\conE_1,\conE_2,\cdots, \conE_t$ be all the configurations in this run such that their counter values are zero. There exists words $u_1,u_2,\cdots,u_{t+1}\in\Sigma^*$ such that $z=u_1u_2\cdots u_{t+1}$ and
\[
\con \xrightarrow{u_1}\conE_1\xrightarrow{u_2} \conE_2 \xrightarrow{u_3}\cdots\xrightarrow{u_t} \conE_t \xrightarrow{u_{t+1}}\conD
\]
Note that $\con\xrightarrow{u_1}\conE_1$, $\conE_t \xrightarrow{u_{t+1}}\conD$ and $\conE_i\xrightarrow{u_{i+1}}\conE_{i+1}$ for all $i\in[1,t-1]$ are floating runs (refer \Cref{fig:nonfloatingrun}). % with $n_{\conD}>max_\ce(\pi(u,\con_1))$ or $n_{\conH}>max_\ce(\pi(u,\con_1))$.

We show that the counter values are bounded during these floating runs. First, we consider the floating run $\con\xrightarrow{u_1}\conE_1$.
Given a configuration $\con$, we use $\vc x_\con$ to denote $\wgtvec{\con}$.
Let $\mata\in\Sring^{\mq\times\mq}$ be such that $\vc x_{\conD}={\vc x}_{\conE_1} \mata$. From \Cref{lem:comp} we know that the set $\lsu=\{ \vc y \in \Sring^{\mq} \mid {\vc y} \mata \in \lsv \}$
is a vector space and hence the vector ${\vc x}_{\conE_1}\in\clsu$. From \Cref{newvs}, we know that $u_1$ is a minimal reachability witness for $(\con, \clsu, \{p_{\conE_1}\} , \{0\})$ and therefore by
\Cref{lengthboundfloating} we know that the maximum counter value encountered during the run $\pi(u_1,\con)$ is less than $n+(\mq\cdot\mc)^2$.

Similarly for the floating run $\conE_t \xrightarrow{u_{t+1}} \conD$, the maximum counter value is bounded by $m+(\mq\cdot\mc)^2$.
Now consider the floating runs $\conE_i \xrightarrow{u_{i+1}} \conE_{i+1}$ for all $i\in[1,t-1]$. Again by applying \Cref{lengthboundfloating} we get that the maximum counter value encountered during these sub-runs is less than $(\mq\cdot\mc)^2$. Therefore, the maximum counter value encountered during the run
$\con \xrightarrow{z} \clsv\times S\times \{m\}$ is less than $max(n,m)+ (\mq\cdot\mc)^2$.
 \end{point}
 \begin{point}
 From the previous point, we know that the maximum counter value encountered during the run $\con\xrightarrow{z} \clsv\times S \times \{m\}$ is less than $max(n,m)+ (\mq\cdot\mc)^2$. Therefore, there are at most $max(n,m)+ (\mq\cdot\mc)^2$ many distinct counter values encountered during this run. Now from \Cref{lengthboundcounter} we get that $|z|\leq (\mq\cdot\mc)\cdot(max(n,m)+ (\mq\cdot\mc)^2)$.
\end{point}
\end{proof}
The following lemma helps us show that the length of a minimal witness for co-VS coverability is polynomially bounded in the number of states.
\begin{figure}[htbp]
\centering
\begin{minipage}{.55\textwidth}
\centering
\resizebox{.8\columnwidth}{!}{%
\begin{tikzpicture}
\tikzset{every path/.style={line width=.3mm}}
\draw [line width=0.35mm] [-stealth](.5,.5) -- (7,.5)node[anchor=south, xshift=-4cm, yshift=-.8cm] {word length};
\draw [line width=0.35mm] [-stealth](.5,.5) -- (.5,6.5) node[anchor=south,rotate=90, yshift=.2cm, xshift=-3.5cm] {counter value};
\draw plot [smooth, tension=1] coordinates {(1,1) (1.3,1.7) (2,2) (2.5,2.5)};
\draw [dashed] plot[smooth, tension=1] coordinates { (2.5,2.5) (3,3.7) (4,3.2) (4.3,4.3) };
\draw plot [smooth, tension=.5] coordinates {(4.3,4.3) (4.4,4.6) (4.7,4.7) (5.2,4.9) (5.7,5.7)};
\filldraw (1,1) circle[radius=1.5pt] node[anchor=east,xshift=.7cm, yshift=-.1cm] {$\con_1$};
\filldraw (1.3,1.7) circle[radius=1.5pt] node[anchor=east,xshift=.7cm, yshift=-.1cm] {$\con_{i_0}$};
\filldraw (2,2) circle[radius=1.5pt] node[anchor=west,xshift=0cm, yshift=-.1cm]{$\con_{i_1}$};
\filldraw (2.5,2.5) circle[radius=1.5pt] node[anchor=west,xshift=0.2cm]{$\con_{i_{l}}$};
\filldraw (4.3,4.3) circle[radius=1.5pt] node[anchor=north,xshift=0.2cm]{$\con_{i_k}$};
\filldraw (4.7,4.7) circle[radius=1.5pt] node[anchor=north,xshift=0.2cm]{$\con_{i_\mq}$};
\filldraw (5.7,5.7) circle[radius=1.5pt] node[anchor=west,xshift=0.2cm]{$\con_\ell$};
\draw [|<->|](4.1,4.5) -- (5.5,5.9) node[anchor=east, xshift=-.7cm, yshift=-.4cm]{$\mata_k$};
\end{tikzpicture}
}
%}
\caption{The figure shows a run from configuration $\con_1$ to $\con_\ell= (\vc x_{\con_\ell},p_{\con_\ell},n_{\con_\ell})$ such that $\vc x_{\con_\ell}\in \clsv$. The configurations $\con_{i_{l}}$ and $\con_{i_k}$ are where the counter values $n_{\con_{i_{l}}}$ and $n_{\con_{i_k}}$ are encountered for the last time. Also the configurations ${\con_{i_l}}$ and ${\con_{i_{k}}}$ have the same counter state. The dashed lines denote a part of that run that can be removed to get a shorter witness for $(\con,\clsv,\{p_{\con_\ell} \},\N)$.}
%\captionof{figure}{cut on a floating run\prince{elaborate here}}
\label{cut}
\end{minipage}%
\begin{minipage}{.45\textwidth}
\centering
\resizebox{.95\columnwidth}{!}{%
\begin{tikzpicture}
\tikzset{every path/.style={line width=.3mm}}
\draw plot [smooth, tension=.6] coordinates {(.1,.5) (.2,1)};
\draw [dashed] plot [smooth, tension=.6] coordinates{(.2,1) (.4,2) (.7,1.2) (1,2.5)};
\draw plot [smooth, tension=.6] coordinates{(1,2.5) (1.3,5) (1.7,3) (2.1,3.5) (2.3,2.5)};
\draw [dashed] plot [smooth, tension=.6] coordinates{(2.3,2.5) (2.6,1)};
%\draw (3.5,3.1) node {$v_j$};
\draw plot [smooth, tension=.6] coordinates{ (2.6,1) (2.8,0)(2.9,.5)};
\draw [dashed] plot [smooth, tension=.6] coordinates{(2.9,.5) (3,1)(3.1,.7)(3.2,1.3)};
\draw plot [smooth, tension=.6] coordinates{ (3.2,1.3) (3.3,2) (3.4,1.3) };
\draw [dashed] plot [smooth, tension=.6] coordinates{(3.4,1.3)(3.45,.5)};
\draw plot [smooth, tension=.6] coordinates{(3.45,.5)(3.5,0)(3.6,1)(3.63,.6)(3.69,.8)(3.9,0)(4,.4)(4.2,1)(4.3,.5)(4.4,.3)(4.5,.6)(4.6,.3)(4.65,.4)};
%\draw (4.3,.5) node {$\cdots$};
\draw plot [smooth, tension=.5] coordinates{(4.65,.4)(4.7,0)(4.9,2)(5,1)(5.1,2.5)(5.2,3.5)(5.4,3)(5.5,4)(5.7,4.5)};
\draw [dashed]plot [smooth, tension=.5] coordinates{(5.7,4.5)(5.8,5.5)};
\draw plot [smooth, tension=.5] coordinates{(5.8,5.5)(5.9,6.5)(6,6.7)(6.1,5.5)};
\draw [dashed]plot [smooth, tension=.5] coordinates{(6.1,5.5)(6.2,5)(6.3,4.7)(6.4,5)(6.6,5.4)(6.67,4.5)};
\draw plot [smooth, tension=.5] coordinates{(6.67,4.5)(6.8,3.5)(6.9,3)};

%\draw plot [smooth, tension=.6] coordinates{
%\filldraw (6.9,2) circle[radius=1.5pt] node[anchor=south] {$\conF$} node[xshift=.15cm, yshift=-1.2cm]{$z_3$};
\filldraw (2.8,0) circle[radius=1.5pt] node[anchor=north,yshift=0cm] {$\conE_1$};
\filldraw (3.5,0) circle[radius=1.5pt] node[anchor=north,yshift=0cm] {$\conE_2$};
\filldraw (3.9,0) circle[radius=1.5pt] node[anchor=north,yshift=0cm] {$\conE_3$};
\filldraw (.1,.5) circle[radius=1.5pt] node[anchor=west,yshift=0cm] {$\con$};
\filldraw (4.7,0) circle[radius=1.5pt] node[anchor=south,yshift=-0.6cm] {$\conE_4$};
\filldraw (6.9,3) circle[radius=1.5pt] node[anchor=north] {$\conD$};% node[xshift=-.45cm, yshift=-.6cm]{$u_i$};%rotate=70,
% \filldraw (5.7,1) circle[radius=1.5pt] node[anchor=west] {$\con^\prime_{i}$} node[xshift=.45cm, yshift=-.6cm]{$w_i$};%yshift=-.3cm
% \filldraw (2,2.5) circle[radius=1.5pt] node[anchor=east] {$\con_j$};
% \filldraw (5.2,2.5) circle[radius=1.5pt] node[anchor=west] {$\con^\prime_{j}$};
\draw [line width=0.4mm] [-stealth](0,0) -- (7.7,0)node[anchor=south, xshift=-4.5cm, yshift=-1cm] {word length};
%\draw [gray] [|<->|](-.2,0) -- (-.2,5)node[anchor=south,rotate=90, xshift=-2.3cm] {$>\K^4$};
% \draw [dotted] (-1,5)node[anchor=east]{} -- (7,5) ;%dashed
% \draw[dotted](-1,0) node[anchor=east]{}-- (7,0);
% \draw[dotted](-1,1)node[anchor=east]{} -- (7,1);
% \draw[dotted](-1,2.5)node[anchor=east] {}-- (7,2.5);
\draw [line width=0.4mm] [-stealth](0,0)node[anchor=south,rotate=90, xshift=3.5cm, yshift=.3cm] {counter value} -- (0,7);
\end{tikzpicture}
}
\caption{The figure shows a run from configuration $\con$ to $\conD= (\vc x_\conD, p_\conD, n_\conD)$ such that $\vc x_\conD\in \clsv$. Configurations $\conE_1, \conE_2, \conE_3, \conE_4$ are where the counter value zero is encountered during the run. The dashed lines denote the parts that can be removed to obtain a shorter witness for $(\con, \clsv, \{p_\conD\}, \{n_\conD\})$.}
\label{fig:nonfloatingrun}
\end{minipage}
\end{figure}
\begin{lemma}[cut lemma]%[cut]
\label{lem:cut}  
Let $z\in\Sigma^*$ be a witness for $(\con, \clsv,S,\N)$, where $\con$ is a configuration with $\cntval{\con}=n$ for some $n\in\N$, and $\con\xrightarrow{z} \clsv \times S\times \{m\}$ is a floating run for some $m\in\N$. If $m- n> K$, then 
there exists $z_{sub} \in \Sigma^*$ such that $z_{sub}$ is a subword of $z$, $\con \xrightarrow{z_{sub}} \clsv \times S \times \{m^\prime\}$ is a floating run and $m^\prime-n<m-n$.
\end{lemma}
%\sav{Alternate way to write the lemma. Which do we want?}
%\begin{lemma}
%Let $c, d$ be configurations of A such that $\vc x_{d} \in \clsv$, and there exists a floating run $\con \xrightarrow{*} d$. Then there exists a word $u$, and a configuration $e$ such that $\con \xrightarrow{u} e$ is a floating run with $\vc x_{e} \in \clsv$, $p_{d}= p_{e}$, and $|n_{e}- n_{\con}| \leq \K^2$.
%\end{lemma}

\begin{proof}
Let $z\in\Sigma^*$ be a witness for $(\con, \clsv,S,\N)$ and $\con\xrightarrow{z} \clsv \times S\times \{m\}$ is a floating run.
Let $n$ be the counter value of configuration $\con$ and $m> n+ \mq\cdot\mc$. 
Let $\con_1=\con$ and $\pi(z, \con_1) = \con_1 \tau_1 \con_2 \cdots \tau_{\ell-1} \con_\ell$ be such that configuration ${\con_\ell}$ has counter value $m$. Consider the sequence of transitions $T=\tau_0\tau_1\cdots\tau_{\ell-1}$ in $\pi(z, \con_1)$.

Since there are only $\mc$ counter states, by the Pigeon-hole principle, there exists a strictly increasing sequence $I=0< i_0< i_1 < \cdots < i_\mq\leq \ell$ such that  for all $j,j^\prime \in I$ (Condition 1) $\cntstate{\con_{j}}=\cntstate{\con_{j^\prime}}$ and (Condition 2) if $j<j^\prime$ then $\cntval{\con_{j}}<  \cntval{\con_{j^\prime}}$ and for all $d\in[j+1,j^\prime-1]$, $\cntval{\con_j}<\cntval{\con_d}<\cntval{\con_{j^\prime}}$. % lies between the counter values of ${\con_j}$ and ${\con_{j^\prime}}$.
Given a configuration $\con$, let $\vc x_\con$ denote $\wgtvec{\con}$. 
Consider the set of configurations $\con_{i_0}, \con_{i_1}, \ldots , \con_{i_\mq}$ (see \Cref{cut}). For any $j \in [0,\mq]$, let $\mata_{j}$ denote the matrix such that ${\vc x}_{\con_{i_j}} \mata_j = \vc x_{\con_{\ell}}$.
%From \Cref{lem:vec} we get that there exists $d\leq \K$, such that $\vc x_{\con_{i_d}}$ is dependent on $\vc x_{\con_{i_0}}, \vc x_{\con_{i_1}}, \ldots , \vc x_{\con_{i_{d-1}}}$.
Since ${\vc x}_{\con_{i_d}} \mata_d \in \clsv$ for all $d\in [0,\mq]$, from \Cref{lem:vec} we get that there exists {$l,k\in[0,\mq]$ with $l<k$ such that ${\vc x}_{\con_{i_{l}}} \mata_k \in \clsv$.}
{Consider a configuration $\conE=\configodc{x}{p}{n}$. If $\pi(u,\conE)$ is a floating run with $min_\ce(\pi(u,\conE))>0$, then for all $m\in\N$ and $\vc y \in \Sring^\mq$, $\pi(u,\configodc{y}{p}{m})$ is a run.}
Consider the sequence of transitions $T^\prime=\tau_{i_k \cdots {\ell-1}}$ and let $u=\word(T^\prime)$. Because of Condition 2, $min_\ce(\pi(u,\con_{i_k}))>0$. Therefore the run $T^{\prime\prime}(\con_1)$ where $T^{\prime\prime}=\tau_{1\cdots {i_{l}-1}}\tau_{i_k \cdots {\ell-1}}$ is a run shorter than $\pi(z,\con_1)$ with smaller counter effect. 
%\begin{point}
%Since ${\vc x}_{\con_{i_d}} \mata_d \in \clsv$ for all $d\in [0,\mq]$, from \Cref{lem:vec} we get that there exists \final{$l,k\in[0,\mq]$ with $l>k$ such that ${\vc x}_{\con_{i_{l}}} \mata_k \in \clsv$.}
%Consider the run $T^{\prime\prime}(\con_1)$ where $T^{\prime\prime}=\tau_{1\cdots {i_{k}-1}}\tau_{i_l \cdots {\ell-1}}$ is a valid run longer than $\pi(z,\con_1)$ with bigger counter effect. 
%\end{point}
%Let $w= \word(T^{\prime\prime})$. The run $\con_1\xrightarrow{w} \clsv\times S \times \{m^\prime\}$ for some $m^\prime\in\N$. 
%This contradicts the minimality of $z$.}
%we now apply \Cref{lem:vec} to get the desired result. \prince{Can be elaborated if required.}
\end{proof}

Note that if the run of a minimal coverability witness $z$ for $(\con, \clsv,S,\N)$ is a floating run, then the number of distinct counter values encountered during the run $\con\xrightarrow{z}\clsv\times S \times \N$ is polynomially bounded in the number of states of the machine.% and $n_\con$.
Now we show that for any run (need not be floating)
of a minimal coverability witness $z$ for $(\con,\clsv,S,\N)$, the maximum counter value encountered during the run $\con\xrightarrow{z} \clsv \times S \times \N$ is bounded by a polynomial in the number of states of the machine and the initial counter value. %The proof is given in the Appendix.

\begin{corollary}
\label{counterboundagnostic}
%The counter values (and therefore length) of a run to a complement to a vector space can be bounded.
If $z\in \Sigma^*$ is a minimal witness for $(\con, \clsv,S,\N)$, where $\con$ is a configuration with counter value $n$, then
%\begin{enumerate}
the maximum counter value encountered during the run $\con\xrightarrow{z}\clsv\times S\times\N$ is less than $max(n,\mq\cdot\mc)+ (\mq\cdot\mc)^2$.
%\item $|w| \leq \K^2\cdot(n_\con+ \K^4)$.
%\end{enumerate} %\sav{counter bound, say what is $ \clsv$ in all the cuts.}
\end{corollary}
\begin{proof}
%[proof of \Cref{smpreachability} Point \ref{agnosticsmp}]
Let $z\in \Sigma^*$ be a minimal witness for
$(\con,\clsv, S, \N)$, where $\con$ is a configuration with counter value $n$.
%Let $\con\xrightarrow{w}\conD$ be the run of the minimal word $w$ from $\con$ .
%\begin{point}
Consider the run of word $z$ from $\con$. Let $\conD\in\clsv\times S\times \N$ such that $\con \xrightarrow{z} \conD$. If $\con \xrightarrow{z} \conD$ is a floating run, then by \Cref{lem:cut} the maximum counter value encountered during this run will be less than $n+ \mq \cdot \mc$. Now if $\con \xrightarrow{z} \conD$ is not a floating run,
then there exists $u_1,u_2\in\Sigma^*$ such that $z=u_1u_2$ and $\con \xrightarrow{u_1} \conE \xrightarrow{u_{2}}\conD$ 
where, counter value of configuration ${\conE}$ is zero and $\conE \xrightarrow{u_{2}}\conD$ is a floating run. % with $n_{\conD}>max_\ce(\pi(u,\con_1))$ or $n_{\conH}>max_\ce(\pi(u,\con_1))$.

Given a configuration $\con$, let $\vc x_\con$ denote $\wgtvec{\con}$.
Let $\mata\in\Sring^{\mq\times\mq}$ be such that $\vc x_{\conD}={\vc x}_{\conE} \mata$. From \Cref{lem:comp}, we know that the the set $\lsu=\{ \vc y \in \Sring^{\mq} \mid {\vc y} \mata \in \lsv \}$
is a vector space and hence the vector ${\vc x}_{\conE}\in\clsu$. Note that for all $\vc y\in\clsu$, the vector $\vc y \mata \in \clsv$. From \Cref{newvs}, we know that $u_1$ is a minimal reachability witness for $(\con, \clsu, \{p_{\conE}\} ,\{0\})$, where $p_{\conE}$ is the counter state of configuration $\conE$, and therefore by \Cref{lengthbound}, we know that the maximum counter value encountered during the run $\pi(u_1,\con)$ is less than $n+(\mq\cdot\mc)^2$.
Now since $\conE \xrightarrow{u_{2}} \conD$ is a floating run and
$u_{2}$ is the minimal such word, from \Cref{lem:cut}, we get that the counter value of configuration $\conD$ is less than or equal to $\mq\cdot\mc$, and by \Cref{lengthboundfloating}, we know that the maximum counter value encountered during this run is less than $\mq\cdot\mc+(\mq\cdot\mc)^2$.
Therefore, we get that the maximum counter value encountered during the run $\con\xrightarrow{z}\conD$ is less than $max(n,\mq\cdot\mc)+ (\mq\cdot\mc)^2$.
%\end{inlinepoint}
% \begin{point}
%Let $T(\con_1)= \con_1 \tau_1 \con_2 \cdots \tau_{h-1} \con_h$ be the run on word $w$ from $\con_1$ and $T$ the corresponding sequence of transitions. From \Cref{maxcounterbound}, we know that the maximum counter value encountered during this run is less than $n_{\con_1}+\K^4$. Now from \Cref{lengthboundcounter} we get that $|z|\leq\K^2\cdot(n_{\con_1}+\K^4)$.
%\end{point}
%\prm{We have to say what $ \clsv$ is in each case. lemma 5 only dealt with floating runs. the argument is correct though but there are some reasoning to do.}
\end{proof}

Our next objective is to show that the counter values are polynomially bounded during the run of a minimal coverability witness. The problem is similar to co-VS reachability, except that now we are not given a final counter value. A crucial ingredient in proving this is \Cref{lem:cut} which will help us in proving that if the run of a minimal coverability witness $z$ for $(\con, \clsv,S,\N)$ is a floating run, then the number of distinct counter values encountered during the run $\con\xrightarrow{z}\clsv\times S \times \N$ is polynomially bounded in the number of states of the machine and the initial counter value. Using this and the ideas presented earlier for co-VS reachability, we can prove the existence of a polynomial length witness for the co-VS coverability problem.

\begin{corollary}
Let $\con$ be a configuration with counter value $n$. If $z$ is a minimal witness for $(\con, \clsv,S,\N)$ then %$|z|\leq \K\cdot(max(n_\con,\K)+ \K^2)$. %is bounded by a polynomial in the number of states of the machine and $n$. % 
$|z| \leq (\mq\cdot\mc)\cdot(max(n,(\mq\cdot\mc))+ (\mq\cdot\mc)^2)$.
\end{corollary}
\begin{proof}
Let $z\in \Sigma^*$ be a minimal reachability witness for $(\con, \clsv, S, \N)$.
From \Cref{counterboundagnostic}, we know that the maximum counter value encountered during the run $\con\xrightarrow{z} \clsv\times S \times \N$ is less than $max(n,(\mq\cdot\mc))+ (\mq\cdot\mc)^2$. Therefore, there are at most $max(n,(\mq\cdot\mc))+ (\mq\cdot\mc)^2$ many distinct counter values encountered during this run. Now from \Cref{lengthboundcounter} we get that $|z|\leq (\mq\cdot\mc)\cdot(max(n,(\mq\cdot\mc))+ (\mq\cdot\mc)^2)$.
\end{proof}

Now, we prove that the co-VS reachability and co-VS coverability problems of weighted \odca are in $\CF{P}$ by demonstrating a small model property. 
We have already established using \Cref{lem:uturn}, \Cref{smp}, and \Cref{lem:cut} that the maximum and minimum counter values encountered during the run of the minimal witness do not exceed some polynomial bound. 
This, in turn, implies a polynomial bound on the length of the witness by \Cref{lengthboundcounter}. As a result, we get the following theorem.
%\end{toappendix}
%\begin{theoremrep}
%
%co-VS coverability problem is decidable in $\CF{P}$.
%\end{theoremrep}
\begin{theorem}%[reachability in $\CF{PTIME}$] 
\label{reachabilityandcoverability}
\label{ptimereachability}
The co-VS reachability and co-VS coverability problems for weighted \odca can be decided in polynomial time when the counter values are given in unary notation. % decidable in $\CF{P}$.%polynomial time.
\end{theorem}
\begin{proof}
%[proof of \Cref{ptimereachability}]
% \prm{reduction to wa reachability}

Assume we are given a weighted \odca $\Autom=((C, \delta_0,\delta_1, p_0),(Q, \vc \lambda, \Delta, \vc\eta))$, initial configuration $\con= (\vc x,p,n)$, vector space $\lsv$, set of counter states $S$ and counter value $m$ as inputs for the co-VS reachability problem. For solving this reachability problem,
we first consider the $max(n,m)+ (\mq\cdot\mc)^2$-unfolding weighted automata $\Autom^{max(n,m)+ (\mq\cdot\mc)^2}= (C^\prime, \delta^\prime, p^\prime_0;\ Q^\prime, \vc \lambda^\prime, \Delta^\prime, \vc\eta^\prime_F)$ of $\Autom$ as described in \Cref{unfolding}. From \Cref{lengthbound}, we know that the maximum counter value encountered during the run of the minimal reachability witness $z$ for $(\con, \clsv, S , \{m\})$ is less than $max(n,m)+ (\mq\cdot\mc)^2$. We define a vector space $\lsu\subseteq \Sring^{|Q^\prime|}$ as follows:
A vector $\vc z \in\Sring^{|Q^\prime|}$ is in $\lsu$ if there exists $\vc y \in \lsv$ such that for all $i\in [0,\mq-1]$, $\vc z[\mq\cdot m+i]= \vc y[i]$ and for all $m^\prime \neq m$ and $i\in [0,\mq-1]$, $\vc z[\mq\cdot m^\prime+i]=0$.

Given a configuration $\con=(\vc x,p,n)$ of a weighted $\odca$, we define the vector $\vc z_\con\in\Sring^{|Q^\prime|}$.
\[
\vc z_\con[i]= \begin{cases} \vc x[i \bmod \mq], \text{ if }\frac{i}{\mq}=n\\ 0, \text{ otherwise} \end{cases}
\]
Now, consider the configuration $\uconN = (\vc z_\con, (p,n))$ of $\Autom^{max(n,m)+ (\mq\cdot\mc)^2}$
and check whether $\uconN \xrightarrow{*} \clsu \times S \times \{0\}$. This is a co-VS reachability problem of weighted automata. Using \Cref{wacovs}, this can be solved in polynomial time.

For solving co-VS coverability problem when a weighted \odca$\Autom$ with an initial configuration $\con=(\vc z, p,n)$, a vector space $\lsv$ and a set of counter states $S$ are given as inputs, we consider the $max(n,(\mq\cdot\mc))+ (\mq\cdot\mc)^2$-unfolding weighted automata $\Autom^{max(n,(\mq\cdot\mc))+ (\mq\cdot\mc)^2}= (C^\prime, \delta^\prime, p^\prime_0;\ Q^\prime, \vc \lambda^\prime, \Delta^\prime, \vc\eta^\prime_F)$ of $\Autom$. From \Cref{counterboundagnostic}, we know that the maximum counter value encountered during the run of a minimal reachability witness $z$ for $(\con, \clsv, S , \N)$ is less than $max(n,(\mq\cdot\mc))+ (\mq\cdot\mc)^2$. We define a vector space $\lsu\subseteq \Sring^{|Q^\prime|}$ as follows:
A vector $\vc x \in\Sring^{|Q^\prime|}$ is in $\lsu$ if there exists $\vc y \in \lsv$ and $m\in\N$ such that for all $i\in [0,\mq-1]$, $\vc x[\mq\cdot m+i]= \vc y[i]$ and for all $m^\prime\neq m$ and $i\in [0,\mq-1]$, $\vc x[\mq\cdot m^\prime+i]=0$.
Given a configuration $\con=(\vc x,p,n)$ of a weighted $\odca$, we define the vector $\vc z_\con\in\Sring^{|Q^\prime|}$.
\[
\vc z_\con[i]= \begin{cases} \vc x[i \bmod \mq], \text{ if }\frac{i}{\mq}=n\\ 0, \text{ otherwise} \end{cases}
\]
Now, consider the configuration $\uconN = (\vc z_\con, (p,n))$ of $\Autom^{max(n,(\mq\cdot\mc))+ (\mq\cdot\mc)^2}$
and check whether $\uconN \xrightarrow{*} \clsu \times S\times\{0\}$. This is a co-VS reachability problem of a weighted automaton. From \Cref{wacovs}, we know that this can be solved in polynomial time.
\end{proof}

%{From \Cref{counterboundagnostic} we get that the maximum counter value encountered during the run of a minimal witness is bounded by a polynomial value $M$. Therefore, we can reduce the co-VS coverability problem to co-VS reachability problem of the M-unfolding weighted automata. This can be solved in polynomial time by \Cref{wacovs}.}

%\begin{proof}
% From \Cref{lengthbound}, we get that the length of the minimal word $w$ such that $\con\xrightarrow{w} \clsv \times \{q\} \times \{n\}$ is less than $\K^2\cdot(max(n_{\con},m)+ \K^4)$.
% \prm{Unfold $\Autom$ upto the max. counter value. Checking for reachability can now be done in poly time.}
% \end{proof}
%Now, we prove that if the number of distinct counter values encountered during the run of a minimal witness is polynomially bounded, then we can bound the length of that witness. %The proof is given in the Appendix. %We prove this in \Cref{lengthboundcounter}.
%\prm{new lemma- length bounding}

%\prm{Remaining- to be moved elsewhere.}

%\prm{Explain why this section is needed. Mention that the minimal witness has a canonical form.}
%The upcoming subsection will show that there is a canonical minimal witness. %However, unlike deterministic \textsc{oca}, we cannot get this result, if we start from some arbitrary minimal word. 

\subsection{Lexicographically minimal witness}\label{sec:lex_min_witness}

This section will show that the lexicographically minimal witness has a distinct structure. 
We assume a total order on the symbols in $\Sigma$.
Given two words $u,v \in \Sigma^*$, we say that $u$ precedes $v$ in the \emph{lexicographical ordering} if $|u| < |v|$ or if $|u|=|v|$ and there exists an $i\in[0,|u|-1]$ such that $u[0,i-1]=v[0,i-1]$ and $u[i]$ precedes $v[i]$ in the total ordering assumed on $\Sigma$.
A word $z\in \Sigma^*$ is called the lexicographically minimal witness for $(\con, \clsv,S,\{m\})$, if $\con\xrightarrow{z} \clsv \times S \times X$ and for all $u\in\Sigma^*$ with $\con\xrightarrow{u} \clsv \times S \times X$, $z$ precedes $u$ in the lexicographical ordering. We show that the lexicographically minimal witness $z$ for $(\con, \clsv,S,\{m\})$ has a canonical form. First, we prove this for floating runs. 
\begin{figure}%{.42\textwidth}
 %\centering
 \centering
\resizebox{.4\columnwidth}{!}{%
\begin{tikzpicture}
\tikzset{every path/.style={line width=.2mm}}
%\draw plot [smooth, tension=.6] coordinates { (0,0) (.3,1) (.7,2) (1.35,1) (1.64,1.5) };
%\draw [dashed] plot [smooth, tension=.6] coordinates{(1.3,3.5) (1.5,2.5)(2.25,3.3)};
\draw plot [smooth, tension=.6] coordinates{(1.6,5.5) (1.65,5)};
\draw [dashed] plot [smooth, tension=.6] coordinates{(1.65,5) (1.8,4)(2.1,2.5) (2.7,5.2) (3,3.3) (3.5,1) (4,3) (4.2,2.8)(4.25,2.3)};
\draw (3.5,3.5) node [xshift=.4cm]{$z[l_i,r_i-1]$};
\draw plot [smooth, tension=.6] coordinates{(4.25,2.3)(4.35,1.5) }; %(6.3,-.4)(6.7,1) (6.9,2)
%\filldraw (6.9,2) circle[radius=1.5pt] node[anchor=south] {$\conF$} node[xshift=.15cm, yshift=-1.1cm]{$z_3$};
%\filldraw (3,5) circle[radius=1.5pt] node[anchor=north,yshift=0.6cm] {$\conD$};
\filldraw (1.6,5.5) circle[radius=1.5pt] node[anchor=south,yshift=0cm] {$\con$};
\filldraw (4.35,1.5) circle[radius=1.5pt] node[anchor=south,yshift=-0.6cm] {$\conG$};
\filldraw (1.65,5) circle[radius=1.5pt] node[anchor=east] {$\conE_{i}$};%rotate=70,
\filldraw (4.25,2.3) circle[radius=1.5pt] node[anchor=west] {$\conF_{i}$};%yshift=-.3cm
% \filldraw (2.25,3.3) circle[radius=1.5pt] node[anchor=east] {$\con_j$};
%\filldraw (3.8,3.3) circle[radius=1.5pt] node[anchor=west] {$\con^\prime_{j}$};
\draw [line width=0.25mm] [-stealth](1,.8) -- (5,.8);%node[anchor=south, xshift=-1.5cm, yshift=-.7cm] {\scriptsize word length};
%\draw [gray] [|<->|](-.2,0) -- (-.2,5)node[anchor=south,rotate=90, xshift=-2.3cm] {$>\K^4$};
\draw [dotted] (1,5.5)node[anchor=east]{$n$} -- (5,5.5) ;%dashed
\draw[dotted](1,1.5) node[anchor=east]{$m$}-- (5,1.5);
\draw[dotted](1,2.3)node[anchor=east]{$n-i+d$} -- (5,2.3);
\draw[dotted](1,5)node[anchor=east]{$n-i$} -- (5,5);
\draw [line width=0.25mm] [-stealth](1,.8)node[anchor=south,rotate=90, xshift=2.8cm, yshift=.5cm] {\scriptsize counter value} -- (1,6);
\draw[gray!5] [-] (0,.3)-- (0,.3);
\end{tikzpicture}
}
\caption{The figure shows the floating run from a configuration $\con$ (where $\cntval{\con}=n$) to a configuration $\conG= (\vc x, p,m)$ such that $\vc x\in\clsv$. The configurations $\conE_{i}$ and $\conF_{i}$ are where the counter values $n-i$ and $n-i+d$ are encountered for the first (resp. last) time during this run. The dashed line is the part of the run due to factor $z[l_i,r_i-1]$ and has a counter effect $d$.}
\label{fig:window}
% \resizebox{1\columnwidth}{!}{%
%}
%\captionof{figure}{u-turn cut on a floating run \prince{elaborate here}}
%\label{fig:ucut}
\end{figure}
\begin{lemma} \label{specialword}
There exist $poly_1:\N \rightarrow \N$, and $poly_2:\N^2 \rightarrow \N$ such that, 
if $z\in \Sigma^*$ is the lexicographically minimal witness for $(\con, \clsv,S,\{m\})$ and
$\con\xrightarrow{z}\clsv\times S \times \{m\}$ is a floating run, then there exist $u,y,w \in \Sigma^*$ and $r\in\N$ such that $z=uy^rw$ and 
%\rem{for all configurations $\conD_k$, $k\in[0,r]$, where $\con\xrightarrow{uy^k}\conD_k$} 
the following are true:%conditions hold:
%Given a configuration $\con$ of an \odca $\Autom$, and $\lsv \subseteq \Sring^\K$,
%if there exists $\conD$ such that there is a floating run $\con_1 \xrightarrow + \con_\ell$ with $\vc x_{\con_\ell} \in \clsv$, then there exists a word $u=u_1u_2^ru_3$ (with $r\geq 0$) such that $\con_1\xrightarrow{u} {\con_{\ell^\prime}}$ is a floating run with $n_{\con_{\ell^\prime}}=n_{\con_\ell},\ p_{\con_{\ell^\prime}}=p_{\con_\ell}$ and the following conditions hold:
\begin{enumerate}
%\item for all $w\in \Sigma^{*}$, if $\con_1 \xrightarrow{w} \con_j$, with $\vc x_{\con_j} \in \clsv$ then $u \ll w$.
\item $|uyw| \leq poly_1(K)$, and % are polynomially bounded by the number of states of the machine, and%$|u|,|y|\leq 3\cdot\mq^3\cdot\mc^4$ and $|w| <6\cdot\mq^3\cdot\mc^4$,
%\item either $n_{\conD_i}> n_{\conD_j}$ for all $i,j$ such that $0\leq i<j\leq r$  or\\
%$n_{\conD_i}<n_{\conD_j}$ for all $i,j$ such that $0\leq i<j\leq r$, 
%\item for all $i,j\in[0,r]$, ${\conD_i}$ and ${\conD_j}$ have the same counter state, and%There exists $p\in C$ such that for all $i\in[0,r]$, $p_{\conD_i}=p$ and for all $i,j\in[0,r]$ with $i<j$, $n_{\conD_i}>n_{\conD_j}$ .
\item $r \leq poly_2(K,|\cntval{\con}-m|)$.% is bounded by a polynomial in the number of states of the machine and the difference between the counter value of $\con$ and $m$. %\in [0,(\mq\cdot\mc)\cdot|n_\con-m|+(\mq\cdot\mc)^3]$.
%\item either $y=\epsilon$ or it has a counter effect in $[-\K^4,\K^4]$.
%\item If $n_{\con_\ell}<\K$ then, $|u|= \alpha n_{\con_i} + \beta$ where \prm{$\alpha, \beta$ are polybounded. Fix the correct poly values.}
%\item \label{newpoint} $|u| \leq (max\{|\Autom|, n_c\}+|\Autom|^2)|\Autom|$ and the maximum counter value encountered during the run $\con \xrightarrow{u} \con_k$ is less than $max\{|\Autom|, n_c\}+|\Autom|^2$ %(we can make this a new lemma and prove this separately)
\end{enumerate}\end{lemma}

\begin{proof}
Let $z$ be the lexicographically minimal witness for $(\con, \clsv,S,\{m\})$, and $\conG\in\clsv\times S \times \{m\}$ such that $\con\xrightarrow{z}\conG$ is a floating run. Let $n$ be such that $\cntval{c} = n$. 
 %\prm{$u_2=\epsilon$ and $r=0$}
We consider the case $n > m$. The case where $m>n$ is analogous. 
Let $t=n- m$. 
\begin{clam}
$|z| \leq 2K^3 + t \cdot K$. \label{floatinglength} %(\mq\cdot\mc)\cdot(t+ 2\cdot\mq^2\cdot\mc^2)$.
\end{clam}
\begin{clamproof}
From Point \ref{pumpdown} of \Cref{lengthboundfloating}, it follows that the maximum counter value during the run $\con \xrightarrow{z} \conG$ is less than $n+K^2$. By a symmetric argument, it follows that the minimum counter value during the run is greater than $m-K^2$. Hence, there are at most $t+2K^2$ distinct counter values during the run. From \Cref{lengthboundcounter} it follows that $|z| \leq 2K^3 + t \cdot K$.
%
%
%
%The minimum counter value during the run $\con\xrightarrow{z}\clsv\times S \times \{m\}$ is greater than $min(n,m)-\mq^2\cdot\mc^2$. The proof of this is symmetric to that of Point \ref{pumpdown} of \Cref{lengthboundfloating} and can be proven analogously.
%Using this and \Cref{lengthboundfloating}, we get that the counter values encountered during the run $\con\xrightarrow{z}\clsv \times S \times \{m\}$ lies between $max(n,m)+(\mq\cdot\mc)^2$ and $min(n,m)-(\mq\cdot\mc)^2$.  There are at most $t+2\cdot(\mq\cdot\mc)^2$ distinct counter values during this run. Now, from \Cref{lengthboundcounter} we get that $|z|\leq(\mq\cdot\mc)\cdot(t+2\cdot(\mq\cdot\mc)^2)$.
\end{clamproof}

If $t\leq K^2$, then from \Cref{floatinglength}, we get that $|z| \leq3 K^3$, and the lemma is trivially true.
Let us assume $t> K^2$ and let $d = K^2-t$.
Let $\con_1=\con$ and $\pi(z,\con_1) = \con_1 \tau_1 \con_2 \cdots \tau_{\ell-1} \con_\ell$ denote the run on word $z$ from $\con$.
For any $i \in [0,K^2]$, we denote by $l_i$ the index such that the counter value $n-i$ is encountered for the first time,  and $r_i$ the index such that the counter value $n-i+d$ is encountered for the last time in $\pi(z,\con_1) $ (see \Cref{fig:window}). Let $X=\{(l_i,r_i)\}_{i\in[0,K^2]}$ be the set of these pairs of indices, and let $W=\{z[l,r-1]\mid (l,r)\in X\}$ be the set of corresponding factors. Note that $|X| > K^2$. We argue that these factors $z[l_i,r_i-1]$ for $ i\in[0,K^2]$ need not all be distinct. 
%
%
%Note that, since $t>K^2$, there will be more than $K^2$ many factors with counter counter effect $d$ in this run.
%Hence, there are at least $(\mq\cdot\mc)^2+1$ pairs of positions $(l_i,r_i), i\in[0,(\mq\cdot\mc)^2]$ such that for all $ i\in[0,(\mq\cdot\mc)^2]$ the factor $z[l_i,r_i-1]$ has counter effect $d$ with respect to counter state of configuration ${\con_{l_i}}$. Note that these factors $z[l_i,r_i-1]$ for $ i\in[0,(\mq\cdot\mc)^2]$ need not be all distinct. 
% such that for all $i,j\in[0,\K^4]$, if $i<j$ then $l_i<l_j$.
%\prm{define counter effect of a word (floating run). defined only for runs now.}
\begin{clam}
\label{factors}% \prm{claim should be that one word repeats for $t=n_{\con_{\ell^\prime}}- n_{\con_{1}}$ with $|t|>\K^4$ and $d=t-\K^4-1$}
$|W| \leq K^2$.
%There are only $\K^4$ distinct factors with counter-effect $d$ in $T$% and hence there is a factor $v$ in $z$ with counter-effect $d$ that repeats.
\end{clam}
\begin{clamproof}
Assume for contradiction that $|W| >(\mq\cdot\mc)^2$.
%Let $\conG\in\clsv\times S\times \{m\}$ be such that $\con \xrightarrow{z} \conG$. 
Since the number of counter states is $\mc$,
by Pigeon-hole principle {there exists $Y\subseteq X$ with $|Y|=\mq^2+1$ such that for all $(l,r),(l^\prime,r^\prime) \in Y$, configurations ${\con_{l}}$ and ${\con_{l^\prime}}$ have the same counter state, configurations ${\con_{r}}$ and ${\con_{r^\prime}}$ have the same counter state, and $z[l,r-1]\neq z[l^\prime,r^\prime-1]$. We say $(l,r)<(l^\prime,r^\prime)$ if $z[l,r-1]$ precedes $z[l^\prime,r^\prime-1]$ in the lexicographical order. Therefore, the elements in $Y$ have an ordering as follows: $(l_0,r_0)<(l_1,r_1)< \cdots<(l_{\mq^2},r_{\mq^2})$. }%there exists $\K^2+1$ pairs $(l_{i_0},r_{i_0}),(l_{i_1},r_{i_1}), \ldots, (l_{i_{\K^2}},r_{i_{\K^2}}) \in X$ such that for $j\in[0,\K^2]$, $z[l_{i_j},r_{i_j}]$ are all distinct words, and for any $m,n \in[0,\K^2]$, $p_{\con_{l_{i_m}}}= p_{\con_{l_{i_n}}}$ and $p_{\con_{r_{i_m}}}= p_{\con_{r_{i_n}}}$. For each $j\in[0,\K^2]$ let $x_j=z[l_{i_j},r_{i_j}]$ and are ordered in such a way that for all $j,k\in[0,\K^2]$ with $j<k$, $x_j$ precedes $x_k$ in the lexicographical ordering.
%Let $v_0,v_1,v_3, \ldots v_{\K^2}$ be the words corresponding to these factors such that
For any configuration $\conH$, let $\vc x_\conH = \wgtvec{\conH}$.
For all $i\in[0,\mq^2]$, {let $u_i=z[1,l_i-1], x_i=z[l_i,r_i-1],w_i= z[r_i,\ell-1]$}, configurations $\conE_{i},\conF_{i}$ be such that $\con\xrightarrow{u_i}\conE_{i}\xrightarrow{x_i}\conF_{i}\xrightarrow{w_i}\conG$ and matrices $\mata_i,\matm_i,\matb_i$ be such that
$\vc x_{\conE_{i}} = \vc x_{\con}\mata_i\ ,
\vc x_{\conF_{i}} = \vc x_{\conE_{i}}\matm_i\ ,
\vc x_{\conG} = \vc x_{\conF_{i}}\matb_i$.

We know that for all $k\in[0,\mq^2]$, $\vc{x}_{\con}\mata_k\matm_k\matb_k\in\clsv$.
Consider the sequence of matrices $\matm_0,\matm_1,\cdots,\matm_{\mq^2}$. Since there can be at most $\mq^2$ independent matrices, we get that there exists $i\in[0,\mq^2]$ such that $\matm_{i}$ is a linear combination of $\matm_{0}, \dots, \matm_{i-1}$.
Hence, we get that there exists a $j$ where $j<i$ such that $\vc{x}_{\con}\mata_i\matm_j\matb_i\in\clsv$. Since $x_j = z[l_j, r_j-1]$ precedes $x_i = z[l_i,r_i-1]$, the word $u_ix_jw_i$ precedes $z$ in the lexicographical ordering. Therefore the run $\pi(u_ix_jw_i,\con)$
%\con\xrightarrow{u_ix_jw_i}\clsv\times S \times \{m\}$ 
contradicts the lexicographical minimality of $z$. %This contradicts the minimality of $z$. %Hence there exists a factor of $z$ with counter-effect $d$ that repeats.
\end{clamproof}

Since $|W| \leq K^2$ and $|X|>K^2$, there exists $i, j\in[0,K^2]$, with $i< j$ and $x\in\Sigma^*$ such that $(l_i,r_i)\in X,(l_j,r_j)\in X$ and $x=z[l_i,r_i-1]= z[l_j,r_j-1]$ (see \Cref{looprepeat}).
Let $u_1,w_1,u_2,w_2\in\Sigma^*$ such that $z=u_1xw_1=u_2xw_2$. Since $u_1\neq u_2$, either $u_1$ is a prefix of $u_2$ or $u_2$ a prefix of $u_1$. Without loss of generality, let us assume $u_1$ is a prefix of $u_2$. Therefore, there exists $v\in\Sigma^*$ such that $u_2=u_1v$. Let $\conE$ be a configuration such that $\con \xrightarrow{u_1} \conE$.

\begin{clam}
\label{shortrep}
$|u_1|,|v|,|w_1| \leq 3K^3$.
\end{clam}
\begin{clamproof}
Consider the set $X$.
For any $i,j\in[0,K^2]$, the difference between the counter values of configurations ${\con_{l_i}}$ and ${\con_{l_j}}$ and the difference between the counter values of the configurations ${\con_{r_j}}$ and ${\con_{r_i}}$ is at most $K^2+1$.
%Since we have considered $\K^4+1$ factors with counter-effect $d$ starting from configurations where counter value $n_{\con_1}-i$, $i\in[0,\K^4]$ is encountered for the first time during the run of $z$ from $\con_1$,
Therefore the counter-effect of $u_2$, $w_2$, and $v$ can be at most $K^2$.
Since $\pi(v,\conE)$ is a floating run from \Cref{floatinglength}, we get that $|v|\leq 3K^3$.
By similar arguments, the counter-effect of $u_1$ and $ w_1$ can be at most $K^2$, and again by \Cref{floatinglength}, we get that their lengths are at most $3K^3$.
\end{clamproof}
\begin{figure}[htbp]
\centering
\resizebox{.85\columnwidth}{!}{%
\begin{tikzpicture}
\tikzset{every path/.style={line width=.25mm}}
\draw [|-|](0,1) -- (1,1)node[anchor=south, xshift=-.5cm, yshift=.3cm]{$u_1$};
\draw [|-|](1,1) -- (7.6,1)node[anchor=south, xshift=-3.2cm, yshift=.3cm]{$x$};
\draw [|-|](7.6,1) -- (9,1)node[anchor=south, xshift=-.7cm, yshift=.3cm]{$w_1$};
\draw [|-|](0,0) -- (2,0)node[anchor=north, xshift=-.8cm, yshift=-.3cm]{$u_2$};
\draw [|-|](2,0) -- (8.6,0)node[anchor=north, xshift=-3.2cm, yshift=-.3cm]{$x$};
\draw [|-|](8.6,0) -- (9,0)node[anchor=north, xshift=-.2cm, yshift=-.3cm]{$w_2$};

\draw[gray] [|-|](0,2.3) -- (6.6,2.3)node[anchor=south, xshift=-3.2cm, yshift=-.6cm]{$z[l_0,r_0-1]$};
\draw[gray] [|-|](6.6,2.3) -- (9,2.3)node[anchor=south, xshift=-3.2cm, yshift=.3cm]{};
\draw(-.2,2.2) node {$l_0$};
\draw (6.6,1.9) node {$r_0$};

\draw[gray] [|-|](0,-1.3) -- (2.4,-1.3)node[anchor=south, xshift=-3.2cm, yshift=.3cm]{};
\draw[gray] [|-|](2.4,-1.3) -- (9,-1.3)node[anchor=north, xshift=-3.2cm, yshift=.6cm]{$z[l_{K^2}, r_{K^2}-1]$};
\draw(2.7,-.9) node {$l_{K^2}$};
\draw (9.4,-1.2) node {$r_{K^2}$};

\draw [dotted](1,0) -- (1,1);
\draw [dotted](2,0) -- (2,1);\
\draw [dotted](3,0) -- (3,1);
\draw [dotted](4,0) -- (4,1);
\draw [dotted](5,0) -- (5,1);
\draw [dotted](6,0) -- (6,1);
\draw [dotted](7,0) -- (7,1);
\draw [dashed](2,0) -- (1,1);\
\draw [dashed](3,0) -- (2,1);
\draw [dashed](4,0) -- (3,1);
\draw [dashed](5,0) -- (4,1);
\draw [dashed](6,0) -- (5,1);
\draw [dashed](7,0) -- (6,1);
\draw [dashed](8,0) -- (7,1);

\draw[gray] (1.5,1.2) node {$v$};
\draw[gray] (2.5,1.2) node {$v$};
\draw[gray] (3.5,1.2) node {$v$};
\draw[gray] (4.5,1.2) node {$v$};
\draw[gray] (5.5,1.2) node {$v$};
\draw[gray] (6.5,1.2) node {$v$};
\draw(1,1.35) node {$l_i$};
\draw (7.6,.7) node {$r_i$};
\draw[gray] (7.3,1.2) node {$v^\prime$};
\draw(2,-.35) node {$l_j$};
\draw[gray] (2.5,-.2) node {$v$};
\draw[gray] (3.5,-.2) node {$v$};
\draw[gray] (4.5,-.2) node {$v$};
\draw[gray] (5.5,-.2) node {$v$};
\draw[gray] (6.5,-.2) node {$v$};
\draw[gray] (7.5,-.2) node {$v$};
\draw(8.65,.25) node {$r_j$};
\draw[gray] (8.3,-.2) node {$v^\prime$};

\end{tikzpicture}
}
\caption{The figure shows the factorisation of a word $z=u_1xw_1=u_2xw_2$, where $x=z[l_i,r_i-1]= z[l_j,r_j-1]$, and $u_1\neq u_2$. The factor $v$ is a prefix of $x$ such that $u_2=u_1v$. The word $z$ can be written as $u_1v^iv^\prime w_2$ for some $i\in\N$ and $v^\prime$ prefix of $v$. For $k\in[0,{K^2}]$, $l_k$ is the index such that the counter value $n - k$ is encountered for the first time and $r_k$ the index such that the counter value $n-k+d$ is encountered for the last time during the run $\con\xrightarrow{z}\conG$. }
\label{looprepeat}
\end{figure}
\begin{clam}
There exist $v^\prime \in\Sigma^*$ and $r\in[0,K^3+t \cdot K]$ such that $x=v^rv^\prime$ with $|v^\prime| \leq |v|$.
\label{shortsuffix}
\end{clam}
\begin{clamproof}
Let $r\in\N$ be the largest number such that $x$ is of the form $v^rv^\prime$ for some $v^\prime \in \Sigma^*$ (see \Cref{looprepeat}). We know that $z=u_2xw_2$ and $u_2=u_1v$. Therefore,
$ z= u_1vxw_2 = u_1vv^rv^\prime w_2 = u_1v^rvv^\prime w_2$. Furthermore,
$z= u_1xw_1 = u_1v^rv^\prime w_1$. Now since $u_1v^rvv^\prime w_2 = u_1v^rv^\prime w_1$, we get that $vv^\prime w_2= v^\prime w_1$. Hence, if $|v^\prime| \geq |v|$, then $v$ is a prefix of $v^\prime$. This is a contradiction since $r$ was chosen to be the largest number such that $x$ is of the form $v^rv^\prime$.

To show the bound on the value $r$, we observe the following. We know that the counter effect of the run $\pi(x,\conE)$ is $d$. Therefore from \Cref{floatinglength}, we get that $|x| \leq 2 K^3 + |d| \cdot K$. Hence, $r\leq 2 K^3 + |d| \cdot K$.%the difference between the maximum and minimum counter values encountered during the run of $z$ from $\con$ is at most $|n_\con-m| + 2 \K^4$. Therefore,
\end{clamproof}
%\prm{counter drops different for different portions. easier to argue using length.}
%Our aim is to show that the word $v$ has a periodic structure such that for all $i,j\in [0,\frac{|v|}{|y|}-2]$ we have $v_{[i\cdot|y|,i\cdot|y|+|y|-2]}=v_{[j\cdot|y|,j\cdot|y|+|y|-1]}$ and $v_{[\frac{|v|}{|y|}-1|,|v|-1]}$ is a prefix of $y$. \prince{bad notation.}
% \prm{we claim that the word has a periodic structure, assume not... $z_{[i,i+p]}, z_{[i+p+1, i+2p]}$}
%Let $x_i$ denote the word $u_1x^i$ for all $i\in\N$ where $x\in\Sigma^*$ such that $x_1=u_1x$.
%Since $z=u_1vw_1=u_2vw_2$, we get that $y$ is a prefix of $v$ as well. Now since $y$ is a prefix of $v$ and $u_2=u_1v$, we get that $y^2$ is a prefix of $v$.

% This process can be repeated and $v$ can be written in the form $v=y^ry^\prime$ for some $r\in\N$ and $y^\prime$ prefix of $y$ as shown in \Cref{looprepeat}.
%\prm{show $r$ is in this bound.}
%Now we show that $|u_1|$ and $|y^\prime\cdot w_1|$ is also polynomially bounded in $\K$.
From Claim \ref{shortsuffix} and Claim \ref{shortrep}, we get that $|u_1vv^\prime w_1|\leq 12 K^3$ and $z=u_1v^rv^\prime w_1$ for some $r\in[0, 2 K^3 + |d| \cdot K)]$. 
%\rem{Note that the factor $v$ might start and end in different counter states during the run and, therefore need not always have a negative counter effect. However, we also know that the word $v^r$ has a negative counter effect. For $i\in[1,2\mc]$, let $\conG_i$ be the configuration such that $\conE\xrightarrow{v^i}\conG_i$. By Pigeon-hole principle there exists $j,k\in[1,2\mc]$ with $j<\mc$ and $k-j \leq \mc$ such that $p_{\conG_j} = p_{\conG_k}$. Also, note that the word $y=v^{k-j}$ has a negative counter-effect from the counter state $p_{\conG_j}$. Let $r^\prime={\frac{r-j}{k-j}}$ and $j^\prime=(r-j)\pmod{(k-j)}$. Now consider the word $z=u_1v^{j}y^{r^\prime}v^{j^\prime}v^\prime w_1$. Since $|u_1|,|w_1|,|v|\leq 3(\mq\cdot\mc)^3$, $j<\mc$ and $k-j \leq \mc$, we get that $|u_1v^{j}| \leq 3\cdot\mq^3\cdot\mc^4$, $|v^{j^\prime}v^\prime w_1| < 6\cdot\mq^3\cdot\mc^4$, $|y| \leq 3\cdot\mq^3\cdot\mc^4$ and $r^\prime \in [0, (\mq\cdot\mc)\cdot(|d|+ 2(\mq\cdot\mc)^2)]$.}
\end{proof}

{We now establish that the lexicographically minimal witness $z$ (whose run need not be floating) for a co-VS reachability problem has the form $uy_1^{r_1} vy_2^{r_2}w$. Here, lengths of the words $u,y_1,y_2,v$, and $w$ are polynomially bounded in the number of states, and $r_1$ and $r_2$ are polynomial values dependent on the number of states and the input counter values. 
%The technically challenging aspect of this proof was proven in \Cref{specialword}.}
%\prm{Use $poly_i(\mq,\mc)$ for these polynomials in the lemma statement?}
\begin{lemma}[special-word lemma] \label{specialwordnonfloating}
{If $z\in \Sigma^*$ is the lexicographically minimal witness for $(\con, \clsv,S,\{m\})$, then there exists $u, y_1, v_1, v_2, v_3, y_2, w \in \Sigma^*$ and $r_1, r_2 \in \N$ such that $z = uy_1^{r_1}vy_2^{r_2}w$ and the following  are true:
\begin{enumerate}
\item $|uy_1vy_2w|$ is polynomially bounded in the number of states of the machine.%\leq 25\cdot\mq^3\mc^4$, 
\item $r_1$ and $r_2$ are polynomially bounded in the number of states of the machine, $m$, and $\cntval{\con}$.%\leq max\{m,n_{\con}\}\cdot \mq\cdot\mc+(\mq\cdot\mc)^3$, 
%\item $\pi(uy_1^{r_1}v_1,\con)$, $\pi(v_3y_2^{r_2}w,\conD)$ are floating runs for configuration $\conD$ where $\con\xrightarrow{ uy_1^{r_1}v_1v_2}\conD$,  and
%\item $\ce(\pi(uy_1^{r_1}v_1, \con)) = \ce(\pi(uy_1^{r_1}v_1v_2, \con)) = -n_{\con}$.
\end{enumerate}}
%
%
%$z_1,z_2,z_3,u_1,u_3,v_1,v_3,y_1,y_3 \in \Sigma^*$ and $r_1,r_3\in\N$ such that $z=z_1z_2z_3$, $z_1=u_1y_1^{r_1}v_1$, $z_3=u_3y_3^{r_3}v_3$, $|u_1|, |u_3| \leq 3\K^7$, $|v_1|,|v_3| \leq 6\K^7$, $|y_1|,|y_3| \leq 3\K^7$, $|z_2|\leq \K^6$, $r_1\in[0,n_\con\cdot\K^2+\K^6]$ and $r_3\in[0,m\cdot\K^2+\K^6]$.}
\end{lemma}
\begin{proof}
Let $z\in \Sigma^*$ be the lexicographically minimal reachability witness for $(\con, \clsv,S,\{m\})$, where $\con$ is a configuration with counter value $n$.
Consider the run of word $z$ from $\con$. Let $\conD\in\clsv\times S\times \{m\}$ such that $\con \xrightarrow{z} \conD$. Let $\con=\con_1$ and $T(\con_1)= \con_1 \tau_1 \con_2 \cdots \tau_{\ell-1} \con_\ell$ denote the run on word $z$ from the configuration $\con_1$ and $T$ the corresponding sequence of transitions.
Let $\conE_{1}$ be the first configuration with counter value zero and $\conE_{2}$ be the last configuration with counter value zero during this run. Let $z_1,z_2,z_3\in\Sigma^*$ be such that $\con\xrightarrow{z_1}
\conE_1\xrightarrow{z_2}\conE_2\xrightarrow{z_3}\con_\ell$ and $z=z_1z_2z_3$. Observe that $\con\xrightarrow{z_1}\conE_1$ and $\conE_2\xrightarrow{z_3}\con_\ell$ are floating runs. % with counter value zero.

From \Cref{specialword}, we know that there exists $u_1,u_3,v_1,v_3,y_1,y_3 \in \Sigma^*$ and $r_1,r_3\in \N$ such that $z_1=u_1y_1^{r_1}v_1$, $z_3=u_3y_3^{r_3}v_3$, $|u_1|, |u_3| \leq 3\cdot\mq^3\cdot\mc^4$, $|v_1|,|v_3| \leq 6\cdot\mq^3\cdot\mc^4$, $|y_1|,|y_3| \leq 3\cdot\mq^3\cdot\mc^4$, $r_1\in[0,n\cdot\mq\cdot\mc+(\mq\cdot\mc)^3]$ and $r_3\in[0,m\cdot\mq\cdot\mc+(\mq\cdot\mc)^3]$. Also, from \Cref{smp} we get that $|z_2|\leq (\mq\cdot\mc)^3$.
\end{proof}
%\vspace{-0.5cm}
%We now prove that the binary co-VS reachability and co-VS coverability problems are in $\CF{NP}$. From \Cref{specialwordnonfloating} we observe there is a polynomial-size encoding of the lexicographically minimal word (where $r_1$ and $r_2$ are in binary). A non-deterministic machine can guess this encoding and verify the reachability in polynomial time since $\matm^{r}$ can be computed in $\log(r)$ time (see \Cref{lalemmas}). A detailed proof of \Cref{binaryreachability} is given in the Appendix. %give proof for \Cref{binaryreachability} and show that the binary reachability problems are in $\CF{NP}$.

%\prm{general theorem for non-floating run and for reachability (end configuration has high counter value)}
% \prince{
% - $\backslash/\backslash/\backslash/\backslash/\backslash/\backslash/\backslash$ - bounded length lemma ( counter inc. and decrease)}\\

%\section{Equivalence, Regularity and Covering}
%\prm{Equivalence, Regularity, Covering, Coverable equivalence}

%% file: TeX/equiv.tex
\section{Equivalence of weighted ODCA}

%\prm{Equivalence, Regularity, Covering to be removed.}
\label{sec:ptime} \label{sec:polywitness}
In this section, we present a polynomial time algorithm to decide the equivalence of two weighted \odcas whose weights come from a fixed field. The techniques developed in the previous section in conjunction with those presented in Valiant and Paterson~\cite{doca}, and  B\"ohm et al.~\cite{droca} for deterministic real-time \textsc{oca} give us the algorithm.
{The idea here is to prove that the maximum counter value encountered during the run of a minimal witness is polynomially bounded. We use this to reduce the equivalence problem to that of weighted automata. }
%\sav{Before we fix 2 ODCAs, we should tell what this section is all about. We should also try to explain why we were doing reachability earlier. Are we directly reducing the problem to reachability etc? }

In the remainder of this section, we fix two non-equivalent weighted \odcas $\Autom_1$ and $\Autom_2$ over an alphabet $\Sigma$ and a field $\Sring$. For $i\in\{1,2\}$,  \[\Autom_i=(C_i, \delta_{0_i},\delta_{1_i}, p_{0_i};\ Q_i, \vc \lambda_i, \Delta_i, \vc\eta_i).\]% \text{ and, } \]\[\Autom_2=(C_2, \delta_2, p_{0_2};\ Q_2, \vc \lambda_2, \Delta_2, \vc\eta_2).\]
Without loss of generality assume $\K = |C_1| = |Q_1|= |C_2| = |Q_2|$. We will reason on the synchronised runs on pairs of configurations.
Given two weighted \odcas, $\Autom_1$ and $\Autom_2$ and $i\in\N$, we denote a \emph{configuration pair} as $\conH_i=\pconfig{\con_i}{\conD_i}$ where $\con_i$ is a configuration of $\Autom_1$ and $\conD_i$ is a configuration of $\Autom_2$.
We similarly consider \emph{transition pairs} of $\Autom_1$ and $\Autom_2$, and consider \emph{synchronised runs} as the application of a sequence of transition pairs to a configuration pair.
We fix a minimal word $z$ (also called witness) that distinguishes $\Autom_1$ and $\Autom_2$ and $\ell=|z|$. Henceforth we will denote by
\[
\mrun = \conH_0 \tau_0 \conH_1 \cdots \tau_{\ell-1} \conH_\ell
\]
the synchronisation of runs over $z$ in $\Autom_1$ and $\Autom_2$ from their initial configurations, where $\conH_i$ are pairs of configurations and $\tau_i$ are pairs of transitions.
 We denote by $T = \tau_0\cdots\tau_{\ell-1}$ the sequence of transition pairs of this run pair.
%We now state the main result of our paper.
%\begin{theorem}
%\label{equivcheck}
%There is a polynomial time algorithm that decides if two \odcas are equivalent or outputs a minimal witness.
%\end{theorem}
The main idea to prove \Cref{main} is to show that the length of $z$ is polynomially bounded in the size of the two weighted \odcas.
%To prove \Cref{main}, we use the following lemma, which states that the counter values in $\mrun$ are bounded by a polynomial $\Pthree(\K)$.

%%%%%%%%%%%%%%%%%%%%%%%%%%%%%%%%%%%%%%%%
\begin{lemma}\label{counterbound}
There is a polynomial $\Pthree:\N \rightarrow \N$ such that if two weighted $\odcas$ $\Autom_1$ and $\Autom_2$ are not equivalent, then there exists a witness $z$ such that the counter values encountered during $\mrun$ are less than $\Pthree(\K)$.
\end{lemma}

We use \Cref{counterbound} to show that the length of a minimal witness $z$ is bounded by a polynomial $\Pzero(\K)=2\K^5\Pthree(\K)$.
\begin{lemma} \label{lengthboundodca}
There is a polynomial $\Pzero:\N \rightarrow \N$ such that if two weighted $\odcas$ $\Autom_1$ and $\Autom_2$ are not equivalent, then there exists a witness $z$ such that $|z|$ is less than or equal to $\Pzero(\K)$.
\end{lemma}
\begin{proof}
Assume for contradiction that the length of a minimal witness $z$ is greater than $\Pzero(\K)$.
From \Cref{counterbound}, we know that the counter values encountered during the run $\mrun$ in less than $\Pthree(\K)$.
Since $|z|> \Pzero(\K)$, by the Pigeonhole principle, we get that there exist indices $0 \leq i_0 < i_2< \cdots < i_{2\K}\leq\ell$ such that for all configuration pairs $\conH_{i_j}, j\in[1,2\K]$, 
${\con_{i_j}}$ and ${\con_{i_{j-1}}}$ have the same counter value and counter state and ${\conD_{i_j}}$ and ${\conD_{i_{j-1}}}$ have the same counter state and counter value.
%$n_{\con_{i_j}}=n_{\con_{i_{j-1}}}$, $n_{\conD_{i_j}}=n_{\conD_{i_{j-1}}}$, $p_{\con_{i_j}}=p_{\con_{i_{j-1}}}$ and $p_{\conD_{i_j}}=p_{\conD_{i_{j-1}}}$.

For all $j\in[0,2\K]$ we define the vector $\vc x_j\in \Sring^{2\K}$ such that $\vc x_j[r]=\vc x_{\con_{i_j}}[r], \text{ if } r<\K$ and $\vc x_{\conD_{i_j}}[r-\K], \text{ otherwise}$. We also define the vector $\vc\eta\in \Sring^{2\K}$ such that $\vc\eta[r]= \vc \eta_1[r], \text{ if } r<\K$ and $\vc \eta_2[r-\K], \text{ otherwise}$. For all $j\in[0,2\K]$, let $\mata_j$ denote the matrix such that $\vc x_j \mata_j = \vc x_\ell$. Since $z$ is a minimal witness, we know that for all $j\in[0,2\K]$, $\vc x_j \mata_j \trans{\vc\eta} \neq 0$.
From \Cref{lem:mat}, we get that there exists $r,r^\prime \in [0,2\K]$, with $r^\prime<r$ such that ${\vc x}_{r^\prime}\mata_r\trans{\vc \eta} \neq 0$. The sequence of transitions $\tau_{i_r+1} \cdots \tau_\ell$ can be taken from $\conH_{i_r^\prime}$ since the counter values and counter states are the same for both configurations.
Consider the sequence of transitions $T^\prime=\tau_0\cdots \tau_{i_r^\prime}\tau_{i_r+1} \cdots \tau_\ell$ and let $w= \word(T^\prime)$. The word $w$ is a shorter witness than $z$ and contradicts its minimality.
\end{proof}
%The proof is given in the Appendix. 
Thus we can reduce the equivalence problem of weighted \odca over fields to that of weighted automata over fields (which is in $\CF{P}$~\cite{prob})  by
``simulating'' the runs of weighted \odcas $\Autom_1$ and $\Autom_2$ up to length $\Pzero(\K)$ by two weighted automata. The naive algorithm will only give us a $\CF{PSPACE}$ procedure, but there is a polynomial time procedure to do this, and the proof is given below.

\begin{proof}[Proof of \Cref{main}]
{We consider the two weighted \odcas $\Autom_1$ and $\Autom_2$. From \Cref{lengthboundodca}, we know that the length of the minimal witness $z$ is less than $\Pzero(\K)$. Let $M=\Pzero(\K)$. We construct the $M$-unfolding weighted automata $\Autom^{M}_1$ and $\Autom^{M}_2$ as described in \Cref{unfolding}. It follows that, $\Autom_1$ is non-equivalent to $\Autom_2$ if and only if there exists a word $w \in \Sigma^{\leq M}$ such that $f_{\Autom_1^M}(w)\neq f_{\Autom_2^M}(w)$. Tzeng \cite[Lemma 3.4]{prob} gives a polynomial time algorithm to output a minimal word that distinguishes two probabilistic automata. We conclude the proof by noting that the algorithm can be extended to the case of weighted automata.}
%
%Note that for $i\in\{1,2\}$ and for every word $w\in\Sigma^{\leq M}$, $f_{\Autom_i}(w) = f_{{\Autom^M_i}}(w)$. %, and for every word $w\in\Sigma^{>M}$, $f_{\Autom^M}(w)=0$.
%
%The two weighted \odcas $\Autom_1$ and $\Autom_2$ are not equivalent if and only if there is a word $w\in\Sigma^{\leq M}$ such that $f_{{\Autom_1}^M}(w)\neq f_{{\Autom_2}^M}(w)$ . %and thus if and only if $\Autom^M_1$ and $\Autom^M_2$ are not equivalent.
%Assume two weighted automata are not equivalent. In that case, there is a linear-sized word to distinguish them, and the equivalence check can be done in polynomial time \cite[Lemma 3.4]{prob}.
%Tzeng \cite{prob} shows this for two probabilistic automata. The proof can easily be extended to weighted automata over fields.
%
%Using this, we decide the equivalence of $\Autom^M_1$ and $\Autom^M_2$ in polynomial time, which returns a word $w$ that distinguishes them if they are not equivalent. The \odcas $\Autom_1$ and $\Autom_2$ are not equivalent if and only if $|w|<M$. Thus we have a polynomial time procedure to decide the equivalence of two weighted \odcas, provided that \Cref{counterbound} holds.
\end{proof}

{The rest of this section is dedicated to proving \Cref{counterbound}. We adapt techniques developed by B\"ohm et al.~\cite{droca} for \textsc{oca}s. We start by labelling some configuration pairs as background points (see \Cref{fig1}). Consider the case where there is no background point in $\Pi$. By reducing the problem to co-VS reachability/coverability we show that the counter values in $\Pi$ are polynomially bounded. Now consider the case where there is a background point $\conH_j$ in $\Pi$. We show that the counter values encountered during the run of $\Pi$ till $\conH_j$ is polynomially bounded. This is shown by \cref{beltreturn} and \cref{firstbgpoint}. We conclude by arguing that the length of the run from $\conH_j$ is polynomially bounded. }
%We label some configuration pairs as background points (see \Cref{fig1}) for this purpose.
%First, we define a configuration space and partition the configuration pairs into initial, belt, and background points. 
%{We prove that if the run of a minimal witness does not reach a background point, then the counter values encountered during this run are polynomially bounded. This is proved by reducing to co-VS reachability/coverability problems of a weighted \odca.  We show that the counter values of the first background point encountered during the run of a minimal witness is polynomially bounded in $\K$. We also show that the remaining length of the run after the first background point is polynomially bounded in $\K$. Hence, we get that the maximum counter value encountered during the run of a minimal witness is polynomially bounded in $\K$. We adapt the techniques developed by B\"ohm et al.~\cite{droca} (also see \cite{bisim} \cite{bisimjour}) in the context of real-time \textsc{oca}.}%\sav{First should always have a second. We should explain what each of the upcoming subsections is going to do.
\[
\overbrace{\conH_0 \tau_0\conH_1\tau_1\conH_2 \cdots \conH_{j-1}\tau_{j-1} }^{ \text{counters poly-bounded}} \underbrace{\conH_j =\pconfig{\con_j}{\conD_j}}_{\substack{1^{st} \text{ configuration pair in }\\ \text{ background space }}} \overbrace{\tau_{j} \cdots \tau_{\ell-1} \conH_\ell}^{\substack{\text{counters} \\ \text{poly-bounded}}}
\]

Following B\"ohm et al.~\cite{droca}, we define a partition of the set of configuration pairs to facilitate this. %{We partition the set of configuration pairs into three: initial space, belt space, and background space (see \Cref{fig1} in \Cref{apxequiv}).

\subsection{Configuration Space}

Each pair of configuration $\conH=\pconfig{\con}{\conD}$ is mapped to a point in the space $\N \times \N \times (C_1 \times C_2) \times \Sring^\K \times \Sring^\K$, henceforth referred to as the \textit{configuration space}. Here, the first two dimensions represent the two counter values, the third dimension $C_1 \times C_2$ corresponds to the pair of counter states, and the remaining dimensions represent the weight vector. The projection of the configuration space onto the first two dimensions is depicted in \Cref{fig1}.
{We partition the configuration space into three: initial space, belt space, and background space.}
The size of the initial space and, thickness and number of belts will be polynomially bounded in $\K$.
%This partition is indexed on two polynomials, $\Pone: \N \to \N$ and $\Ptwo: \N \to \N$ chosen so that all belts are disjoint outside the initial space.
 %The motivation behind choosing these polynomials will be discussed later.
These partitions are indexed on two carefully chosen polynomials $\Pone(\K) = \initialsize$ and $\Ptwo(\K)=\beltsize$, , so that all belts are disjoint outside the initial space. The precise polynomials are required in the proofs of \Cref{beltpoint} and \Cref{distancelemma}. We use some properties of these partitions to show that the length of a minimal witness is bounded.
Given a configuration $\con$, we use $n_\con$ to denote $\cntval{\con}$.

\begin{itemize}
\item \emph{initial space}: All configuration pairs $\pconfig{\con}{\conD}$ such that $n_{\con}, n_{\conD} < \Pone(\K)$.
\item \emph{belt space}: Let $\alpha, \beta \in [1,3\K^7]$ be co-prime. A belt of slope $\frac{\alpha}{\beta}$ consists of those configuration pairs $\pconfig{\con}{\conD}$ outside the initial space that satisfies $|\alpha. n_{\con} - \beta. n_{\conD} | \leq \Ptwo(\K)$.
The belt space contains all configuration pairs $\pconfig{\con}{\conD}$ that is inside belts with slope $\frac{\alpha}{\beta}$.
\item \emph{background space}: All remaining configuration pairs.
\end{itemize}

\begin{figure}[htbp]
\centering
\resizebox{.4\columnwidth}{!}{%
\begin{tikzpicture}
%\tikzset{every path/.style={line width=.5mm}}
\draw [line width=0.4mm] [-stealth](0,0) -- (8,0)node[anchor=south, xshift=-.7cm, yshift=-.5cm] {$\N$};
\draw [line width=0.4mm] [-stealth](0,0) -- (0,8) node[anchor=east, yshift=-.7cm] {$\N$};
\draw[line width=0.4mm](0,3)--(3,3);
\draw[line width=0.4mm](3,3)--(3,0);
\draw[line width=0.4mm](.3,3)--(1.4,7.4);
\draw[line width=0.4mm](1.3,3)--(2.4,7.4);
\draw[line width=0.4mm](2.3,3)--(6.6,7.4);
\draw[line width=0.4mm](3,2.3)--(7.4,6.8);
\draw[line width=0.4mm](3,.3)--(7.4,1.4);
\draw[line width=0.4mm](3,1.3)--(7.4,2.4);
\draw (1,1.5) node[anchor=west, xshift=-.5cm]{initial space};
\draw (5,3.2) node[anchor=west, xshift=-.5cm]{background space};
\draw (4,4) node[anchor=west, xshift=-.5cm, rotate=45, yshift=-.3cm]{belt space};
\draw (4,1.5) node[anchor=west, yshift=-.5cm, rotate=14]{belt space};
\draw (1.5,4) node[anchor=west, xshift=-.5cm, rotate=76]{belt space};
\draw [densely dotted,line width=0.4mm](0,0) -- (0,-.3);
\draw [densely dotted,line width=0.4mm](3,0) -- (3,-.3);
\draw[stealth-stealth](0,-.2)--(3,-.2)node[anchor=north, xshift=-1.5cm]{\scriptsize $\Pone(\K)$};
\draw[stealth-stealth](5,5.7)--(5.68,5.06)node[anchor=south, xshift=.1cm, yshift=.4cm]{\scriptsize $\Ptwo(\K)$};
\end{tikzpicture}
}
\caption{Projection of configuration space}
\label{fig1}
\end{figure}

%The following lemma is a property of how belts and initial space are defined.
The proof of the following lemma is similar to that of the non-weighted case presented in \cite{droca}. % and is given in Appendix.% and is given in Appendix (\Cref{sec:beltpointlemma}).
\begin{lemma}%[\textbf{Lemma 6 in \cite{droca}}]
\label{beltpoint}
If $\pconfig{\con}{\conD}$ and $\pconfig{\conE}{\conF}$ are configuration pairs inside two distinct belts and lie outside the initial space, then there is no $a\in\Sigma$ such that $\pconfig{\con}{\conD} \xrightarrow{a} \pconfig{\conE}{\conF}$.
%it is not a neighbour to any configuration pair $\pconfig{\conE}{\conF}$, inside another belt.
\end{lemma}
% \begin{proofsketch}
%\prince{proof sketch if required.}
%\end{proofsketch}
\begin{proof}
Recall $\Pone(\K) = \initialsize$ and $\Ptwo(\K)=\beltsize$.
Let $B$ and $B^\prime$ be two distinct belts with $\mu$ being the slope of the belt $B$ and $\mu^\prime$ the slope of the belt $B^\prime$. Hence $\mu \neq \mu^\prime$.
Without loss of generality, let us assume that $\mu^\prime > \mu$.
It suffices to show that for all $x>\Pone(\K)$, we have
\[
\mu x + \Ptwo(\K) +1 < \mu^\prime {x} - \Ptwo(\K) - 1.
\]
We know that $\mu^\prime- \mu \geq \frac{1}{3\K^7}$ and $x> \initialsize$.
\[ \text{Therefore, } \frac{\initialsize}{6\K^7}< (\mu^\prime- \mu)\cdot x.\]
\begin{align*}
\implies & \mu x +\frac{86 \K^{14}}{2} < \mu^\prime x - \frac{86 \K^{14}}{2} \\
%\implies & \mu x + {7 \K^4} < \mu^\prime x - {7 \K^4}
\implies & \mu x + \beltsize +\K^{14} < \mu^\prime x - \beltsize - \K^{14}
\\
\implies & \mu x + \beltsize +1 < \mu^\prime x - \beltsize - 1
\end{align*}
%Hence proved.
\end{proof}

\Cref{beltpoint} ensures that the belts are disjoint outside the initial space and that no run can go from one belt to another without passing through the initial space or background space.
To prove \Cref{counterbound}, there are two cases to consider: either there is no background space point in $\mrun$, or there is a background space point in $\Pi$.

\subsection{Case 1: When there is no background space point in \mrun.}
Since there is no background space point in \mrun, all the points in $\mrun$ are either in the initial or belt space. By definition, the counter values of configuration pairs inside the initial space are bounded by $\Pone(\K)$. Now, we look at the sub-run of $\mrun$ inside the belt space.
If a sub-run of $\mrun$ enters and exits a belt from the initial space or if $\mrun$ ends inside a belt, then we show that the counter values encountered during that belt visit are polynomially bounded. {This is shown by reducing to co-VS reachability of an \odca.} For this proof, it is crucial that the belts are disjoint.
% $\Dutom$ as described below.}

%\sav{What is the objective of this subsection.}

Let $\mrun_b= \conH_i \tau_i \conH_{i+1} \cdots \tau_{j-1} \conH_j$ be a sub-run of the run of $z$ inside a belt with slope $\frac{\alpha}{\beta}$. Similar to the technique mentioned in \cite{docaequiv}, each configuration pair $\conH_r= ((\vc x_{{\con_r}}, p_{{\con_r}},n_{{\con_r}}),(\vc x_{{\conD_r}}, p_{{\conD_r}},n_{{\conD_r}}))$, where $r\in [i,j]$ can alternatively be presented as $((\vc x_{{\con_r}}, \vc x_{{\conD_r}}), p_{{\con_r}}, p_{{\conD_r}}, l_r)$
where $l_r$ denotes a line with slope $\frac{\alpha}{\beta}$ inside the given belt that contains the point $(n_{{\con_r}}, n_{{\conD_r}})$. Let $L$ be the set of all lines with slope $\frac{\alpha}{\beta}$ inside the given belt. Note that $|L|= \Ptwo(\K)$. The run $\mrun_b$ is similar to the run of a weighted \odca $\Dutom$ that has the tuple $(p_{{\con_r}}, p_{{\conD_r}}, l_r)$ as the state of the finite state machine and $\vc x_r\in \Sring^{2\K}$ as its weight vector where $\vc x_r[i]=\vc x_{\con_r}[i], \text{ if } i<\K$ and $\vc x_r[i]=\vc x_{\conD_r}[i-\K], \text{ otherwise}$. A formal definition of the \odca $\Dutom$ is given in \Cref{beltmachine}. 

%\begin{toappendix}
\begin{definition}
\label{beltmachine}
Let $\Autom_i=((C_i, \delta_{0_i},\delta_{1_i}, p_{0_i}),(Q_i, \vc \lambda_i, \Delta_i, \vc\eta_i))$ for $i\in\{1,2\}$, be the two \odcas given. Let $L$ be the set of all lines with slope $\frac{\alpha}{\beta}$ inside the given belt. 
We define the \odca $\Dutom = ((C, \delta_0,\delta_1,p_0),(Q,\vc \lambda, \Delta, \vc\eta))$, where the initial state $p_0$ and the initial distribution $\vc \lambda$ are arbitrarily chosen.
%\prm{\textbf{Must} mention that if the machines can be separated by word reaching final and non-final states, by doca result, we already have a poly witness. Assume this is not that case for the rest of the section.}
\begin{itemize}
\item $C= C_1 \times C_2 \times L$ is a non-empty finite set of states.
\item $\delta_1: C \times \Sigma \rightarrow C \times \{-1, 0, +1\}$ is the deterministic counter transition.
Let $p_1,q_1\in C_1,p_2,q_2\in C_2$, $a\in\Sigma$ and $d_1,d_2\in\{-1,0,+1\}$. Let $l_1,l_2\in L$ and $m_1,m_2\in\N$, such that the point $(m_1,m_2)$ lies on the line $l_1$. $\delta_1((p_1,p_2,l_1),a)= ((q_1,q_2,l_2),d_1) $, if $\delta_{1_1}(p_1,a)=(q_1,d_1)$ and $\delta_{1_2}(p_2,a)=(q_2,d_2)$ and the point $(m_1+d_1, m_2+d_2)$ lies on the line $l_2$. It is undefined otherwise. The function $\delta_0: C \times \Sigma \rightarrow C \times \{0, +1\}$ is arbitrarily chosen.
%\item $F= F_1\times F_2\times L$ is a finite non-empty set of final counter states.
\item $Q=Q_1 \cup Q_2$ is a non-empty finite set of states of the finite state machine.
\item \final{$\Delta: \Sigma \times \{0,1\} \to \Sring^{2\K \times 2\K}$ gives the transition matrix for all $a \in \Sigma$ and $d \in \{0,1\}$.
For $l\in L, m\in\N$ and $a\in\Sigma$,
%$$\Delta^\prime((p,m),a,0)= \begin{cases} \Delta(p,a,0)\text{ if } m=0\\
% \Delta(p,a,1) \text{ if } 0<m<M\end{cases}$$

$$ \hspace{-.5cm} \Delta(l,a)[i][j]=\begin{cases} \Delta_1(a,1)[i][j], \text{if } i,j<\K\\
\Delta_2(a,1)[i-\K][j-\K], \text{if } i,j >\K\\% \text{ and } \frac{i}{\K}= \frac{j}{\K}$}\\
0, \text{ otherwise}\end{cases}$$
}

\item $\vc\eta\in \Sring^{2\K}$ is the final distribution. $$\vc\eta[i]=\begin{cases} \vc \eta_1[i], \text{ if } i<\K\\
\vc \eta_2[i-\K], \text{ otherwise}
\end{cases}$$

\end{itemize}
\end{definition}
%\end{toappendix}

The sub-run $\mrun_b$ can now be seen as a floating run of a weighted \odca $\Dutom$. If the run $\mrun$ ends inside a belt, then $\mrun_b= \conH_i \tau_i \cdots \tau_{\ell-1} \conH_\ell$. In this case, we show that the difference between the counter values of the first and last configuration pairs is smaller than a polynomial in $\K$.%, then we show that some portion of the run inside the belt can be pumped out to get a shorter run. This is proved in the following lemma.

\begin{lemma}
\label{beltend}
There is a polynomial $poly:\N\to\N$, such that 
if $\mrun_b= \conH_i \tau_i \cdots \tau_{\ell-1} \conH_\ell$ lies inside a belt where $\conH_r= \pconfig{{\con_r}}{{\conD_r}}$, for $r\in [i,\ell]$, then $|\cntval{\con_\ell}-\cntval{\con_i}| \leq poly(\K)$ and $|\cntval{\conD_\ell}-\cntval{\conD_i}| \leq poly(\K)$. 
\end{lemma}
\begin{proof}
Let $\mrun_b= \conH_i \tau_i \conH_{i+1} \cdots \tau_{\ell-1} \conH_\ell$ be a sub-run of the run of a minimal witness inside a belt and ends in the belt, where $\conH_r= \pconfig{(\vc x_{{\con_r}}, p_{{\con_r}},n_{{\con_r}})}{(\vc x_{{\conD_r}}, p_{{\conD_r}},n_{{\conD_r}})}$, for $r\in [i,\ell]$. As mentioned in \Cref{beltmachine}, we consider this as the run of the weighted \odca $\Dutom$. Since it is the run of a witness, $\vc x_j \trans{\vc \eta} \neq 0$. Consider the vector space $
\lsu=\{ \vc y \in \Sring^{2\K} \mid {\vc y} \trans{\vc \eta} = 0\}$. Our problem now reduces to the co-VS coverability problem in machine $\Dutom$ and asks whether $(\vc x_i, (p_{\con_i},p_{\conD_i},l_i),n_{\con_i}) \xrightarrow{*} \clsu \times \{(p_{\con_\ell},p_{\conD_\ell},l_\ell)\} \times \N $.
From \Cref{smp}, we know that the length of a minimal reachability witness for
$((\vc x_i, (p_{\con_i},p_{\conD_i},l_i),n_{\con_i}), \clsu, (p_{\con_\ell},p_{\conD_\ell},l_\ell), \N)$ is polynomially bounded in $n_{\con_i}$ and $\K$. Hence proved.
\end{proof}
In the following lemma, we show that if $\mrun_b= \conH_i \tau_i \conH_{i+1} \cdots \tau_{j-1} \conH_j$ is a sub-run of $\mrun$ inside a belt and either $\cntval{\con_i} = \cntval{\con_j}$ or $\cntval{\conD_i} = \cntval{\conD_j}$, where $\conH_r= \pconfig{{\con_r}}{{\conD_r}}$, for $r\in [i,j]$,   then the counter values in $\mrun_b$ cannot increase more than a polynomial in $\K$ from $\cntval{\con_i}$ and $\cntval{\conD_i}$.
\begin{lemma}
\label{beltreturn}
There is a polynomial $poly:\N \to \N$ such that, if $\mrun_b= \conH_i \tau_i \conH_{i+1} \cdots \tau_{j-1} \conH_j$ is a run inside a belt with $\cntval{\con_i} = \cntval{\con_j}$ or $\cntval{\conD_i} = \cntval{\conD_j}$, where $\conH_r= \pconfig{{\con_r}}{{\conD_r}}$, for $r\in [i,j]$,  then the counter effect of any sub-run of $\mrun_b$ is less than or equal to $poly(K)$.
\end{lemma}
\begin{proof}
%\prince{reduce to co-VS reachability}
Let $\mrun_b= \conH_i \tau_i \conH_{i+1} \cdots \tau_{j-1} \conH_j$ be a sub-run of the run of a minimal witness inside a belt such that $n_{\con_i}= n_{\con_j}$, where $\conH_r= \pconfig{(\vc x_{{\con_r}}, p_{{\con_r}},n_{{\con_r}})}{(\vc x_{{\conD_r}}, p_{{\conD_r}},n_{{\conD_r}})}$, for $r\in [i,j]$ . We consider this as the run of the weighted \odca $\Dutom$ as mentioned in \Cref{beltmachine}. Since it is the run of a witness, we know that there exists $\mata \in \Sring^{2\K \times 2\K}$ such that $\vc x_j \mata \trans{\vc \eta} \neq 0$. Consider the vector space $
\lsu=\{ \vc y \in \Sring^{2\K} \mid {\vc y} \mata\trans{\vc\eta} = 0\}$.

Our problem now reduces to the co-VS reachability problem in machine $\Dutom$ and asks whether $(\vc x_i, (p_{\con_i},p_{\conD_i},l_i),n_{\con_i}) \xrightarrow{*} \clsu \times \{(p_{\con_j},p_{\conD_j},l_j)\} \times \{n_{\con_i}\} $.
From \Cref{smp}, the length of a minimal reachability witness for
$((\vc x_i, (p_{\con_i},p_{\conD_i},l_i),n_{\con_i}), \clsu, (p_{\con_j},p_{\conD_j},l_j), \{n_{\con_i}\})$ is bounded by a polynomial in $n_{\con_i}$ and $\K$. Hence proved.
\end{proof}

%The proofs of \Cref{beltend} and \Cref{beltreturn} are given in the Appendix. 
Hence, we get that the pair of runs of the minimal witness cannot reach counter values higher than some polynomial bound if it does not enter the background space. Now we look at the case where the run enters the background space.

%The following lemma says that the length of a minimal witness from a configuration pair with unequal distance is polynomial with respect to its counter values and $\K$. A length bound implies a bound on the counter value since the counters cannot increase more than the length of a word.

\subsection{Case 2: When there is a background space point in \mrun.}
We now consider the case where the witness ultimately enters the background space. 
Using co-VS reachability, we prove that the counter values encountered during $\Pi$ till the first background space point are polynomially bounded. We also show that the length of the remaining run is polynomially bounded in the number of states of the machines.

%\subsubsection{Underlying Weighted Automata} \label{underlyingwa}

%\sav{There has to be a number 2 if there is a number 1. Everyone will be looking around for that. If the entire subsection is a subsubsection called underlying weighted automata why do we have a subsection called background space. Why are we not calling it underlying weighted automata then?}
To that end, we need the notion of \emph{underlying uninitialised weighted automaton}.
Roughly speaking, an underlying uninitialised weighted automaton of an \odca $\Autom$ is the uninitialised weighted automaton $\uwa {\Autom}$ that is syntactically equivalent to $\Autom$ without zero tests. In other words, the transition function of $\uwa {\Autom}$ will be determined by the transition functions of $\Autom$ for positive counter values.
Floating runs of $\Autom$ are isomorphic to runs of this weighted automaton $\uwa {\Autom}$. 
%Here, we formalise this so-called notion of \emph{underlying uninitialised weighted automaton}.

%%%%%%%%%%%%%%%%%%%%%%%%%%%%%%%%%%%%%%%%

\begin{definition}\label{uuwa}
{ %Let $\Autom=(C, \delta, p_0;\ Q, \vc \lambda, \Delta, \vc\eta)$ be a weighted automata with $\K=|Q|=|C|$.
{For $l\in\{1,2\}$,} the \emph{underlying uninitialised weighted automaton} of $\Autom$ is the uninitialised weighted automaton $\uwa {\Autom_l}=(Q_l^\prime, \Delta_l^\prime, \vc\eta_l^\prime)$, where $Q_l^\prime= C_l \times Q_l$ and $\vc\eta_l^\prime\in \Sring^{\K^2}$ is the final distribution. For $i<\K^2, \vc \eta_l^\prime[i]=\eta_l[i\bmod\K]$.
The transition matrix is given by $\Delta_l^\prime: \Sigma \to \Sring^{\K^2\times\K^2}$. 
Let $a\in\Sigma$, $d\in \{-1,0,+1\}, i,j <\K^2$,
$$\Delta_l^\prime(a)[i][j]=
\begin{cases}
\Delta_l(p_{\frac{i}{\K}},a,1)[i\bmod \K][j\bmod \K], \text{ if } \delta_{l_1}(p_{\frac{i}{\K}},a)=(p_{\frac{j}{\K}},d)\\
0 \text{ otherwise}
\end{cases}$$}
\end{definition}

%The automaton $\uwa \Autom$ is said to be \emph{uninitialised} because it has no initial distribution.
%{Note that a configuration of $\uwa{\Autom_l}$ is a vector of dimension $\K^2$.}
A configuration $\con$ of a weighted \odca $\Autom$ is said to be $k$-equivalent to a configuration $ \ucon$ of an uninitialised weighted automata $\Butom$, denoted $\con \sim_k \ucon$, if for all $w\in\Sigma^{\leq k}, f_\Autom(w,\con)= f_\Butom(w, \ucon)$. We say that $\con$ is not $k$-equivalent to $\ucon$ otherwise and denote this as $\con \not\sim_k \ucon$.

As we need to test the equivalence of configurations from $\Autom_1$ and $\Autom_2$, we consider the uninitialised weighted automata $\Butom$, which is a disjoint union of $\uwa{\Autom_1}$ and $\uwa {\Autom_2}$. This gives us a single automaton with which we can compare their configurations.
{
%We use $\con$ to denote a configuration of $\Autom_1$ and $\conD$ to denote a configuration of $\Autom_2$.
Let $i\in\{1,2\}$ and $\con$ be a configuration of $\Autom_i$. For all $p\in C_i$ and $m<|\Butom|$, we define the sets 
$\lsw^{p,m}_i$. The set $\lsw^{p,m}_i$ contains vectors $\vc x\in\Sring^\K$ such that the configuration $(\vc x, p, m)$ is $|\Butom|$-equivalent to some configuration of $\Butom$.  The set $\clsw^{p,m}_i$ is the set $\Sring^\K \setminus \lsw^{p,m}_i$. 
Formally, 
\begin{align*}\lsw^{p,m}_i = \{\vc x \in \Sring^\K |&\exists \ucon \in \Sring^{|\Butom|},\con=(\vc x, p, m) \sim_{|\Butom|} \ucon\} 
%\clsw^{p,m}_i = \{\vc x \in \Sring^\K |&\forall \ucon \in \Sring^{2\K^2},\con=(\vc x, p, m) \not\sim_{2\K^2} \ucon\}
\end{align*}}
\begin{lemma}
For any $i\in\{1,2\}$, $p\in C_i$ and $m<|\Butom|$, the set $\lsw^{p,m}_i$ is a vector space.
\end{lemma}
\begin{proof}
{To prove this, it suffices to show that it is closed under vector addition and scalar multiplication. 
We fix a set $\lsw^{p,m}_i$. 
First, we prove that it is closed under scalar multiplication. For any vector $\vc z_1 \in \lsw^{p,m}_i$, we know that there exists a configuration $\con = (\vc z_1,p,m)$ and $\ucon \in \Sring^{|\Butom|}$ such that $ \con \sim_{|\Butom|} \ucon$.
Now, for any scalar $r\in\Sring$, the configuration $(r.\vc z_1, p, m)\sim_{|\Butom|2} r \cdot {\ucon}$. Therefore $r\cdot \vc z_1 \in  \lsw^{p,m}_i$.
 Now, we show that it is closed under vector addition.
Let $\vc z_1, \vc z_2 \in \lsw^{p,m}_i$ be two vectors.
Therefore, there exists configurations $\con_1= (\vc z_1,p,m)$, $\con_2 =(\vc z_2,p,m)$, ${\ucon}_1 \in \Sring^{|\Butom|}$ and ${\ucon}_2 \in \Sring^{|\Butom|}$, such that  $\con_1 \sim_{|\Butom|} {\ucon}_1$ and $\con_2 \sim_{|\Butom|} \ucon_2$. Consider the configuration ${\con}_3=(\vc z_1+\vc z_2,p,m)$, $\con_3 \sim_{|\Butom|} {\ucon}_1+{\ucon}_2$. Therefore, $\vc z_1+\vc z_2 \in  \lsw^{p,m}_i$.}
\end{proof}

%The proof is given in Appendix.
{The distance of a configuration $\con$ of $\Autom_i$ (denoted as $\dist_{\Autom_i}(\con)$) is the length of a minimal word that takes you from $\con$ to a configuration $(\vc x, p, m)$ for some $m<|\Butom|$ and $p\in C_i$ such that $\vc x \in \clsw^{p,m}_i$. We define $\dist_{\Autom_i}(\con)$ as:
$$\min\{|w| \mid \con \xrightarrow{w} (\vc x,p,m)\ \exists p\in C_i,m<|\Butom|, \vc x \in \clsw^{p,m}_i \}$$}
%\Dwak \Autom \Butom k &= \{\con \in \Sring^\K \times C \times \N\ |\ \exists \ucon \in \Sring^\K \times C,\ \con \sim_k \ucon \}
%Let $i\in\{1,2\}$ and $\con$ be a configuration of $\Autom_i$. %\[\dist_{\Autom_2}(\conD) = \min\{|w| \mid \exists p\in C_2, m<2\K^2, \vc x_{\conD^\prime}\in \clsw^{p,m}_2,\]
%\[ \hspace{4cm}\conD \xrightarrow{w} \conD^\prime\}\]}
%\[{\mathcal{C}}(c) = \min\{|w| \mid \exists c'\in \mathcal{C}, c \xrightarrow{w} c'\}\]
%Notice that distance is less than infinity if and only if t is reachable from $\con$. %We denote this by $\con\rightarrow^*\lsw$. By abuse of notation, we denote $c \xrightarrow{w} \lsw$ if there exists a configuration $c'\in \mathcal{C}$ such that $c \xrightarrow{w} c'$.
The notion of distance play a key role in determining which parts of the run of a witness can be pumped out if it is not minimal. Given two configurations $\con$, $\conD$ of $\Autom_1$  and $\Autom_2$ respectively, if $\dist_{\Autom_1}(\con) \neq \dist_{\Autom_2}(\conD)$, then $\con \not\equiv \conD$.
%The set $\inc$ consists of all configurations not equivalent to configurations in the underlying weighted automata. It is defined as follows:
%\[
%\inc = \bigcup_{i \in \{1,2\}} \bigcap_{j \in \{1,2\}} \Dcwak{\Autom_i}{\uwa{\Autom_j}}{\K}
%\]
%\sav{Should we define both $\WA$ and $\NWA$?}
%The set $\WA$, the complement of $\NWA$, contains all configurations equivalent to some configuration in the underlying weighted automata.
%Note that, if a configuration $\con$ is in $\clsw_1$ or $\clsw_2$ then $n_\con<2\K^2$. %The distance that will be interesting in our reasoning is the one to the set $\inc$. Therefore for simplicity, we denote $\dist(\con) = \dist(\con,{\inc})$ for a configuration $\con$ in $\Autom_1$ or $\Autom_2$.
{By special word lemma (\Cref{specialword}), the lexicographically minimal reachability witness has a special form. This is used to show that if a configuration $\con$ of an \odca $\Autom$ has finite distance, then $\dist_{\Autom}(\con)=\frac{a}{b} \cntval{\con}+t$, where $a,b,t\in\N$ and are polynomially bounded in $|\Autom|$.
This helps us in proving that configuration pairs outside the initial space having equal distance lie inside a belt.} %The proof of \Cref{distancelemma} and \Cref{bgpoint} are similar to the non-weighted case presented in \cite{droca} and is given in Appendix.

%%%%%%%%%%%%%%%%%%%%%%%%%%%%%%%%%%%%%%%%
\begin{lemma}
\label{distancelemma}
Let $\con=(\vc x_\con, p_\con,n_\con)$ be a configuration of weighted \odca $\Autom$. If $\dist_\Autom(\con)<\infty$ then, $\dist_\Autom(\con)= \frac{a}{b} n_\con + t$ where $a,b \in[0,3\K^7]$ and $|t| < 42\K^{14}$.
\end{lemma}
% \begin{proofsketch}
%\prince{proof sketch if required.}
%\end{proofsketch}
\begin{proof}
Without loss of generality, let us consider the weighted \odca $\Autom_1$ and a configuration $\con=(\vc x_\con, p_\con,n_\con)$ of $\Autom_1$.
Let us assume that $\dist_{\Autom_1}(\con)<\infty$. {This means that $\con \rightarrow^{*} \conD$, for some configuration $\conD=(\vc x_\conD, p,m)$ with $\vc x_\conD \in \clsw^{p,m}_1$ for some $p\in C_1$ and $m < 2\K^2$. Since $\cntval{\conD}=m$}, by \Cref{specialword}, we know that there is a word $u=u_1u_2^ru_3$ (with $r\geq 0$) such that that $\con \xrightarrow{u} \conD$ where $|u|= \dist_{\Autom_1}(\con), |u_1u_3|\leq 9\K^7,\ |u_2|\leq 3\K^7$ and $u_2$ has a negative counter effect $\ell$. Let $g$ be the combined counter effect of $u_1,u_3$ and $\alpha=\frac{|u_2|}{\ell}$. Since $ |u_1u_3|\leq 9\K^7$, we have $|g| \leq 9\K^7$.
\begin{align*}
\dist_{\Autom_1}(\con)&= \frac{n_\con- m- g}{\ell} |u_2| + |u_1u_3|\\
&= \alpha n_\con -\underbrace{\alpha(m+g)+ |u_1u_3|}_{t}
\end{align*}
Since $1 \leq \alpha \leq 3\K^7$ it follows that $
-42 \K^{14} < t < 42\K^{14}$. Hence proved.
\end{proof}
%The proof is in Appendix.
%The polynomials $\Pone$ and $\Ptwo$ were picked so that the configuration pairs with equal distance always lie in the belt space. Therefore, the background space points either have unequal or infinite distance.

Therefore, the background space points either have unequal or infinite distances. 

\begin{lemma} \label{bgpoint}
For any configuration pair $\pconfig{\con}{\conD}$, in the background space, either $\dist_{\Autom_1}(\con) \neq \dist_{\Autom_2}(\conD)$ or $\dist_{\Autom_1}(\con) = \dist_{\Autom_2}(\conD) = \infty$.
\end{lemma}
\begin{proof}
Assume for contradiction that there is a configuration pair $\pconfig{\con}{\conD}$, in the background space such that $\dist_{\Autom_1}(\con) = \dist_{\Autom_2}(\conD) < \infty$. Let $n_\con= \cntval{\con}$ and $n_\conD=\cntval{\conD}$.
%Let $\Pone(\K) = \initialsize$ and $\Ptwo(\K)=\beltsize$.\\
Since $\dist_{\Autom_1}(\con) = \dist_{\Autom_2}(\conD)$. From \Cref{distancelemma}, there exists $a_1, b_1, a_2, b_2 \in [0,3\K^7]$ and $d_1,d_2 < 42\K^{14}$ such that
\[
\frac{a_1}{b_1} n_\con + d_1 = \dist_{\Autom_1}(\con) = \dist_{\Autom_1}(\conD) = \frac{a_2}{b_2} n_\conD + d_2
\]
Therefore $|\frac{a_1}{b_1} n_\con - \frac{a_2}{b_2} n_\conD| \leq |d_2 - d_1| < 42\K^{14}$. This satisfies the belt condition and is a configuration pair in the belt space. This contradicts our initial assumptions.
\end{proof}

Similar to that in~\cite{droca}, we can show that the length of the run $\mrun$ in the background space is polynomially bounded in $\K$, and the counter values of the first background point in $\mrun$.
%The following lemma shows that the length of the run $\mrun$ in the background space is polynomially bounded in $\K$ and the counter values of the first background point in $\mrun$. The proof is similar to that in \cite{droca}. % and is given in Appendix.
\begin{lemma} \label{shortbackgroundwitness}
If $\conH_j=\pconfig{\con_j}{\conD_j}$ is the first configuration pair in the background space during $\mrun$, then $\ell-j$ is bounded by a polynomial in $\cntval{\con_j}$, $\cntval{\conD_j}$, and $\K$.
\end{lemma}
\begin{proof}
Let $\conH_j=\pconfig{\con_j}{\conD_j}$ be the first configuration pair in the background space during the run $\mrun$, then from \Cref{bgpoint}, either $\dist_{\Autom_1}(\con_j)= \dist_{\Autom_2}( \conD_j) = \infty$ or $\dist_{\Autom_1}(\con_j) \neq \dist_{\Autom_2}(\conD_j)$.
We separately consider the two cases.

\emph{Case-1,} $\dist_{\Autom_1}(\con_j)= \dist_{\Autom_2}( \conD_j) = \infty$: then we prove that the remaining length of the witness from $\pconfig{\con_j}{\conD_j}$ is bounded by $2\K^2$. Assume for contradiction that this is not the case and $\con_j \sim_{2\K^2}\conD_j$ but $\con_j \not\equiv \conD_j$. Let $v \in \Sigma^{>2\K^2}$ be the word which distinguishes
$\con$ and $\conD$.
Therefore, there exists a prefix of $v$, $u \in \Sigma^{|v|-2\K^2}$, and $i=\ell-2\K^2$ such that $\pconfig{\con_j}{\conD_j} \xrightarrow{u} \pconfig{\con_i}{\conD_i}$ and $\con_i \not \equiv_{2\K^2} \conD_i$.

{Since $v$ is a minimal witness $\con_i \equiv_{2\K^2-1} \conD_i$ and $\con_i \not\equiv_{2\K^2} \conD_i$. Since $\dist_{\Autom_1}(\con_j)= \dist_{\Autom_2}( \conD_j) = \infty$, there exists configurations ${\ucon_i}$ and ${\uconD_i}$ in the underlying automaton $\Butom$ such that $\con_i \sim_{2\K^2} {\ucon_i}$ and $\conD_i\sim_{2\K^2} {\uconD_i}$. Since $\con_i\equiv_{2\K^2-1} \conD_i$, it follows that ${\ucon_i} \sim_{2\K^2-1} {\uconD_i}$. From the equivalence result of weighted automata, we know that if two configurations of a weighted automata with $k$ states are non-equivalent, then there is a word of length less than $k$ which distinguishes them. 
Therefore, this is sufficient to prove that the underlying weighted automata with ${\ucon_i}$ and ${\uconD_i}$ as initial distributions are equivalent, and thus ${\ucon_i} \sim_{2\K^2} {\uconD_i}$.} This allows us to deduce that $\con_i\equiv_{2\K^2} \conD_i$, which is a contradiction. Therefore, the remaining length of the witness from $\pconfig{\con_j}{\conD_j}$ is bounded by $2\K^2$.

\emph{Case-2,} $\dist_{\Autom_1}(\con_j) \neq \dist_{\Autom_2}(\conD_j)$:
Without loss of generality, we suppose $\dist_{\Autom_1}(\con_j) > \dist_{\Autom_2}(\conD_j)$.
{By definition of $\dist_{\Autom_2}$, there exists $u \in \Sigma^{\dist_{\Autom_2}(\conD_j)}$, $i>j$ and a configuration $\ucon$ of the underlying automaton $\Butom$ such that $\con_j \xrightarrow{u} \con_i$, $\conD_j \xrightarrow{u} \conD_i$, $\con_i \sim_{2\K^2} \ucon_i$ and $\conD_i \not\sim_{2\K^2} \ucon_i$. Therefore $\con_i \not \equiv_{2\K^2} \conD_i$.}
By definition, there exists $v \in \Sigma^{\leq {2\K^2}}$ such that $ f_{\Autom_1}(v,\con_i) \neq f_{\Autom_2}(v, \conD_i)$ and hence $ f_{\Autom_1}(uv,\con_j) \neq f_{\Autom_2}(uv, \conD_j)$.
As $uv \in \Sigma^{\dist_{\Autom_2}(\conD_j)+{2\K^2}}$, we get that $\con_j \not\equiv_{\dist_{\Autom_2}(\conD_j)+{2\K^2}} \conD_j$.
Therefore, there is $w\in\Sigma^{\leq \min\{\dist_{\Autom_1}(\con_j),\dist_{\Autom_2}(\conD_j)\}+{2\K^2}}$ that distinguishes $\con_j$ and $\conD_j$.
\end{proof}

Let $\alpha, \beta \in [1,3\K^7]$ be co-prime. We say configuration pairs $\pconfig{(\vc x_\con,p_\con,n_\con)}{(\vc x_\conD,p_\conD, n_\conD)}$ and $\pconfig{(\vc x_\conE,p_\conE,n_\conE)}{(\vc x_\conF,p_\conF,n_\conF)}$ are
\emph{$\rep$ related} if $p_\con=p_\conE$, $p_\conD=p_\conF$ and $\alpha\cdot n_\con -\beta\cdot n_\conD = \alpha \cdot n_\conE -\beta\cdot n_\conF$. Roughly speaking, two configuration pairs are $\rep$ related
if they have the same state pairs and lie on a line with slope $\frac{\alpha}{\beta}$.
An \emph{$\rep$ repetition} is a run ${\bar\pi_1} =c_i \tau_i c_{i+1} \tau_{i+1} \cdots \tau_{j-1} c_j$ that lies inside a belt with slope $\frac{\alpha}{\beta}$ such that $c_i$ and $c_j$ are $\rep$ related.% (see \Cref{fig2}). %The projection of an $\rep$ repetition onto the counters is given in the Appendix (see \Cref{fig2}).
%\begin{toappendix}
\begin{figure}
\begin{minipage}{.5\textwidth}
\centering
\resizebox{.85\columnwidth}{!}{%
\begin{tikzpicture}
\tikzset{every path/.style={line width=.3mm}}
\draw [line width=0.4mm] [-stealth](2,2) -- (8,2)node[anchor=south, xshift=-.7cm, yshift=-.5cm] {$\N$};
\draw [line width=0.4mm] [-stealth](2,2) -- (2,8) node[anchor=east, yshift=-.7cm] {$\N$};
% \draw[line width=0.3mm](0,3)--(3,3);
% \draw[line width=0.3mm](3,3)--(3,0);
% \draw[line width=0.3mm](.3,3)--(1.4,7.4);
% \draw[line width=0.3mm](1.3,3)--(2.4,7.4);
\draw[line width=0.4mm](2.3,3)--(6.6,7.4);
\draw[line width=0.4mm](3,2.3)--(7.4,6.8);
% \draw[line width=0.3mm](3,.3)--(7.4,1.4);
% \draw[line width=0.3mm](3,1.3)--(7.4,2.4);
\draw [densely dotted](4,4) -- (5.5,4) node[anchor=west]{$\pconfig{\con}{\conD}$};
\draw [densely dotted](6,6) -- (7.5,6) node[anchor=west]{$\pconfig{\conE}{\conF}$};
\filldraw (4,4) circle[radius=1.5pt] ;
\filldraw (6,6) circle[radius=1.5pt] ;
\draw [dotted](2,2) -- (7,7);
\draw plot [smooth, tension=1] coordinates {(2.7,2)(2.8,3)(3.2,3.4)(3.5,3.2)(4,4)};
\draw [dashed] plot[smooth, tension=1] coordinates {(4,4) (4.3,4.4)(4.6,5.2)(5,5.4)(5.5,5.2)(6,6)};
\draw plot [smooth, tension=1] coordinates {(6,6)(6.2,6.7)(7,7.4)};
\end{tikzpicture}
}
\caption{The figure depicts an $\rep$ repetition inside a belt with slope $\frac{\alpha}{\beta}$. The configuration pairs $\pconfig{\con}{\conD}$ and $\pconfig{\conE}{\conF}$ are $\rep$-related. i.e., they lie on a line with slope $\frac{\alpha}{\beta}$. Note that $\cntstate{\con}=\cntstate{\conE}$ and $\cntstate{\conD}=\cntstate{\conF}$ if $\pconfig{\con}{\conD}$ and $\pconfig{\conE}{\conF}$ are $\rep$-related.}
\label{fig2}
\end{minipage}
\begin{minipage}{.5\textwidth}
\centering
\resizebox{.78\columnwidth}{!}{%
\begin{tikzpicture}
\tikzset{every path/.style={line width=.3mm}}
\draw [line width=0.4mm] [-stealth](0,0) -- (8,0)node[anchor=south, xshift=-.7cm, yshift=-.5cm] {$\N$};
\draw [line width=0.4mm] [-stealth](0,0) -- (0,8) node[anchor=east, yshift=-.7cm] {$\N$};
\draw[line width=0.4mm](0,3)--(3,3);
\draw[line width=0.4mm](3,3)--(3,0);
\draw[line width=0.4mm](.3,3)--(1.4,7.4);
\draw[line width=0.4mm](1.3,3)--(2.4,7.4);
\draw[line width=0.4mm](2.3,3)--(6.6,7.4);
\draw[line width=0.4mm](3,2.3)--(7.4,6.8);
\draw[line width=0.4mm](3,.3)--(7.4,1.4);
\draw[line width=0.4mm](3,1.3)--(7.4,2.4);
\draw [densely dotted](4,4) -- (6.8,4);% node[anchor=west]{$\pconfig{\con}{\conD}$};
\draw [densely dotted](5.2,5.2) -- (6.5,5.2);% node[anchor=west]{$\pconfig{\conE}{\conF}$};
\draw [densely dotted](3.3,3.3) -- (6.8,3.3);
\draw [densely dotted](4.55,4.55) -- (6.65,4.55);
\draw [densely dotted](4.3,4.3) -- (6.65,4.3);
\draw [dotted](3,3) -- (6.8,6.8);
\filldraw (4,4) circle[radius=1.5pt] ;
\filldraw (6.63,4.55) circle[radius=1.5pt] ;
\filldraw (3.3,3.3) circle[radius=1.5pt] ;
\filldraw (4.3,4.3) circle[radius=1.5pt] ;
\filldraw (4.55,4.55) circle[radius=1.5pt] ;
\filldraw (6.74,4) circle[radius=1.5pt] ;
\filldraw (5.2,5.2) circle[radius=1.5pt] ;%
\filldraw (6.5,5.2) circle[radius=1.5pt] ;%
\filldraw (6.8,3.3) circle[radius=1.5pt] ;%
\filldraw (6.68,4.3) circle[radius=1.5pt] ;
\draw plot [smooth, tension=1] coordinates {(0,0)(2,.7)(1,2)(.5,1.4)(.7,2.2)(2,2.5)(2.5,1)(2.7,2)(2.8,3)(3.2,3.4)(3.5,3.2)(4,4)};
\draw [dashed] plot[smooth, tension=1] coordinates {(4,4) (4.15,4.6) (4.3,4.3) (4.4,4.1) (4.55,4.55) (4.6,5.2)(5,5.35) (5.2,5.2)};
\draw plot [smooth, tension=1] coordinates {(5.2,5.2)(5.8,5.3)(6,6)(6,6)(6.1,6.4)(6.4,6.4)(6.5,5.2)};
\draw [dashed] plot[smooth, tension=1] coordinates {(6.5,5.2) (6.73,4)};
\draw plot [smooth, tension=1] coordinates { (6.73,4)(6.8,3.3) (6.9,0)};
\draw plot [smooth, tension=1] coordinates {(6.9,0)(7.2,.5)};
\end{tikzpicture}
}
\caption{The figure shows the run of a word that enters the background space from the belt such that the counter values of the first configuration pair in the background space exceed a polynomial bound. Some $\rep$ repetitions inside the belt can be removed to show the existence of a shorter witness. }
\label{fig3}
\end{minipage}
\end{figure}
%\end{toappendix}
The following lemma bounds the counter values of the first configuration pair in the background space, if it exists, during the run $\mrun$.
%\prince{We show that if $\conH_j$ is the first configuration pair in the background space during the run $\mrun$ and the counter values of $\conH_j$ are greater than $\K^5\cdot\beltsize$ then, we can ``pump out" some portion inside the belt to reach a configuration pair $\conH_{j^\prime}= \pconfig{\con_{j^\prime}}{\conD_{j^\prime}}$ in the background space. Furthermore, The configuration pair $\conH_{j^\prime}$ will have smaller counter values than $\conH_j$ and either $\con_{j^\prime} \not\equiv_{2\K^2} \conD_{j^\prime}$ or $\dist_{\Autom_1}(\con_{j^\prime}) \neq \dist_{\Autom_2}(\conD_{j^\prime})$.} 

%%%%%%%%%%%%%%%%%%%%%%%%%%%%%%%%%%%%%%%%
\begin{lemma} \label{firstbgpoint}
If $\conH_j = \pconfig{\con_j}{\conD_j}$ is the first background point in $\mrun$ then, $\cntval{\con_j}$ and $\cntval{\conD_j}$ are less than $\K^5\cdot\beltsize$.
\end{lemma}
\begin{proof}
%\begin{figure}
%\end{figure}
Let $\conH_j= \pconfig{\con_j}{\conD_j}$ be the first point in the background space during the run $\mrun$. Given any configuration $\con$, let $n_{\con}$ denote $\cntval{\con}$.
Assume for contradiction that $n_{\con_j}$ is greater than $\K^5\cdot\beltsize$. %Without loss of generality, let us assume that $n_{\con_j}>\K^5\cdot\beltsize$.
Let $\mrun=\conH_0 \tau_{0}\cdots \conH_{j-1}\tau_{j-1} \conH_j\cdots \conH_\ell$ be a run of a minimal witness. Since $\conH_j$ is the first point in the background space in this run and $n_{\con_j}>\K^5\cdot\beltsize$, there exists $0<i<j$ such that the sub-run $\mrun_b= \conH_i \tau_{i}\conH_{i+1}\cdots \tau_{j-2}\conH_{j-1}$ lies inside a belt $B$ with slope $\frac{\alpha}{\beta}$ for some $\alpha,\beta \in[1,3\K^7]$.
Since we are looking at the run of a minimal witness, from \Cref{bgpoint} either $\con_j \not\equiv_{2\K^2} \conD_j$ or $\dist(\con_j) \neq \dist(\conD_j)$.
We separately consider the two cases.

\emph{Case-1:} $\dist_{\Autom_1}(\con_j) \neq \dist_{\Autom_2}(\conD_j)$: Without loss of generality, let us assume $\dist_{\Autom_1}(\con_j) < \dist_{\Autom_2}(\conD_j)$. Therefore there exists $t\in\N$ with $j<t\leq\ell$ and configuration pair $\conH_t$ such that $\conH_t= \pconfig{(\vc x_{\con_t}, p, m)}{( \vc x_{\conD_t},p_{\conD_t}, n_{\conD_t})}$, 
{ $m < 2 \K^2$ and $\vc x_{\con_t}\in \clsw^{p,m}_1$.}
We show that we can pump some portion out from $\mrun_b$ to reach a configuration in the background space with unequal distance and smaller counter values.

Since $n_{\con_j} > \K^5\cdot\beltsize$, by Pigeonhole principle, there exists indices $i_0<i_1<i_2< \cdots, i_{\K^2}<i^\prime_0<i^\prime_1<i^\prime_2, \cdots, <i^\prime_{\K^2}$ such that for all $r\in[1,\K^2]$, (1) $\conH_{i_{r-1}}$ and $\conH_{i_{r}}$ are $\rep$ related and lie in belt $B$, (2) $n_{\con_{i_{r-1}}} < n_{\con_{i_{r}}}=n_{\con_{i^\prime_{r}}}$, (3) $p_{\con_{i^\prime_{r}}}= p_{\con_{i^\prime_{r-1}}}$, (4) for all $t\in\N$ with $i_{r}<t<j$, $n_{\con_t}>n_{\con_{i_{r}}}$, and (5) for all $t\in\N$ with $j<t<{i^\prime_{r}}$, $n_{\con_t}<n_{\con_{i^\prime_{r}}}$. % \prince{First occurence of counter value}.

Given any configuration $\con$ let $\vc x_{\con}$ denote $\wgtvec{\con}$.
For $r\in[0,\K^2]$ let $\mata_{r}\in \Sring^{\K\times\K}$ denote the matrix such that $\vc x_{\con_{i_r}}\mata_r = \vc x_{\con_{i^\prime_r}}$ and $\matb_{r}\in \Sring^{\K\times\K}$ denote the matrix such that {$\vc x_{\con_{i^\prime_r}}\matb_r = \vc x_{\con_{t}} \in \clsw^{p,m}_1$}. Therefore for all $r\in[0,\K^2]$, we have {$\vc x_{\con_{i_r}}\mata_r\matb_r\in\clsw^{p,m}_1$}. From \Cref{lem:mat}, we have that there exists $s,r \in [0,\K^2]$ with $s<r$ such that {$\vc x_{\con_{i_s}}\mata_r\matb_s\in\clsw^{p,m}_1$}. Consider the sequence of transitions $T^\prime= \tau_0, \cdots,\tau_{i_s-1}\tau_{i_r},\cdots,\tau_{j-1}$ and let $w= \word(T^\prime)$. Let $\conH_{j^\prime}$ be the configuration such that $\conH_0\xrightarrow{w}\conH_{j^\prime}$. Since we have removed an $\rep$ repetitions inside the belt, the configuration pair $\conH_{j^\prime}$ is a point in the background space (see \Cref{fig3}). Moreover, $n_{\con_{j^\prime}}< n_{\con_j}$ and $\dist_{\Autom_1}(\con_{j^\prime}) < \infty$. Since it is a point in the background space, from \Cref{bgpoint}, we get that $\dist_{\Autom_1}(\con_{j^\prime}) \neq \dist_{\Autom_2}(\conD_{j^\prime})$. Therefore, there is a shorter path to a configuration in background space with smaller counter values and unequal distance. This is a contradiction.

\emph{Case-2:} $\con_j \not\equiv_{2\K^2} \conD_j$:
Consider the sub-run $\mrun_b$. Since it is a run inside a belt, we can consider this as the run of the weighted \odca $\Dutom$.
Since $n_{\con_j}> \K^4 \cdot \beltsize$, by Pigeon-hole principle, there exists indices $i_0,i_1,i_2, \cdots, i_{\K^2}$ such that for all $r\in[1,\K^2]$, $\conH_{i_{r-1}}$ and $\conH_{i_{r}}$ are $\rep$ related with $n_{\con_{i_{r-1}}}< n_{\con_{i_{r}}}$ and for all $t\in\N$ with $i_{r}<t<j$, $n_{\con_t}>n_{\con_{i_{r}}}$. %\prince{last occurance of counter value before bg point.}

Since it is the run of a minimal witness, we know that there exists $\mata \in \Sring^{2\K \times 2\K}$ such that $\vc x_{j-1} \mata \trans{\eta}_F \neq 0$. Consider the vector space $
\lsu=\{ \vc y \in \Sring^{2\K} \mid {\vc y} \mata\trans{\eta}_F = 0\}$.
For $r\in[0,\K^2]$, let $\mata_{r}$ denote the matrices such that $\vc x_{i_r} \mata_{r}= \vc x_j \in \lsu$.
Since ${\vc x}_{{i_r}} \mata_r \in \clsu$ for all $r\in [0,\K^2]$, from \Cref{lem:mat}, we get that there exists $r^\prime \in [0,r-1]$ such that ${\vc x}_{\con_{{i_r^\prime}}} \mata_r \in \clsv$. The sequence of transitions $\tau_{i_r+1} \cdots \tau_\ell$ can be taken from $\conH_{i_r^\prime}$ since the counter values always stay positive.
Consider the sequence of transitions $T^\prime=\tau_0\cdots \tau_{i_r^\prime}\tau_{i_r+1} \cdots \tau_\ell$ and let $w= \word(T^\prime)$. The word $w$ is a shorter witness than $z$ and contradicts its minimality.
\end{proof}

%\sav{Proof of the two lemmas corresponding to the background space can be put here. The underlying automata definition, distance etc will also come here.}
Finally, we prove that the counter values encountered during the run $\mrun$ are polynomially bounded in $\K$ using  above lemmas.
\begin{proof}[Proof of \Cref{counterbound}]
Consider the run $\mrun$. From \Cref{beltend}, \Cref{beltreturn} and \Cref{firstbgpoint}, we get that the counter values of configuration pairs inside the belt space during this run in polynomially bounded in $\K$. Therefore, if it exists, the first background point in $\mrun$ has polynomially bounded counter values. From \Cref{shortbackgroundwitness}, the length of $\mrun$ after the first background point is polynomially bounded in $\K$. Since the counter values of configuration pairs inside the initial space are also bounded by a polynomial in $\K$, the maximum counter value in $\mrun$ is polynomially bounded in $\K$.
\end{proof}

%% file: TeX/regularity.tex
% !TEX root = ../odca.tex

\section{Regularity of ODCA is in $\CF{P}$}
\label{sec:regularity}
We say that a weighted \odca $\Autom$ is regular if there is a weighted automaton $\Butom$ that is equivalent to it.
We look at the regularity problem - the problem of deciding whether a weighted \odca is regular. We fix a weighted \odca $\Autom= ((C, \delta_0,\delta_1, p_0),(Q, \vc \lambda, \Delta, \vc\eta))$ and use $\KN$ to denote $|C|\cdot|Q|$.

%\sav{Explain what is special about 'our' proof here. Mention from where our work is inspired from. State in simple english the main difficulty is that one needs to pump up (this is my terminology. you will have to change it to suit the writing) large enough words. Say because of our special word property in conjunction with dependency of matrices we have the capability to pump up.}
{ The proof technique is adapted from the ideas developed by B\"ohm et al. \cite{bisimjour} in the context of real-time \textsc{oca}. The crucial idea in proving regularity is to check for the existence of infinitely many equivalence classes. The proof relies on the notion of distance of configurations. Distance of a configuration is the length of a minimal word to be read to reach a configuration that does not have an $\KN$  equivalent configuration in the underlying automata.  The challenge is to find infinitely many configurations reachable from the initial configuration, so that no two of them have same distance. Our main contribution is in designing a ``pumping" like argument to show this (\Cref{lem:uturn}, Point \ref{pumpup}).} %We show that given a run from the initial configuration, which reaches a configuration with a sufficiently large counter value having finite distance, we can find another run from the initial configuration, which reaches a configuration with a higher counter value, having finite distance. }
%We prove \Cref{regularity} in this section.

Recall the definition of $\uwa{\Autom}$ from \Cref{uuwa}. { We use $\con$ to denote some configuration of $\Autom$ and $\ucon$ to denote some configuration of $\uwa{\Autom}$. {For a $p\in C$ and $m\in\N$, we define }
\begin{align*}\lsw^{p,m} = \{\vc x \in \Sring^{|Q|} |&\exists \ucon \in \Sring^{\KN},\con=(\vc x, p, m) \sim_{\KN} \ucon\}\ . 
%\clsw^{p,m} = \{\vc x \in \Sring^{|Q|} |&\forall \ucon \in \Sring^{\KN},\con=(\vc x, p, m) \not\sim_{\KN} \ucon\} 
 \end{align*}
 The set $\clsw^{p,m}$ is $\Sring^{|Q|} \setminus \lsw^{p,m}$.
The distance of a configuration $\con$ (denoted by $\dist(\con))$ is
\[\min\{|w| \mid \con \xrightarrow{w} (\vc x, p, m)\ \exists p\in C,m<\KN 
\text{, and }\vc x \in \clsw^{p,m}\}\ . \]
% \min\{|w| \mid \exists \con^\prime \in \clsw, \con \xrightarrow{w} \con^\prime\}\]
}
The following lemma shows when $\Autom$ is not regular. % us in proving \Cref{regularity}.
Given any configuration $\con$, we use $\vc x_{\con}$ to denote $\wgtvec{\con}$, $p_\con$ to denote $\cntstate{\con}$ and $n_\con$ to denote $\cntval{\con}$.
%\prm{Clean up, move to appendix, proof sketch.}
\begin{lemma}\label{increasingdist}
Let $\con$ be an initial configuration {of an \odca $\Autom$}. Then the following are equivalent.
\begin{enumerate}
\item \label{notreg} $\Autom$ is not regular.
{
\item \label{arbitt} for all $t\in\N$, there exists configurations $\conD,\conE$ s.t. $n_\conE <\KN$,$\con\xrightarrow*\conD\xrightarrow* \conE$, $\vc x_\conE\in \clsw^{p_\conE,n_\conE}$ and $t < \dist(\conD)<\infty$.
\item \label{polytest} there exists configurations $\conD,\conE$ and a run $\con\xrightarrow{*}\conD\xrightarrow{*}\conE$ s.t. $\KN^2+\KN \leq n_\conD \leq 2\KN^2+\KN$, $\vc x_\conE\in \clsw^{p_\conE,n_\conE}$ with $n_\conE <\KN$.}
\end{enumerate}
\end{lemma}
\begin{proof}
\ref{polytest} $\rightarrow$ \ref{arbitt}: Consider an arbitrary $q \in C$, $m < \KN$ and vector space $\lsv = \lsw^{q,m}$. Let us assume for contradiction the complement of Point \ref{arbitt}. That is, there exists a $t \in \N$ such that for all configurations $\conD^\prime$ where $\con \xrightarrow{*} \conD^\prime \xrightarrow{*} \clsv \times \{q\} \times \{m\}$, $\dist(\conD^\prime)\leq t$. Note that for all $\conD^\prime$ where $n_{\conD^\prime} > \KN$, $\dist(\conD^\prime) \geq n_{\conD^\prime} - \KN$. Hence there exists an $M\in\N$ such that for all $\conD^\prime$ where $\con \xrightarrow{*} \conD^\prime \xrightarrow{*} \clsv \times \{q\} \times \{m\}$, $n_{\conD^\prime} \leq M$. 

Consider a configuration $\conD$ where $n_\conD > \KN^2+\KN$ and a run $\con \xrightarrow{*} \conD \xrightarrow{*} \clsv \times \{q\} \times \{m\}$. Point \ref{polytest} shows the existence of such a run. From \Cref{lem:uturn}, Point \ref{pumpup}, we get that there exists a $\conD^\prime$ such that $\con \xrightarrow{*} \conD^\prime \xrightarrow{*} \clsv \times \{q\} \times \{m\}$ and $n_{\conD^\prime} > n_\conD$, which is a contradiction.

\ref{arbitt} $\rightarrow$ \ref{notreg}:
Assume for contradiction that for all $t\in\N$, there exists {
configurations $\conD,\conE$ such that $\con\xrightarrow*\conD\xrightarrow* \conE$, $\vc x_\conE\in \clsw^{p_\conE,n_\conE}, n_\conE <\KN$ and $t < \dist(\conD)<\infty$} but $\Autom$ is regular. Let $\Butom$ be the weighted automaton equivalent to $\Autom$. We use $|\Butom|$ to represent the number of states of $\Butom$.

Let $t_1,t_2, \ldots t_{|\Butom|+1} \in \N$ such that for $i \in [1,|\Butom|]$, $t_i < t_{i+1}$, and {$\conD_{t_i}$ be such that $\con\xrightarrow*\conD_{t_i}\xrightarrow*(\vc x_{i},p_\conE,n_\conE)$, $\vc x_i \in \clsw^{p_\conE,n_\conE}$ and $t_i < \dist(\conD_{t_i})< t_{i+1}<\infty$. Clearly $\conD_{t_i}\not\equiv \conD_{t_j}$ }for $i \neq j$ and hence corresponds to two different states of $\Butom$. Since we can find more than $|\Butom|$ pairwise non-equivalent configurations, it contradicts the assumption that $\Butom$ is equivalent to $\Autom$.

\ref{notreg} $\rightarrow$ \ref{polytest}: We prove the contrapositive of the statement. Let us assume that there is no {configurations
$\conD,\conE$ and a run $\con\xrightarrow{*}\conD\xrightarrow{*}\conE$ such that $\KN^2+\KN \leq n_\conD \leq 2\KN^2+\KN$, $\vc x_\conE\in \clsw^{p_\conE,n_\conE}$ with $n_\conE <\KN$. This implies that there does not exists a configuration $\conD^\prime$ such that $n_{\conD^\prime}>2\KN^2$, $\con \xrightarrow{*}\conD^\prime \xrightarrow{*} (\vc y,p_\conE,n_\conE)$ for some $\vc y\in \clsw^{p_\conE,n_\conE}$.
Assume for contradiction that there is such a run, then there exists a portion in this run that can be ``pumped down" to get a run $\con \xrightarrow{*}\conD^{\prime\prime} \xrightarrow{*} (\vc y^\prime,p_\conE,n_\conE)$ for some configuration $\conD^{\prime\prime}$ such that $\KN^2+\KN \leq n_{\conD^{\prime\prime}} \leq 2\KN^2+\KN$ and $\vc y^\prime\in \clsw^{p_\conE,n_\conE}$. This is a contradiction. Therefore
} all runs starting from configuration with counter value greater than or equal to $\KN^2+\KN$ ``looks'' similar to a run on a weighted automaton. This allows us to simulate the runs of $\Autom$ using a weighted automaton. 
\end{proof}
\begin{theorem} \label{regularity}
There is a polynomial time algorithm to decide whether a weighted \odca is equivalent to some weighted automata.
\end{theorem}
\begin{proof}
%[Proof of \Cref{regularity}]
Let $\Autom$ be a weighted \odca. \Cref{increasingdist} shows that if $\Autom$ is not regular, then there are words $u,v \in \Sigma^*$ { and configurations $\conD,\conE$ such that there is a run of the form $\con\xrightarrow{u}\conD\xrightarrow{v}\conE$ such that $\KN^2+\KN \leq \cntval{\conD} \leq 2\KN^2+\KN$, $\wgtvec{\conE}\in \clsw^{\cntstate{\conE},\cntval{\conE}}$ with $\cntval{\conE} <\KN$.} The existence of such words $u$ and $v$ can be decided in polynomial time since the minimal length of such a path if it exists, is polynomially bounded in the number of states of the weighted \odca by \Cref{smp}. This concludes the proof.
\end{proof}
%{The proof of the remaining cases and of \Cref{regularity} are in Appendix.}
%\prince{The direction \ref{polytest} $\rightarrow$ \ref{arbitt} is the non-trivial one. A detailed proof is given in the Appendix. The other directions are similar to the non-weighted case presented in \cite{bisim}.}
%We now prove that the regularity problem for weighted \odca is decidable in polynomial time.

%% file: TeX/covering.tex
% !TEX root = ../odca.tex
\section{Covering}
\label{sec:covering}
%{In this section, we show the existence of polynomial time algorithms for covering and coverable equivalence problems of uninitialised weighted \odcas.}
%\sav{Again, what is special about this section? Mention what is going to be our proof technique in simple English.}
%
%For an uninitialised weighted \odca $\Autom$ and alphabet $\Sigma$, and a configuration $\con_0= (\vc x_0, p_0,0)$, we define $\un{\Autom}{\con_0}$ as the weighted \odca with transitions and final state as defined in $\Autom$ and the initial configuration $\con_0$.
%

Let $\Autom_1$ and $\Autom_2$ be two uninitialised weighted \odcas. We say $\Autom_2$ \emph{covers} $\Autom_1$ if for all initial configurations $\con_0$ of $\Autom_1$ there exists an initial configuration $\conD_0$ of $\Autom_2$ such that 
$\un{\Autom_1}{\con_0}$ and $\un{\Autom_2}{\conD_0}$ are equivalent.
%
%For an initial configuration $\con_0$ of $\Autom_1$, we say that $\Autom_2$ {covers} $\un{\Autom_1}{\con_0}$ if there exists a configuration $\conD_0$ of $\Autom_2$ such that $\un{\Autom_1}{\con_0}$ and $\un{\Autom_2}{\conD_0}$ are equivalent. The \emph{covering} problem asks whether $\Autom_2$ covers $\un{\Autom_1}{\con_0}$ for all initial configurations $\con_0$ of $\Autom_1$.
%We say that $\Autom_1$ covers $\Autom_2$ if for any initial configuration $c=(p,0,s)$ in $\Autom_1$, there is an initial configuration $\bar c = (q,0,t)$ in $\Autom_2$ such that $\Autom_1(c)$ and $\Autom_2(\bar c)$ are equivalent.
We say $\Autom_1$ and $\Autom_2$ are \emph{coverable equivalent} if $\Autom_1$ covers $\Autom_2$, and $\Autom_2$ covers $\Autom_1$.
We show that the covering and coverable equivalence problems for uninitialised weighted \odcas are decidable in polynomial time.
The proof relies on the algorithm to check the equivalence of two weighted \odcas. % and is given in Appendix.
% (\Cref{main}) developed in \Cref{sec:ptime}. %To check whether $\Autom_2$ covers $\Autom_1$, we check for the existence of equivalent initial distributions of $\Autom_2$ for all initial distributions of $\Autom_1$ for a fixed counter states, where only a single state has an initial weight one. If there exists such an initial distribution of $\Autom_1$ for which $\Autom_2$ does not have an equivalent initial distribution, then $\Autom_2$ does not cover $\Autom_1$. Otherwise, we say that $\Autom_2$ covers $\Autom_1$.}

%First, we prove the following lemma.
%\begin{lemmarep}
%\label{coversplit}
%There is a polynomial time algorithm to decide whether $\Autom_2$ covers $\un{\Autom_1}{\conH_{j,q}}$ for any $j\in[1,\K]$ and $q\in C_1$.
%\end{lemmarep}
%\begin{proof}
%\end{proof}
%The proof is in the Appendix. We now prove that covering and coverable equivalence are decidable in polynomial time. 
\begin{theorem}
Covering and coverable equivalence problems of uninitialised weighted \odcas are in $\CF{P}$.
\end{theorem}
\begin{proof}
We fix two uninitialised weighted \odcas $\Autom_1=(C_1, \delta_1;\ Q_1, \Delta_1, \vc\eta_1)$ and $\Autom_2=(C_2, \delta_2;\ Q_2, \Delta_2, \vc\eta_2)$ for this section. Without loss of generality, assume $\K = |C_1| = |Q_1|= |C_2| = |Q_2|$.
For $i\in[1,\K]$ we define the vector $\vc e_i\in\Sring^\K$ as follows:
\[ \vc e_i[j]= \begin{cases} 1, \text{ if } i=j\\
0, \text{ otherwise}
\end{cases}
\]
For $j\in[1,\K]$, $q\in C_1$, we use $\conH_{j,q}$ to denote the configuration $(\vc e_j,q,0)$ of $\Autom_1$ and for $i\in[1,\K]$, $p\in C_2$, we use $\conG_{i,p}$ to denote the configuration $(\vc e_i,p,0)$ of $\Autom_2$.
\begin{clam}
\label{coversplit}
There is a polynomial time algorithm to decide whether $\Autom_2$ covers $\un{\Autom_1}{\conH_{j,q}}$ for any $j\in[1,\K]$ and $q\in C_1$.
\end{clam}
\begin{clamproof}
First, we check, in polynomial time (equivalence with a zero machine), whether $\un{\Autom_1}{\conH_{j,q}}$ accepts all words with weight $0 \in \Sring$. If that is the case, then $\un{\Autom_1}{\conH_{j,q}}$ and $\un{\Autom_2}{\conG_0}$ are equivalent for the configuration $\conG_0 = (\{0\}^\K,p,0)$, for any $p\in C_2$.
Otherwise, there is some word $w_0$ accepted by $\un{\Autom_1}{\conH_{j,q}}$ with non-zero weight $s$ (returned by the previous equivalence check). Without loss of generality, we consider the smallest one, whose size is polynomial in $\K$.

We pick a $p\in C_2$ and check whether there exists an initial distribution from counter state $p$ that makes the two machines equivalent. Assume that such an initial distribution exists and for all $i\in[1,\K]$, let $\alpha_i$ denote the initial weight on state $q_i \in Q_2$. We use $\vc \alpha$ to denote the resultant initial distribution. We initialise an empty set $B$ to store a system of linear equations.

The following steps will be repeated at most $\K$ times to check the existence of an initial distribution with initial state $p\in C_2$.
Let $w$ be the counter-example returned by the equivalence query in the previous step.
For all $i\in[1,\K]$, we compute $f_{\un{\Autom_2}{\conG_{i,p}}}(w)$.
We add the linear equation $\sum_{i=1}^\K \alpha_i \cdot f_{\un{\Autom_2}{\conG_{i,p}}}(w) = f_{\un{\Autom_1}{\conH_{j,q}}}(w)$ to $B$ and compute values for $\alpha_i$, $i\in[1,\K]$, such that it satisfies the system of linear equations in $B$. We check whether $\un{\Autom_1}{\conH_{j,q}}$ and $\un{\Autom_2}{({\alpha},{p},{0})}$ are equivalent or not. If they are not equivalent, we get a new counter example that distinguishes them. Now we repeat the procedure to compute a new initial distribution.

Note that the above procedure is executed at most $\K$ times to find an initial distribution if it exists. This is because we can find only $\K$ many linearly independent linear equations in $\K$ variables. Suppose the above procedure fails to find an initial distribution for which the machines are equivalent. In that case, there is an initial distribution of $\Autom_1$ with initial counter state $q$, for which $\Autom_2$ with initial counter state $p$ does not have an equivalent initial distribution.
We now pick a different counter state of $C_2$ and repeat the process until we find a $p\in C_2$ for which the algorithm finds an equivalent initial distribution. If for all $p\in C_2$, the algorithm returns false, then $\Autom_2$ does not cover $\un{\Autom_1}{\conH_{j,q}}$.
\end{clamproof}
First, we show the existence of a polynomial time procedure to check whether $\Autom_2$ covers $\Autom_1$.
For all $j\in[1,\K]$, we check whether there exists an initial state $p\in C_2$ such that $\Autom_2$ with initial counter state $p$ has an initial distribution that makes it equivalent to $\un{\Autom_1}{\conH_{j,q}}$ using Claim \ref{coversplit}. If we fail to find such a state in $C_2$ then we return false. We repeat this procedure for all $q\in C_1$. If for all $q\in C_1$ there exists a $p\in C_2$ such that $\Autom_2$ with initial counter state $p$ has an initial distribution that makes it equivalent to $\un{\Autom_1}{\conH_{j,q}}$ for all $j\in [1,K]$, then we say that $\Autom_2$ covers $\Autom_1$ otherwise we say that $\Autom_2$ does not cover $\Autom_1$. 
{Let us see why this is true. 
For simplifying the arguments we fix a $q\in C_1$. Assume that for all $j\in[1,\K]$, there exits  $p\in C_2$ such that $\un{\Autom_1}{\conH_{j,q}}$ is equivalent to the configuration $(\vc x_{j,q}, p, 0)$ for some $\vc x_{j,q}\in\Sring^\K$. Now, any initial configuration $(\vc \lambda, q, 0)$ of $\Autom_1$ is equivalent to the configuration $(\sum_{j=1}^\K \vc \lambda[j] \cdot \vc x_{j,q}, p ,0)$  of $\Autom_2$.}

The coverable equivalence problem can now be solved by checking whether $\Autom_1$ covers $\Autom_2$ and $\Autom_2$ covers $\Autom_1$, which can be done in time polynomial in $\K$.
\end{proof}
% \prince{add conclusion. short summary of results, future work (learning, non-realtime \dots}

%% file: TeX/nodca.tex
% !TEX root = ../odca.tex
\section{Non-deterministic ODCA}
\label{sec:nondet}
%The weights in a weighted \odca need not necessarily be from a field. A natural model to consider is non-deterministic \odca, where the weights are from the boolean semiring.
In this section, we consider the counter-determinacy restriction over weightless \textsc{oca}s (equivalently, with weights from the boolean semiring). These results do not follow from previous sections, as booleans are not a field.
\begin{example} 
\label{examplesnodca}
The following languages are defined over the alphabet $\Sigma=\{a,b\}$ and are recognised by non-deterministic \textsc{oca} with counter-determinacy. 
\begin{enumerate}[(a)]
\item The language $\mathcal{L}_1=\{a^nba^n\mid n >0\}$. 
\item The language $\mathcal{L}_2=\{(a+b)^*\mid$ number of a's is greater than number of b's$\}$.
\item {The language $\mathcal{L}_3= \{a^n(b+c)^m b (b+c)^k \mid m,n\in\N \text{ and } m>n\}$. }
\end{enumerate}
%\label{eg:oca-cd2}
\end{example}
Note that none of the above languages are definable by visibly pushdown automata. The \odcas corresponding to these languages are given in \Cref{fig:examples}.

\begin{figure}
   \centering
   %   \hfill
   \fbox{
   \begin{subfigure}[b]{0.45\textwidth}
    \centering\scalebox{.6}{
\begin{tikzpicture}[shorten >=1pt,node distance=3cm,on grid,auto,
every state/.style={draw,thick},
    every edge/.style={draw,thick}]
    \tikzset{every path/.style={line width=.3mm}}
\node[state,initial,initial text=] at (0,0) (q_0) {$q_0$};
\node[state] at (5.5,0) (q_1) {$q_1$};
\node[state] at (5.5,-2) (q_2) {$q_2$};
\path[->]
(q_0) edge [loop above] node {$(\{a\},\{0,1\})$} ()
%edge [loop below] node [xshift=-.4cm]{$(\{b,c\},\{1\},-1)$} ()
edge [above] node {$(\{b\},\{1\})$} (q_1)
(q_1) edge [loop above] node {$(\{a\},\{1\})$} ()
edge [above] node[xshift=.9cm] {$(\{a\},\{0\})$} (q_2);
\draw[<-] (4.6,-2)-- (5.1,-2);
\draw (-.6,-2) node {{Finite state machine}};
\draw (-.6,-3) node {{Counter structure}};
%\draw[<-] (.6,-4)-- (1.1,-4) node[anchor=east,xshift=-.5cm]{1};
%
\node[state,initial,initial text=] at (0,-5) (p_0) {$p_0$};
\node[state] at (5.3,-5) (p_1) {$p_1$};
\node[state] at (5.3,-7) (p_2) {$p_2$};
\path[->]
(p_0) edge [loop above] node {$(\{a\},\{0,1\},+1)$} ()
edge [above] node {$(\{b\},\{1\},-1)$} (p_1)
(p_1) edge [loop above] node {$(\{a\},\{1\},-1)$} ()
edge [left] node {$(\{a\},\{0\},0)$} (p_2);
\end{tikzpicture}}
\caption{$\mathcal{L}_1=\{a^nba^n\mid n >0\}$}
\label{figureex2}
   \end{subfigure}
   }
   \hfill
   \fbox{
   \begin{subfigure}[b]{0.45\textwidth}
    \centering\scalebox{.6}{
\begin{tikzpicture}[shorten >=1pt,node distance=3cm,on grid,auto,
every state/.style={draw,thick},
    every edge/.style={draw,thick}]
    \tikzset{every path/.style={line width=.3mm}}
    \node[state, initial,initial text=] at (-3,0) (q_0) {$q_0$};
\node[state] at (0,0) (q_1) {$q_1$};
\node[state] at (4.5,0) (q_2) {$q_2$};

\path[->]
(q_0) edge [above] node {$(\{a\},\{0\},+1)$} (q_1)
(q_1) edge [loop above] node {$(\{a\},\{0,1\})$} ()
edge [loop below] node {$(\{b\},\{1\})$} ()
%edge [loop below] node [xshift=-.4cm]{$(\{b,c\},\{1\},-1)$} ()
edge [bend left] node {$(\{b\},\{0\})$} (q_2)
(q_2) edge [loop above] node {$(\{a\},\{1\})$} ()
edge [bend left] node[xshift=.9cm] {$(\{a\},\{0\})$} (q_1);
\draw[->] (-.3,-.3)-- (-.7,-.7);
\draw (-.6,-2) node {{Finite state machine}};
\draw (-.6,-3.8) node {{Counter structure}};
%\draw[<-] (.6,-4)-- (1.1,-4) node[anchor=east,xshift=-.5cm]{1};
%
\node[state,initial,initial text=] at (-2,-5.8) (p_0) {$p_0$};
\node[state] at (3.3,-5.8) (p_1) {$p_1$};
\path[->]
(p_0) edge [loop above] node {$(\{a\},\{0,1\},+1)$} ()
edge [loop below] node {$(\{b\},\{1\},-1)$} ()
edge [bend left] node {$(\{b\},\{0\},+1)$} (p_1)
(p_1) edge [loop above] node {$(\{a\},\{1\},-1)$} ()
edge [loop below] node {$(\{b\},\{0,1\},+1)$} ()
edge [bend left] node {$(\{a\},\{0\},+1)$} (p_0);
\end{tikzpicture}}
\caption{$\mathcal{L}_2=\{(a+b)^*\mid$ \#a's $>$ \#b's$\}$}
\label{figureex3}
   \end{subfigure}
   }
    \hfill
\fbox{
   \begin{subfigure}[b]{0.45\textwidth}
    \centering\scalebox{.5}{
\begin{tikzpicture}[shorten >=1pt,node distance=3cm,on grid,auto,
every state/.style={draw,thick},
    every edge/.style={draw,thick}]
\node[state,initial,initial text=] at (0,0) (q_0) {$q_0$};
\node[state] at (5.5,0) (q_1) {$q_1$};
\node[state] at (2,-2) (q_2) {$q_2$};
\node[state] at (5.5,-2) (q_3) {$q_3$};
\node[state] at (5.5,-4) (q_4) {$q_4$};
\node[state] at (1.5,-4) (q_5) {$q_5$};
\path[->]
(q_0) edge [loop above] node {$(\{a\},\{0,1\})$} ()
%edge [loop below] node [xshift=-.4cm]{$(\{b,c\},\{1\},-1)$} ()
edge [above] node {$(\{b,c\},\{0\})$} (q_1)
edge [above] node[xshift=.8cm,yshift=.2cm] {$(\{b,c\},\{1\})$} (q_2)
(q_1) edge [loop above] node {$(\{b,c\},\{0\})$} ()
edge [above] node[xshift=.9cm] {$(\{b\},\{0\})$} (q_3)
(q_2) edge [loop below] node {$(\{b,c\},\{1\})$} ()
edge [below] node[xshift=0cm,yshift=-.4cm] {$(\{b,c\},\{0\})$} (q_1)
(q_3) edge [above] node[xshift=1.1cm] {$(\{b,c\},\{0\})$} (q_4)
(q_4) edge [below] node {$(\{b,c\},\{0\})$} (q_5);
\draw (-2.2,-2) node[rotate=90] {{Finite state machine}};
\draw (-2.2,-8) node[rotate=90] {{Counter structure}};
\draw[<-] (.6,-4)-- (1.1,-4);
%\draw[<-] (.6,-4)-- (1.1,-4) node[anchor=east,xshift=-.5cm]{1};
%
\node[state,initial,initial text=] at (0,-8.95) (p_0) {$p_0$};
\node[state] at (5.3,-8.95) (p_1) {$p_1$};
\path[->]
(p_0) edge [loop below] node {$(\{b,c\},\{1\},-1)$} ()
edge [loop above] node {$(\{a\},\{0,1\},+1)$} ()
edge [above] node {$(\{b\},\{0\},0)$} (p_1)
(p_1) edge [loop above] node {$(\{b,c\},\{0,1\},0)$} ();
\end{tikzpicture}}
\caption{The language $\mathcal{L}_3= \{a^n(b+c)^m b (b+c)^k \mid m,n \in \N$ and $m>n\}$ for $k=2$ is recognised by the above non-deterministic \odca.}
\label{figureex}

   \end{subfigure}
   }
    \caption{The figure gives \odcas corresponding to examples (a), (b), and (c) given in \Cref{examplesnodca}. Let $A\subseteq \Sigma, R\subseteq\{0,1\}$ and $D \in\{-1,0,+1\}$ are non-empty sets. For $i,j\in\N$, if a transition from $q_i$ to $q_j$ is labelled $(A,R)$ and $(a,r)\in A \times R$, then there is a transition from $q_i$ to $q_j$ on reading the symbol $a$. The current counter value should be $0$ if $r=0$ and greater than $0$ if $r=1$. Similarly, if a transition from $p_i$ to $p_j$ is labelled $(A,R,D)$ and $(a,r,d)\in A \times R \times \{D\}$, then there is a transition from $p_i$ to $p_j$ on reading the symbol $a$ with counter action $d$. The current counter value should be $0$ if $r=0$ and greater than $0$ if $r=1$.}
    \label{fig:examples}
\end{figure}

We observe that the relationship between non-deterministic and deterministic \odcas is similar to that between non-deterministic and deterministic finite automata. By definition, deterministic \odcas have at most one unique path for any fixed word. Therefore, they are deterministic \textsc{oca}s with counter-determinacy. It is also easy to observe that deterministic \textsc{oca}s are deterministic \odcas. It follows that deterministic \odcas and deterministic \textsc{oca}s are expressively equivalent. Similar to non-deterministic finite automata, we observe that non-deterministic \odcas can be determinised by a subset construction of the states of the finite state machine. However, this results in an exponential blow-up. In \Cref{examplesnodca}, the deterministic \odca that recognises the language $\mathcal{L}_3$ has to check whether every $b$ encountered after reading the word $a^n(b+c)^{n+1}$ is at the $k^{th}$ position from the end. This will require at least $\mathcal{O}(2^k)$ states. On the other hand, there is a non-deterministic \odca with $\mathcal{O}(k)$ states recognising the same language. Similar to finite automata, non-deterministic \odcas are a ``succinct'' way to represent deterministic \textsc{oca}s. %\sav{See if the example is correct.}

%\begin{toappendix}
\label{appendixNonDet}
A deterministic/non-deterministic \odca $\Autom$ is an \odca over the boolean semiring $\Semi= (\{0,1\},\lor, \land)$. The language recognised by $\Autom$ is given by $\mathcal{L}(\Autom)= \{w \mid f_\Autom(w)=1\}$.
%\prm{Need to change. Different acceptence condition.} 
We say an \odca $\Autom =((C, \delta_0,\delta_1, p_0),(Q, \vc \lambda, \Delta, \vc\eta))$ is \emph{deterministic} if for every transition sequence $T = \tau_0\cdots \tau_{\ell-1}$, the vector $\vc\lambda\we(T)$ contains exactly one $1$ %\lVert\vc\lambda {\we(T)}\rVert \leq 1$, here $\lVert\vc\lambda {\we(T)}\rVert$ is defined as xxx (sum of the entries in the vector) 
 and non-deterministic otherwise.
 
For every language recognised by a non-deterministic \odca, there is a deterministic \odca of at most exponential size that recognises it.
The idea is a simple subset construction (see \Cref{detoca}).
%\end{toappendix}

%For every language recognised by a non-deterministic \odca, there is a deterministic \odca of at most exponential size that recognises it.
\begin{theorem}[Determinisation]
\label{detoca}
Given a non-deterministic \odca, a polynomial space machine can output an equivalent deterministic \odca of exponential size.
\end{theorem}
\begin{proof}
Let $\Autom=((C, \delta_0, \delta_1, p_0),(Q, \vc \lambda, \Delta, \vc\eta))$ be a non-deterministic \odca.
{
Given a vector $\vc x \in \Semi^k$ for some $k\in\N$, we define the function IsDet: $\Semi^k \to \{true,false\}$ as follows:
$$\text{IsDet}(\vc x)= \begin{cases} \text{true, if } \exists i<k \text{ s.t } \vc x[i]=1 \text{ and } \forall j\neq i,  \vc x[i]=0\\
\text{false, otherwise.}
\end{cases}$$
Given a transition matrix $\mata$ corresponding to the states $Q$, we define its determinisation $\text{det}(\mata)$ as follows.  There are rows and columns corresponding to each set in $2^Q$. For any $q_i \in Q$, let $\mathcal{M}(q_i,\mata) = \{ q_j \mid \mata[i][j] = 1 \}$ be the set of all states in the row of $q_i$ whose entries are $1$. With the notation that ${\det}(\mata)[s][s']$ corresponds to the entry of the cell corresponding to the sets $s,s' \in 2^Q$,  we let ${\det}(\mata)[s][s'] = 1$ if and only if $s' = \bigcup_{q \in s}  \mathcal{M}(q_i,\mata)$. We claim that $\Autom_{\det} =((C, \delta_0,\delta_1, p_0),(Q, \vc \lambda, \Delta', \vc\eta'))$, with $\vc\eta'$ such that for any $S \in 2^Q$,$\vc\eta'[S] = \bigvee_{s \in S} \vc\eta[s]$ and {for all $a\in\Sigma$ and $d\in\{0,1\}$,}  $\Delta' (a,d ) = \det(\Delta(a,d ))$ is such that it is deterministic and $\mathcal{L}(\Autom) = \mathcal{L}(\Autom_{\det}) $.

For this,  for any sequence of operations $T = \tau_0\cdots \tau_{\ell-1}$, let $\vc v_T, \vc v^\prime_T$ be the vectors corresponding to $\vc\lambda {\we(T)}$ in $\Autom$ and $\Autom_{\det}$ respectively. Then we have {IsDet$(\vc v^\prime_T) = 1$} and for any $S \in 2^Q$, $\vc v^\prime_T[S] = 1$ if and only if for all $q_i \in S$, $\vc v_T[i] = 1$.}
\end{proof}
%The idea in proving the above theorem is a simple subset construction (proof in \Cref{appendixNonDet}).
The above result and the fact that equivalence of deterministic \odca is in NL gives us the upper bound in the following theorem. The lower bound follows from that of NFAs~\cite{eqnfa}. 
\begin{restatable}{theorem}{restatetest}
The equivalence problem for non-deterministic \odca is $\CF{PSPACE}$-complete.
\end{restatable}
The equivalence of non-deterministic \textsc{oca} is undecidable~\cite{doca}. Our theorem shows that undecidability is due to non-determinism in the component that modifies the counter.